\documentclass[onecolumn,amsmath,showpacs,11pt]{revtex4}
\usepackage{graphicx}
\usepackage{dcolumn}
\usepackage{bm}
\usepackage{color}
\begin{document}
\newcommand{\hs}{\hspace*{0.5cm}}
\newcommand{\vs}{\vspace*{0.5cm}}
\newcommand{\be}{\begin{equation}}
\newcommand{\ee}{\end{equation}}
\newcommand{\bea}{\begin{eqnarray}}
\newcommand{\eea}{\end{eqnarray}}
\newcommand{\ben}{\begin{enumerate}}
\newcommand{\een}{\end{enumerate}}
\newcommand{\bde}{\begin{widetext}}
\newcommand{\ede}{\end{widetext}}
\newcommand{\nn}{\nonumber}
\newcommand{\crn}{\nonumber \\}
\newcommand{\Tr}{\mathrm{Tr}}
\newcommand{\non}{\nonumber}
\newcommand{\noi}{\noindent}
\newcommand{\al}{\alpha}
\newcommand{\la}{\lambda}
\newcommand{\bet}{\beta}
\newcommand{\ga}{\gamma}
\newcommand{\va}{\varphi}
\newcommand{\om}{\omega}
\newcommand{\pa}{\partial}
\newcommand{\+}{\dagger}
\newcommand{\fr}{\frac}
\newcommand{\bc}{\begin{center}}
\newcommand{\ec}{\end{center}}
\newcommand{\Ga}{\Gamma}
\newcommand{\de}{\delta}
\newcommand{\De}{\Delta}
\newcommand{\ep}{\epsilon}
\newcommand{\varep}{\varepsilon}
\newcommand{\ka}{\kappa}
\newcommand{\La}{\Lambda}
\newcommand{\si}{\sigma}
\newcommand{\Si}{\Sigma}
\newcommand{\ta}{\tau}
\newcommand{\up}{\upsilon}
\newcommand{\Up}{\Upsilon}
\newcommand{\ze}{\zeta}
\newcommand{\ps}{\psi}
\newcommand{\Ps}{\Psi}
\newcommand{\ph}{\phi}
\newcommand{\vph}{\varphi}
\newcommand{\Ph}{\Phi}
\newcommand{\Om}{\Omega}

\title{Phenomenology of the 3-3-1-1 model}

\author{P. V. Dong}
\email {pvdong@iop.vast.ac.vn} \affiliation{Institute of Physics, Vietnam Academy of Science and Technology, 10 Dao Tan, Ba Dinh, Hanoi, Vietnam}
\author{D. T. Huong}
\email {dthuong@iop.vast.ac.vn} \affiliation{Institute of Physics, Vietnam Academy of Science and Technology, 10 Dao Tan, Ba Dinh, Hanoi, Vietnam}
\author{Farinaldo S. Queiroz}
\email{fdasilva@ucsc.edu}\affiliation{Department of Physics and Santa Cruz Institute for Particle Physics, University of California, Santa Cruz, CA 95064, USA}
\author{N. T. Thuy}
\email {ntthuy@iop.vast.ac.vn} \affiliation{Department of Physics and IPAP, Yonsei University, Seoul 120-479, Korea}

\date{\today}

\begin{abstract}
In this work we discuss a new $SU(3)_C\otimes SU(3)_L \otimes U(1)_X \otimes U(1)_N$ (3-3-1-1) gauge model that overhauls the theoretical and phenomenological aspects of the known 3-3-1 models. Additionally, we sift the outcome of the 3-3-1-1 model from precise electroweak bounds to dark matter observables. We firstly advocate that if the $B-L$ number is conserved as the electric charge, the extension of the standard model gauge symmetry to the 3-3-1-1 one provides a minimal, self-contained framework that unifies all the weak, electromagnetic and $B-L$ interactions, apart from the strong interaction. The $W$-parity (similar to the $R$-parity) arises as a remnant subgroup of the broken 3-3-1-1 symmetry. The mass spectra of the scalar and gauge sectors are diagonalized when the scale of the 3-3-1-1 breaking is compatible to that of the ordinary 3-3-1 breaking. All the interactions of the gauge bosons with the fermions and scalars are obtained. The standard model Higgs ($H$) and gauge ($Z$) bosons are realized at the weak scales with consistent masses despite of their mixings with the heavier particles, respectively. The 3-3-1-1 model  provides two dark matters which are stabilized by the $W$-parity conservation: one fermion which may be either a Majorana or Dirac fermion and one complex scalar. We conclude that in the fermion dark matter setup the $Z_2$ gauge boson resonance sets the dark matter observables, whereas in the scalar one the Higgs portal dictates them. The standard model GIM mechanism works in the model because of the $W$-parity conservation. Hence, the dangerous flavor changing neutral currents due to the ordinary and exotic quark mixing are suppressed, while those coming from the non-universal couplings of the $Z_2$ and $Z_N$ gauge bosons are easily evaded. Indeed, the $K^0-\bar{K}^0$ and $B^0_s-\bar{B}^0_s$ mixings limit $m_{Z_{2,N}}>2.037$~TeV and $m_{Z_{2,N}}>2.291$~TeV, respectively, while the LEPII searches provide a quite close bound $m_{Z_{2,N}}>2.737\ \mathrm{TeV}$. The violation of the CKM unitarity due to the loop effects of the $Z_{2}$ and $Z_{N}$ gauge bosons is negligible.

\end{abstract}


\maketitle

\section{\label{intro}Introduction}

The standard model \cite{pdg} has been extremely successful. However, it describes only about 5\%~mass-energy density of our universe. There remain around 25\% dark matter and 70\% dark energy that are referred as the physics beyond the standard model. In addition, the standard model cannot explain the nonzero small masses and mixing of the neutrinos, the matter-antimatter asymmetry of the universe, and the inflationary expansion of the early universe. On the theoretical side, the standard model cannot show how the Higgs mass is stabilized against radiative corrections, what makes the electric charges exist in discrete amounts, and why there are only the three generations of fermions observed in the nature.   

Among the standard model's extensions for the issues, the recently-proposed $SU(3)_C\otimes SU(3)_L\otimes U(1)_X\otimes U(1)_N$ (3-3-1-1) gauge model has interesting features~\cite{3311}. (i) The theory arises as a necessary consequence of the 3-3-1 models \cite{331m,331r,dongfla} that respects the conservation of lepton and baryon numbers. (ii) The $B-L$ number is naturally gauged because it is a combination of the $SU(3)_L$ and $U(1)_N$ charges. And, the resulting theory yields an unification of the electroweak and $B-L$ interactions, apart from the strong interaction. (iii) The right-handed neutrinos are emerged as fundamental fermion constituents, and consequently the small masses of the active neutrinos are generated by the type I seesaw mechanism. (iv)~The $W$-parity which has the form similarly to the $R$-parity in supersymmetry is naturally resulted as a conserved remnant subgroup of the broken 3-3-1-1 gauge symmetry. (v) The dark matter automatically exists in the model that is stabilized due to the $W$-parity. It is the lightest particle among the new particles that characteristically have wrong lepton numbers transforming as odd fields under the $W$-parity (so-called $W$-particles). The dark matter candidate may be a neutral fermion ($N$) or a neutral complex scalar ($H'$).  

The 3-3-1-1 model includes all the good features of the 3-3-1 models. Namely, the number of fermion families is just three as a consequence of anomaly cancelation and QCD asymptotic freedom condition \cite{anoma}. The third quark generation transforms under $SU(3)_L$ differently from the first two. This explains why the top quark is uncharacteristically-heavy \cite{tquark}. The strong $CP$ problem is solved by just its particle content with an appropriate Peccei-Quinn symmetry~\cite{palp}. The electric charge quantization is due to a special structure of the gauge symmetry and fermion content~\cite{ecq}. Additionally, it also provides the mentioned dark matter candidates similarly to~\cite{farinaldoDM1,farinaldoDM2}. The 3-3-1-1 model can solve the potential issues of the 3-3-1 models because the unwanted interactions and vacuums that lead to the dangerous tree-level flavor changing neutral currents (FCNCs) \cite{ponce} as well as the $CPT$ violation \cite{cpt} are all suppressed due to the $W$-parity conservation \cite{3311}.  

In the previous work \cite{3311}, the proposal of the 3-3-1-1 model with its direct consequence---the dark matter has been given. In the current work, we will deliver a detailed study of this new model. Particularly, we consider the new physics consequences besides the dark matter that are implied by the new extended sectors beyond those of the 3-3-1 model. These sectors include the new neutral gauge boson ($C$) as associated with $U(1)_N$ and the new scalar ($\phi$) as required for the totally $U(1)_N$ breaking with necessary mass generations. The totally $U(1)_N$ breaking that consequently breaks the $B-L$ symmetry, where the $B-L$ is a residual charge related to the $N$ charge and a $SU(3)_L$ generator, can happen closely to the 3-3-1 breaking scale of TeV order. This leads to a finite mixing and interesting interplay between the new neutral gauge bosons such as the $Z'$ of the 3-3-1 model and the $C$ of $U(1)_N$. Notice that our previous work considers only a special case when the $B-L$ breaking scale is very high like the GUT one \cite{gutscale} as an example so that the new physics over the ordinary 3-3-1 symmetry is decoupled, which has neglected its imprint at the low energy \cite{3311}. Indeed, the stability of the proton is already ensured by the 3-3-1-1 gauge symmetry, there is no reason why that scale is not presented at the 3-3-1 scale. Similarly to the new neutral gauge bosons, there is an interesting mixing among the new neutral scalars that are used to break the above symmetry kinds, the 3-3-1 and the $B-L$. 

It is interesting to note that the new scalars and new gauge bosons as well as the new fermions can give significant contributions to the production and decay of the standard model Higgs boson. They might also modify the well-measured standard model couplings such as those of the photon, $W$ and $Z$ bosons with the fermions. There exist the hadronic FCNCs due to the contribution of the new neutral gauge bosons. These gauge bosons can also take part in the electron-positron collisions such as the LEPII and ILC as well as in the dark matter observables.  The presence of the new neutral gauge bosons also induces the apparent violation of the CKM unitarity. In some case, the new scalar responsible for the $U(1)_N$ breaking may act as an inflaton. The decays of some new particles can solve the matter-antimatter asymmetry via leptogenesis mechanisms.   

The scope of this work is given as follows. The 3-3-1-1 model will be calculated in detail. Namely, the scalar potential and the gauge boson sector are in a general case diagonalized. All the interactions of the gauge bosons with the fermions as well as with the scalars are derived. The new physics processes through the FCNCs, the LEPII collider, the violation of the CKM unitarity as well as the dark matter observables are analyzed. Particularly, we will perform a phenomenological study of the dark matter taking into account the current data as well as the new contributions of the physics at $\La\sim \om$ that have been kept in \cite{3311}. The constraints on the new gauge boson and dark matter masses are also obtained. 

The rest of this work is organized as follows. In Sec. \ref{3311}, we give a review of the model. Secs. \ref{sclsec} and \ref{gsec} are respectively devoted to the scalar and gauge sectors. In Sec. \ref{ints} we obtain all the gauge interactions of the fermions and scalars. Sec. \ref{ctsec} is aimed at studying the new physics processes and constraints. Finally, we summarize our results and make concluding remarks in Sec. \ref{concl}.

\section{\label{3311}A review of the 3-3-1-1 model}

The 3-3-1-1 model \cite{3311} is based on the gauge symmetry,
\be SU(3)_C\otimes SU(3)_L \otimes U(1)_X \otimes U(1)_N, \ee where the first three groups are the ordinary gauge symmetry of the 3-3-1 models \cite{331r,331m,dongfla}, while the last one is a necessary gauge extension of the 3-3-1 models that respects the conservation of lepton ($L$) and baryon ($B$) numbers. Indeed, the 3-3-1 symmetry and $B-L$ symmetry do not commute and also nonclose algebraically. To be concrete, for a lepton triplet (see below), we have $B-L=\mathrm{diag}(-1,-1,0)$, which is not commuted with the $SU(3)_L$ generators as $T_i=\fr 1 2 \la_i$ for $i=4,5,6,7$. It is easily checked that  
\bea &&\left[B-L,T_4\pm iT_5\right]=\mp(T_4\pm i T_5)\neq 0,\crn
&&\left[B-L,T_6\pm i T_7\right]=\mp (T_6\pm i T_7)\neq 0.\nn\eea 
The non-closed algebras can be deduced from the fact that in order for $B-L$ to be some generator of $SU(3)_L$, we have a linear combination $B-L=x_i T_i$ ($i=1,2,3,...,8$) and thus $\mathrm{Tr}(B-L)=0$, which is invalid for the lepton triplet, $\mathrm{Tr}(B-L)=-2\neq 0$, even for other particle multiplets. In other words, $B-L$ and $T_i$ by themselves do not make a symmetry under which our theory based on is manifest. Therefore, to have a closed algebra, we must introduce at least a new Abelian charge $N$ so that $B-L$ is a residual symmetry of closed group $SU(3)_L\otimes U(1)_N$, i.e. $B-L=x_i T_i+y N$, where the embedding coefficients $x_i, y\neq0$ are given below (the existence of $N$ can also be understood by a current algebra approach for $T_i$ and $B-L$ similarly to the case of hyper-charge $Y$ when we combine $SU(2)_L$ with $U(1)_Q$ to perform the $SU(2)_L\otimes U(1)_Y$ electroweak symmetry). Note that $N$ cannot be identified as $X$ (that defines the electric charge operator) because they generally differ for the particle multiplets (see below); thus they are independent charges. As a fact, the normal Lagrangian of the 3-3-1 models (including the gauge interactions, minimal Yukawa Lagrangian and minimal scalar potential) always preserves a $U(1)_N$ Abelian symmetry that along with $SU(3)_L$ realizes $B-L$ as a conserved (non-commuting) residual charge, which has actually been investigated in the literature and given in terms of $B=\mathcal{B}$ and $L=b T_8+\mathcal{L}$ where $b$ is 3-3-1 model-class dependent and $N=\mathcal{B}-\mathcal{L}$ \cite{3311,lepto331}. Note also that a violation in $N$ due to some unwanted interaction, by contrast, would lead to the corresponding violation in $B-L$ and vice versa. Because $T_i$ are gauged charges, $B-L$ and $N$ must be gauged charges (by contrast, $T_i\sim (B-L)-yN$ are global which is incorrect). The gauging of $B-L$ is a consequence of the non-commuting between $B-L$ and $SU(3)_L$ (which is unlike the standard model case). And, the theory is only consistent if it includes $U(1)_N$ as a gauge symmetry which also necessarily makes the resulting theory free from all the nontrivial leptonic and baryonic anomalies \cite{3311}. Otherwise, the 3-3-1 models must contain (abnormal) interactions that explicitly violate $B-L$ (or $N$). Equivalently, the 3-3-1 models are only survival if $B-L$ is not a symmetry of such theories, actually recognized as an approximate symmetry, which has explicitly shown in \cite{dongdongdm}. To conclude, assuming that the $B-L$ charge is conserved (that is respected by the experiments, the standard model, even the typical 3-3-1 models \cite{pdg,331m,331r,dongfla}), the Abelian factor $U(1)_N$ must be included so that the algebras are closed that is needed for a self-consistent theory. Apart from the strong interaction with $SU(3)_C$ group, the $SU(3)_L\otimes U(1)_X\otimes U(1)_N$ framework thus presents an unification of the electroweak and $B-L$ interactions, in the same manner of the standard model electroweak theory for the weak and electromagnetic ones.      

The two Abelian factors of the 3-3-1-1 symmetry associated with the $SU(3)_L$ group correspondingly determine the $Q$ electric charge and $B-L$ operators as residual symmetries, given by
\bea Q=T_3-\fr{1}{\sqrt{3}}T_8+X,\hs
B-L=-\fr{2}{\sqrt{3}}T_8+N,\label{ecqbl}\eea where $T_i\ (i=1,2,3,...,8)$, $X$ and $N$ are the charges of $SU(3)_L$, $U(1)_X$ and $U(1)_N$, respectively (the $SU(3)_C$ charges will be denoted by $t_i$). Note that the above $Q$ and $B-L$ definitions embed the 3-3-1 model with neutral fermions \cite{dongfla} in the considering theory. However, the coefficients of $T_8$ might be different depending on which class of the 3-3-1 models is embedded in \cite{lepto331}.

The $Q$ is conserved responsible for the electromagnetic interaction, whereas the $B-L$ must be broken so that the $U(1)_N$ gauge boson gets a large enough mass to escape from the detectors. Indeed, the $B-L$ is broken down to a parity (i.e., a $Z_2$ symmetry),
\be P=(-1)^{3(B-L)+2s}=(-1)^{-2\sqrt{3}T_8+3N+2s},\ee which consequently makes ``wrong $B-L$ particles'' become stabilized, providing dark matter candidates \cite{3311}. We see that this $R$-parity has an origin as a residual symmetry of the broken $SU(3)_L\otimes U(1)_N$ gauge symmetry, which is unlike the $R$-symmetry in supersymmetry~\cite{susy}. That being said, the parity $P$ is automatically existed, and due to its nature it will play an important role in the model besides stabilizing the dark matter candidates as shown throughout the text.   

The fermion content of the 3-3-1-1 model that is anomaly free is given as \cite{3311}
\bea \psi_{aL} &=& \left(\begin{array}{c}
               \nu_{aL}\\ e_{aL}\\ (N_{aR})^c
\end{array}\right) \sim (1,3, -1/3,-2/3),\\
\nu_{aR}&\sim&(1,1,0,-1),\hs e_{aR} \sim (1,1, -1,-1),
\\
Q_{\al L}&=&\left(\begin{array}{c}
  d_{\al L}\\  -u_{\al L}\\  D_{\al L}
\end{array}\right)\sim (3,3^*,0,0),\hs Q_{3L}=\left(\begin{array}{c} u_{3L}\\  d_{3L}\\ U_L \end{array}\right)\sim
 \left(3,3,1/3,2/3\right),\\ u_{a
R}&\sim&\left(3,1,2/3,1/3\right),\hs d_{a R} \sim
\left(3,1,-1/3,1/3\right),\\ U_{R}&\sim& \left(3,1,2/3,4/3\right),\hs D_{\al R}
\sim \left(3,1,-1/3,-2/3\right),\eea where the quantum numbers located in the parentheses are defined upon the gauge symmetries $(SU(3)_C,\ SU(3)_L,\ U(1)_X,\ U(1)_N)$, respectively. The family indices are $a=1,2,3$ and $\al=1,2$.

The exotic fermions $N_R$, $U$ and $D$ have been included to complete the fundamental representations of the $SU(3)_L$ group, respectively. By the embedding, their electric charges take usual values, $Q(N_R)=0$, $Q(U)=2/3$ and $Q(D)=-1/3$. However, their $B-L$ charges get values, $[B-L](N_R)=0$, $[B-L](U)=4/3$ and $[B-L](D)=-2/3$, which are abnormal in comparison to those of the standard model particles. These exotic fermions including the following bosons of this kind have ordinary baryon numbers, however, possessing anomalous lepton numbers as well as being odd under the parity $P$ (see Table \ref{owparticles} in more detail) \cite{3311}. Such particles are generally called as the wrong-lepton particles (or $W$-particles for short) and the parity $P$ is thus named as the $W$-parity. Whereas, all other particles of the model including the standard model ones (which have both the ordinary baryon and ordinary lepton numbers or only differing from the ordinary lepton number by an even lepton number as just the $\phi$ scalar given below) are even under the $W$-parity, and they can be considered as ordinary particles. 

Let us remind that the neutral fermions $N_{aR}$ might have left-handed counterparts, $N_{aL}$, transforming as singlets under any gauge symmetry group including the $U(1)_N$. By this view, the $N_{aL}$ are truly sterile which is unlike the $\nu_{aR}$ as usually considered in the literature. Interestingly, the sterile fermions $N_{aL}$ are $W$-particles like the $N_{aR}$. If the $N_{aL}$ are not included, the $N_{aR}$ are Majorana fermions. Otherwise, the presence of the $N_{aL}$ yields $N_{aL,R}$ as generic fermions (which may be Dirac ones). Further, we will exploit this matter by deriving the dark matter observables for the cases of the Dirac or Majorana fermions.    

To break the gauge symmetry and generate the masses for the particles in a correct way, the 3-3-1-1 model needs the following scalar multiplets \cite{3311}:
\bea
\eta &=&  \left(\begin{array}{c}
\eta^0_1\\
\eta^-_2\\
\eta^0_3\end{array}\right)\sim (1,3,-1/3,1/3),\hs
\rho = \left(\begin{array}{c}
\rho^+_1\\
\rho^0_2\\
\rho^+_3\end{array}\right)\sim (1,3,2/3,1/3),\crn 
 \chi &=& \left(\begin{array}{c}
\chi^0_1\\
\chi^-_2\\
\chi^0_3\end{array}\right)\sim (1,3,-1/3,-2/3),\hs 
\phi \sim (1,1,0,2),\eea with the VEVs that conserve $Q$ and $P$ being respectively given by
\bea \langle \eta \rangle =\fr{1}{\sqrt{2}}(u,0,0)^T,\hs \langle \rho \rangle = \fr{1}{\sqrt{2}}(0,v,0)^T,\hs \langle \chi \rangle =\fr{1}{\sqrt{2}}(0,0,\om)^T,\hs \langle \phi\rangle =\fr{1}{\sqrt{2}}\La.\label{vevss} \eea
The VEVs of $\eta,\ \rho,\ \chi$ break only $SU(3)_C\otimes SU(3)_L \otimes U(1)_X \otimes U(1)_N$ to $SU(3)_C\otimes U(1)_Q \otimes U(1)_{B-L}$, which leaves the $B-L$ invariant. The $\phi$ breaks $U(1)_N$ as well as the $B-L$ that defines the $W$-parity, $U(1)_{B-L}\rightarrow P$, with the form as given \cite{3311}. It provides also the mass for the $U(1)_N$ gauge boson as well as the Majorana masses for $\nu_{aR}$. Note that the $\rho_3$, $\eta_3$ and $\chi_{1,2}$ are the $W$-particles, while the others including $\phi$ are not (i.e., as the ordinary particles). The electrically-neutral fields $\eta_3$ and $\chi_1$ cannot develop a VEV due to the $W$-parity conservation. To keep a consistency with the standard model, we suppose $u, v\ll \om, \La$.

Up to the gauge fixing and ghost terms, the Lagrangian of the 3-3-1-1 model is given by
\bea \mathcal{L}&=&\sum_{\mathrm{fermion\ multiplets}}\bar{\Psi}i\ga^\mu D_\mu \Psi + \sum_{\mathrm{scalar\ multiplets}}(D^\mu \Phi)^\dagger (D_\mu \Phi)\crn
&& -\fr{1}{4}G_{i\mu\nu}G_i^{\mu\nu} -\fr 1 4 A_{i\mu\nu}A_i^{\mu\nu}-\fr 1 4 B_{\mu\nu} B^{\mu\nu}-\fr{1}{4}C_{\mu\nu} C^{\mu\nu}\crn
&&-V(\rho,\eta,\chi,\phi) + \mathcal{L}_{\mathrm{Yukawa}},\eea with the covariant derivative \be D_\mu = \pa_\mu + ig_s t_i G_{i\mu} + ig T_i A_{i\mu} + ig_X X B_\mu + i g_N N C_\mu,\label{dhhb} \ee and the field strength tensors
\bea G_{i\mu\nu}&=&\pa_\mu G_{i\nu} -\pa_\nu G_{i\mu} -g_s f_{ijk}G_{j\mu}G_{k\nu},\crn
A_{i\mu\nu}&=&\pa_\mu A_{i\nu} -\pa_\nu A_{i\mu} - g f_{ijk}A_{j\mu} A_{k\nu},\crn
B_{\mu\nu}&=&\pa_\mu B_\nu -\pa_\nu B_\mu,\hs C_{\mu\nu}=\pa_\mu C_\nu -\pa_\nu C_\mu.\eea The $\Psi$ denotes fermion multiplets such as $\psi_{aL}$, $Q_{3L}$, $u_{aR}$ and so on, whereas the $\Phi$ stands for scalar multiplets, $\phi$, $\eta$, $\rho$ and $\chi$. The coupling constants ($g_s,\ g,\ g_X,\ g_N$) and the gauge bosons ($G_{i\mu}, A_{i\mu}, B_\mu, C_\mu$) are defined as coupled to the generators ($t_i,\ T_i,\ X,\ N$), respectively. It is noted that in a mass basis the $W^\pm$ bosons are associated with $T_{1,2}$, the photon $\gamma$ is with $Q$, and the $Z$, $Z'$ are with generators that are orthogonal to $Q$. All these fields including the $C$ and gluons $G$ are even under the $W$-parity. However, the new non-Hermitian gauge bosons, $X^{0,0*}$ as coupled to $T_{4,5}$ and $Y^\pm$ as coupled to $T_{6,7}$, are the $W$-particles.

The scalar potential and Yukawa Lagrangian as mentioned above are obtained as follows \cite{3311}
\bea \mathcal{L}_{\mathrm{Yukawa}}&=&h^e_{ab}\bar{\psi}_{aL}\rho e_{bR} +h^\nu_{ab}\bar{\psi}_{aL}\eta\nu_{bR}+h'^\nu_{ab}\bar{\nu}^c_{aR}\nu_{bR}\phi + h^U\bar{Q}_{3L}\chi U_R + h^D_{\al \beta}\bar{Q}_{\al L} \chi^* D_{\beta R}\crn
&&+ h^u_a \bar{Q}_{3L}\eta u_{aR}+h^d_a\bar{Q}_{3L}\rho d_{aR} + h^d_{\al a} \bar{Q}_{\al L}\eta^* d_{aR} +h^u_{\al a } \bar{Q}_{\al L}\rho^* u_{aR} +H.c.,\label{bossung1}\\
V(\rho,\eta,\chi,\phi) &=& \mu^2_1\rho^\dagger \rho + \mu^2_2 \chi^\dagger \chi + \mu^2_3 \eta^\dagger \eta + \la_1 (\rho^\dagger \rho)^2 + \la_2 (\chi^\dagger \chi)^2 + \la_3 (\eta^\dagger \eta)^2\crn
&&+ \la_4 (\rho^\dagger \rho)(\chi^\dagger \chi) +\la_5 (\rho^\dagger \rho)(\eta^\dagger \eta)+\la_6 (\chi^\dagger \chi)(\eta^\dagger \eta)\crn
&& +\la_7 (\rho^\dagger \chi)(\chi^\dagger \rho) +\la_8 (\rho^\dagger \eta)(\eta^\dagger \rho)+\la_9 (\chi^\dagger \eta)(\eta^\dagger \chi) + (f\epsilon^{mnp}\eta_m\rho_n\chi_p+H.c.) \crn
&& + \mu^2 \phi^\dagger \phi + \la (\phi^\dagger \phi)^2 +\la_{10} (\phi^\dagger \phi)(\rho^\dagger\rho)+\la_{11}(\phi^\dagger \phi)(\chi^\dagger \chi)+\la_{12}(\phi^\dagger \phi)(\eta^\dagger \eta).
\eea Because of the 3-3-1-1 gauge symmetry, the Yukawa Lagrangian and scalar potential as given take the standard forms that contain no lepton-number violating interactions. 

If such violating interactions as well as nonzero VEVs of $\eta_3$ and $\chi_1$ were presented as in the 3-3-1 model, they would be the sources for the hadronic FCNCs at tree level \cite{ponce}. The FCNC problem is partially solved by the 3-3-1-1 symmetry and $W$-parity conservation. Also, the presence of the $\eta_3$ and $\chi_1$ VEVs would imply a mass hierarchy between the real and imaginary components of the $X^0$ gauge boson due to their different mixings with the neutral gauge bosons. This leads to the $CPT$ violation that is experimentally unacceptable \cite{cpt}. The $CPT$ violation encountered with the 3-3-1 model is thus solved by the 3-3-1-1 symmetry and $W$-parity conservation too.

Table \ref{owparticles} lists all the model particles with their parity values explicitly provided. The lepton numbers have also been included for a convenience in reading. However, the baryon numbers were not listed since they can be obtained as usual (all the quarks $u$, $d$, $U$ and $D$ have $B=1/3$, whereas the other particles have $B=0$).
\begin{table}[htdp]
\bc
\begin{tabular}{|c|cccccccccccccccccccccc|}
\hline
Particle & $\nu$ & $e$ & $u$ & $d$ & $G$ & $\gamma$ & $W$ & $Z$ & $Z'$ & $C$ & $\eta_{1,2}$ & $\rho_{1,2}$ & $\chi_3$ & $\phi$ & $N$ & $U$ & $D$ & $X$ & $Y$ & $\eta_{3}$ & $\rho_{3}$ & $\chi_{1,2}$\\ \hline
$L$ & 1 & 1 & 0 & 0 &0 & 0 & 0 & 0 & 0 & 0& 0& 0& 0 & $-2$& 0& $-1$& 1& 1 & 1 & $-1$& $-1$ & 1 \\
$P$ & $+$&$+$&$+$&$+$&$+$&$+$&$+$&$+$&$+$&$+$&$+$&$+$&$+$&$+$&$-$&$-$&$-$&$-$&$-$&$-$&$-$&$-$ \\
\hline
\end{tabular}
\caption{\label{owparticles}
The $W$-parity ($P$) separates the model particles into the two classes: (i) $W$-particles that possess $P=-1$, and (ii) Ordinary-particles that have $P=+1$. The first class includes a large portion of the new particles, while the second class is dominated by the standard model particles.} 
\ec
\end{table}
As shown in \cite{3311}, the $X^0$ gauge boson cannot be a dark matter. However, the neutral fermion (a combination of $N_a$ fields) or the neutral complex scalar (a combination of $\eta^0_3$ and $\chi^0_1$ fields) can be dark matter whatever one of them is the lightest wrong-lepton particle (LWP) in agreement with \cite{farinaldoDM2}.

The fermion masses that are obtained from the Yukawa Lagrangian after the gauge symmetry breaking have been presented in \cite{3311} in detail. Below, we will calculate the masses and physical states of the scalar and gauge boson sectors when the $\La$ scale of the $U(1)_N$ breaking is comparable to the $\om$ scale of the 3-3-1 breaking, which has been neglected in \cite{3311}. Also, all the gauge interactions of fermions and scalars as well as the constraints on the new physics are derived. We stress again that in the regime $\La\gg \om$ the $B-L$ and 3-3-1 symmetries decouple; whereas, when those scales become comparable, the new physics associated with the $B-L$ and that of the 3-3-1 model are correlated, possibly happening at  the TeV scale, to be all proved by the LHC or the ILC project.  

\section{\label{sclsec}Scalar sector}

Since the $W$-parity is conserved, only the neutral scalar fields that are even under this parity symmetry can develop the VEVs as given in (\ref{vevss}). We expand the fields around these VEVs as
\bea
\eta = \langle \eta \rangle + \eta^\prime =
 \left(\begin{array}{c}
  \frac{u}{\sqrt{2}}\\
  0\\
  0
\end{array}\right)+ \left(\begin{array}{c}
   \frac{S_1+iA_1}{\sqrt{2}} \\
   \eta^-_2\\
   \frac{S^\prime_3+iA^\prime_3}{\sqrt{2}} \\
\end{array}\right), \hs 
\rho = \langle \rho \rangle + \rho^\prime =
 \left(\begin{array}{c}
  0\\
  \frac{v}{\sqrt{2}}\\
  0
\end{array}\right)+ \left(\begin{array}{c}
 \rho^+_1 \\
  \frac{S_2+iA_2}{\sqrt{2}} \\
   \rho^+_3
\end{array}\right), \label{scl1}
\eea
\bea
\chi = \langle \chi \rangle + \chi^\prime =
 \left(\begin{array}{c}
  0\\
  0\\
  \frac{\om}{\sqrt{2}}\\
\end{array}\right)+ \left(\begin{array}{c}
   \frac{S^\prime_1+iA^\prime_1}{\sqrt{2}} \\
   \chi^-_2\\
   \frac{S_3+iA_3}{\sqrt{2}} 
\end{array}\right), \hs 
\phi =\langle \phi\rangle +\phi'= \frac{\La}{\sqrt{2}} + \frac{S_4+iA_4}{\sqrt{2}}, \label{scl2}
\eea where in each expansion the first term and last term are denoted as the VEVs and physical fields, respectively. Note that $S_{1,2,3,4}$ and $A_{1,2,3,4}$ are $W$-even while those with primed signs, $S^\prime_{1,3}$ and $A_{1,3}^\prime$, are $W$-odd. There is no mixing between the $W$-even and $W$-odd fields due to the $W$-parity conservation. On the other hand, the $f$ parameter in the scalar potential can be complex (the remaining parameters such as $\mu^2$'s and $\la$'s are all real). However, its phase can be removed by redefining the fields $\eta, \rho, \chi$ appropriately. Consequently, the scalar potential conserves the $CP$ symmetry. Assuming that the $CP$ symmetry is also conserved by the vacuum, the VEVs and $f$ can simultaneously be considered as the real parameters by this work. There is no mixing between the scalars ($CP$-even) and pseudoscalars ($CP$-odd) due to the $CP$ conservation. 

To find the mass spectra of the scalar fields, let us expand all the terms of the potential up to the second order contributions of the fields: 
\bea
\mu^2_1 (\rho^\dagger \rho) &=&\mu^2_1(\langle \rho\rangle^\dagger \langle \rho\rangle + \langle \rho\rangle^\dagger \rho^\prime +\rho^{\prime \dagger}\langle \rho\rangle + \rho^{\prime \dagger}\rho^\prime)\crn
&=&\mu^2_1\left(\frac{v^2}{2} + v S_2+ \rho^+_1 \rho^-_1 + \rho^+_3 \rho^-_3 + \frac{S^2_2+A^2_2}{2}\right),\crn
\mu^2_2 (\chi^\dagger \chi)&=&\mu^2_2\left(\frac{\om^2}{2} +\om S_3 + \chi^-_2 \chi^+_2 + \frac{S^{\prime 2}_1+A^{\prime 2}_1 + S^2_3+A^2_3}{2}\right),\crn
\mu^2_3 (\eta^\dagger \eta)&=&\mu^2_3\left(\frac{u^2}{2} +u S_1 + \eta^-_2 \eta^+_2 + \frac{S^2_1+A^2_1 + S^{\prime 2}_3+A^{\prime 2}_3}{2}\right),\crn
\mu^2 (\phi^\dagger \phi) &=& \mu^2\left(\frac{\La^2}{2} +\La S_4 + \frac{S^2_4+A^2_4}{2}\right),\nn
\eea
\bea
\la (\phi^\dagger \phi)^2 &=& \la\left[\frac{\La^4}{4} +\La^2 S^2_4 +\La^3 S_4 + \frac{\La^2}{2}(S^2_4+A^2_4)+ \mathrm{interaction} \right],\crn
\la_1 (\rho^\dagger \rho)^2 &=& \la_1\left[\frac{v^4}{4} + v^2 S^2_2+ v^3 S_2+ v^2 \left(\rho^+_1 \rho^-_1 + \rho^+_3 \rho^-_3+\frac{S^2_2+A^2_2}{2}\right)+ \mathrm{interaction} \right],\crn
\la_2 (\chi^\dagger \chi)^2 &=&\la_2\left[\frac{\om^4}{4} + \om^2 S^2_3 +\om^3 S_3 + \om^2\left(\chi^-_2 \chi^+_2  +\frac{S^{\prime 2}_1+A^{\prime 2}_1 + S^2_3+A^2_3}{2}\right)+ \mathrm{interaction} \right],\crn
\la_3 (\eta^\dagger \eta)^2 &=& \la_3 \left[\frac{u^4}{4} +u^2 S^2_1 + u^3 S_1+u^2\left(\eta^-_2 \eta^+_2 + \frac{S^2_1+A^2_1 + S^{\prime 2}_3+A^{\prime 2}_3}{2}\right)+ \mathrm{interaction} \right],\nn
\eea
\bea 
\la_4 (\rho^\dagger \rho) (\chi^\dagger \chi) &=&\la_4 \left[\frac{v^2\om^2}{4} + \frac{\om v^2}{2} S_3+ \frac{v \om^2}{2} S_2+ v\om S_2 S_3+ \frac{v^2}{2}\left(\chi^-_2 \chi^+_2 + \frac{S^{\prime 2}_1+A^{\prime 2}_1 + S^2_3 + A^2_3}{2}\right)\right.\crn
&&+ \left.\frac{\om^2}{2}\left(\rho^+_1 \rho^-_1 + \rho^+_3 \rho^-_3 + \frac{S^2_2+A^2_2}{2}\right)+\mathrm{interaction} \right],\crn
\la_5 (\rho^\dagger \rho) (\eta^\dagger \eta) &=&\la_5 \left[\frac{v^2 u^2}{4} + \frac{u v^2}{2} S_1+ \frac{v u^2}{2} S_2+ vu S_1 S_2 + \frac{v^2}{2}\left(\eta^-_2 \eta^+_2 + \frac{S^ 2_1+A^2_1 + S^{\prime 2}_3+A^{\prime 2}_3}{2}\right)\right. \crn
&&+ \left.\frac{u^2}{2}\left(\rho^+_1 \rho^-_1 + \rho^+_3 \rho^-_3 + \frac{S^2_2+A^2_2}{2}\right)+\mathrm{interaction} \right],\crn
\la_6 (\chi^\dagger \chi) (\eta^\dagger \eta) &=&\la_6  \left[\frac{\om^2 u^2}{4} + \frac{u \om^2}{2} S_1+ \frac{\om u^2}{2} S_3+ u\om S_1 S_3 + \frac{\om^2}{2}\left(\eta^-_2 \eta^+_2 + \frac{S^ 2_1+A^2_1 + S^{\prime 2}_3+A^{\prime 2}_3}{2}\right)\right. \crn
&&+ \left.\frac{u^2}{2}\left(\chi^+_2 \chi^-_2 + \frac{S^{\prime 2}_1+A^{\prime2}_1+S^ 2_3+A^2_3}{2}\right)+\mathrm{interaction}  \right],\nn
\eea
\bea
\la_7 (\rho^\dagger \chi) (\chi^\dagger \rho) &=&\frac{\la_7}{2}(v\chi^-_2 + \om \rho^-_3)(\om \rho^+_3 + v \chi^+_2)+ \mathrm{interaction}, \crn
\la_8 (\rho^\dagger \eta) (\eta^\dagger \rho) &=&\frac{\la_8}{2}(v\eta^-_2 + u \rho^-_1)(u \rho^+_1 + v \eta^+_2)+ \mathrm{interaction}, \crn 
\la_9 (\chi^\dagger \eta) (\eta^\dagger \chi) &=&\la_9 \left[\frac{\om}{2}(S^\prime _3+iA^\prime_3)+\frac{u}{2}(S^{\prime }_1-iA^{\prime}_1)\right] \left[\frac{u}{2}(S^{\prime }_1+iA^{\prime }_1)+\frac{\om}{2}(S^\prime _3-iA^\prime_3)\right] +\mathrm{interaction},\nn
\eea
\bea 
\la_{10} (\phi^\dagger \phi) (\rho^\dagger \rho) &=&\la_{10}\left[\frac{\La^2 v^2}{4} + \frac{v \La^2}{2} S_2+ \frac{\La v^2}{2} S_4+ v\La S_2 S_4 + \frac{v^2}{2}\left(\frac{S^ 2_4+A^2_4}{2}\right)\right. \crn
 &&+\left.\frac{\La^2}{2}\left(\rho^+_1 \rho^-_1 +\rho^+_3 \rho^-_3+ \frac{S^2_2+A^2_2}{2}\right)+\mathrm{interaction} \right],\crn
\la_{11} (\phi^\dagger \phi) (\chi^\dagger \chi) &=&\la_{11}\left[\frac{\La^2 \om^2}{4} + \frac{\om \La^2}{2} S_3+ \frac{\La \om^2}{2} S_4+ \om\La S_3 S_4 + \frac{\om^2}{2}\left(\frac{S^2_4+A^2_4}{2}\right)\right.\crn
&&+ \left.\frac{\La^2}{2}\left(\chi^+_2 \chi^-_2+ \frac{S'^2_1+ A^{\prime 2}_1+{S^2_3+A^2_3}}{2}\right)+\mathrm{interaction} \right],\crn
\la_{12} (\phi^\dagger \phi) (\eta^\dagger \eta) &=& \la_{12}\left[\frac{\La^2 u^2}{4} + \frac{u \La^2}{2} S_1+ \frac{\La u^2}{2} S_4 + u \La S_1 S_4 + \frac{u^2}{2} \left(\frac{S^2_4+A^2_4}{2}\right)\right.\crn
&&+ \left.\frac{\La^2}{2}\left(\eta^+_2 \eta^-_2+ \frac{S^2_1+A^2_1+{S'^2_3+A'^2_3}}{2}\right)+\mathrm{interaction} \right],\nn
\eea 
\bea
f\ep^{mnp}\eta_m\rho_n\chi_p+H.c. &=& f\left[\fr{uv\om}{\sqrt{2}}+\fr{uv}{\sqrt{2}}S_3 + \fr{u\om}{\sqrt{2}}S_2 + \fr{v\om}{\sqrt{2}}S_1 + \fr{u}{\sqrt{2}}\left(S_2 S_3 - A_2 A_3\right.\right.\crn
&&\left. - \rho^+_3 \chi^-_2 -\rho^-_3\chi^+_2 \right) + \fr{v}{\sqrt{2}}\left(S_1 S_3 - A_1 A_3 - S'_1S'_3 + A'_1 A'_3\right)\crn
&&\left. + \fr{\om}{\sqrt{2}}\left(S_1 S_2 - A_1 A_2 - \eta^-_2 \rho^+_1 -\eta^+_2\rho^-_1 \right)\right]+ \mathrm{interaction}. \nn
\eea

The scalar potential that is summed of all the terms above can be rearranged as  
\be V(\rho, \eta, \chi, \phi) = V_{\mathrm{min}} + V_{\mathrm{linear}} + V_{\mathrm{mass}} + V_{\mathrm{interaction}},\ee where the interactions as stored in $V_{\mathrm{interaction}}$ need not to be explicitly obtained. 
The $V_{\mathrm{min}}$ contains the terms that are independent of the scalar fields,   
\bea 
V_{\mathrm{min}}&=&\mu^2_1\fr{v^2}{2}+\mu^2_2\fr{\om^2}{2}+\mu^2_3\fr{u^2}{2}+\mu^2\fr{\La^2}{2}+\la^2_1\fr{v^4}{4}+\la^2_2\fr{\om^4}{4}+\la^2_3\fr{u^4}{4}+\la^2\fr{\La^4}{4}\crn
&&+ \la^2_4\fr{v^2 \om^2}{4}+\la^2_5\fr{v^2u^2}{4}+\la^2_6\fr{u^2 \om^2}{4}+\la^2_{10}\fr{v^2 \La^2}{4}+\la^2_{11}\fr{\La^2 \om^2}{4}+\la^2_{12}\fr{u^2 \La^2}{4}+f\fr{uv\om}{\sqrt2},\nn
\eea
which contributes to the vacuum energy only. It does not affect to the physical processes. 

The $V_{\mathrm{linear}}$ includes all the terms that linearly depend on the scalar fields, \bea
V_{\mathrm{linear}}&=& S_1\left[u \mu_3^2 +\la_3 u^3+\frac{1}{2}\lambda_5uv^2+\frac{1}{2}\lambda_6u\om^2+\frac{\sqrt{2}}{2}fv\om +\frac{1}{2}\lambda_{12}u\La^2 \right]\crn
&&+S_2\left[v\mu_1^2+\lambda_1 v^3+\frac{1}{2}\lambda_4v\om^2+\frac{1}{2}\lambda_5 u^2v+\frac{\sqrt{2}}{2}fu \om+\frac{\lambda_{10}}{2}v\La^2\right]\crn
&&+ S_3\left[\om\mu_2^2+\lambda_2 \om^3 + \frac{\lambda_4}{2}\om v^2+ \frac{\lambda_6}{2}\om u^2+\frac{\sqrt{2}}{2}fuv+\frac{\lambda_{11}}{2}\om\La^2\right]\crn 
&&+S_4\left[\mu^2\La + \lambda \La^3+\frac{1}{2}\lambda_{10} v^2\La+\frac{1}{2}\lambda_{11}
 \La \om^2+\frac{1}{2}\lambda_{12} \La u^2 \right].
\eea Because of the gauge invariance, the coefficients vanish,  
\bea
v \mu_1^2+\lambda_1 v^3+\frac{1}{2}\lambda_4v\om^2+\frac{1}{2}\lambda_5 u^2v+\frac{\sqrt{2}}{2}fu \om+\frac{\lambda_{10}}{2}v\La^2 &=& 0,  \\
\om\mu_2^2+\lambda_2 \om^3 + \frac{\lambda_4}{2}\om v^2+ \frac{\lambda_6}{2}\om u^2+\frac{\sqrt{2}}{2}fuv+\frac{\lambda_{11}}{2}\om\La^2 &=& 0, \\
u \mu_3^2 +\la_3 u^3+\frac{1}{2}\lambda_5uv^2+\frac{1}{2}\lambda_6u\om^2+\frac{\sqrt{2}}{2}fv\om +\frac{1}{2}\lambda_{12}u\La^2 &= &0, \\
\mu^2 + \lambda \La^2+\frac{1}{2}\lambda_{10} v^2+\frac{1}{2}\lambda_{11}
 \om^2+ \frac{1}{2}\lambda_{12} u^2 &=& 0,
\eea 
which are also the conditions of potential minimization,   
\bea
\fr{\partial V}{\partial u} = \fr{\partial V}{\partial v} = \fr{\partial V}{\partial\om} = \fr{\partial V}{\partial\La} = 0. 
\eea
The 3-3-1-1 gauge symmetry will be broken in the correct way and the potential bounded from below by imposing $\mu^2<0,\ \mu^2_{1,2,3}<0$, $\la>0,\ \la_{1,2,3}>0$, and other necessary conditions for $\la_{4,5,6,...,12}$. In this case, the equations of the potential minimization above give an unique, nonzero solution for the VEVs $(u,\ v,\ \om,\ \La)$.

The $V_{\mathrm{mass}}$ consists of all the terms in the potential that quadratically depend on the scalar fields. It can be decomposed into,   
\be V_{\mathrm{mass}} = V^{\mathrm{charged}}_{\mathrm{mass}}+V^{S}_{\mathrm{mass}} + V_{\mathrm{mass}}^{A} + V_{\mathrm{mass}}^{S'}+V_{\mathrm{mass}}^{A'},
\ee
where the first term includes all the mass terms of charged scalars while the remaining terms belong to the neutral scalars with each term for a distinct group of fields characterized by the two values, $W$- and $CP$- parities, as mentioned before.    

The mass spectrum of the charged scalars is obtained by     
\bea
V^{\mathrm{charged}}_{\mathrm{mass}}&=&\chi_2^+\chi_2^-\left(\mu_2^2+\lambda_2 \om^2 + \frac{\lambda_4}{2} v^2+ \frac{\lambda_6}{2} u^2+\frac{\lambda_{11}}{2}\La^2\right)\crn
&&+ \eta_2^+\eta_2^-\left(\mu_3^2 +\la_3 u^2+\frac{1}{2}\lambda_5v^2+\frac{1}{2}\lambda_6\om^2+\frac{1}{2}\lambda_{12}\La^2\right)\crn
&&+ (\rho_1^+\rho_1^-+\rho_3^+\rho_3^-)\left(\mu_1^2+\lambda_1 v^2+\frac{1}{2}\lambda_4\om^2+\frac{1}{2}\lambda_5 u^2+\frac{\lambda_{10}}{2}\La^2\right)\\
&&+ \frac{\lambda_7}{2}(v\chi^-_2+\om\rho^-_3)(v\chi^+_2+\om\rho^+_3)+\frac{\lambda_8}{2}(v\eta^-_2+u\rho^-_1)(u\rho^+_1+v\eta^+_2)\crn
&&- f\frac{u}{\sqrt{2}}(\rho^+_3\chi^-_2 + \rho^-_3\chi^+_2)-f\frac{\om}{\sqrt{2}}(\eta^-_2\rho^+_1 + \eta^+_2\rho^-_1).\nn\eea From the potential minimization conditions, we extract $\mu^2_1$, $\mu^2_2$, $\mu^2_3$ and substitute them into the above expression to yield
 \bea
V^{\mathrm{charged}}_{\mathrm{mass}}&=& \left(\frac{\lambda_7}{2}-\frac{f u}{\sqrt{2}v\om}\right) (v\chi^-_2+\om\rho^-_3)(v\chi^+_2+\om\rho^+_3)\crn
&&+ \left(\frac{\lambda_8}{2}-\frac{f \om}{\sqrt{2}uv}\right)(v\eta^-_2+u\rho^-_1)(v\eta^+_2+u\rho^+_1)\crn
&=&\left(\frac{\lambda_7}{2}-\frac{f u}{\sqrt{2}v\om}\right)(v^2+\om^2)H^-_4 H^+_4 + \left(\frac{\lambda_8}{2}-\frac{f \om}{\sqrt{2}vu}\right)(v^2+u^2)H^-_5 H^+_5,
\eea
where we have defined, 
\be
H_4^\pm\equiv \fr{v \chi_2^\pm + \om \rho_3^\pm}{\sqrt{v^2+\om^2}},
\hs H_5^\pm \equiv \fr{v \eta_2^\pm + u \rho_1^\pm}{\sqrt{u^2+v^2}}. 
\ee
The fields $H^\pm_4,\ H^\pm_5$ by themselves are physical charged scalars with masses respectively given by 
\bea
m^2_{H_4} &=& \left(\frac{\lambda_7}{2}-\frac{f u}{\sqrt{2}v\om}\right)(v^2+\om^2),\hs m^2_{H_5} = \left(\frac{\lambda_8}{2}-\frac{f \om}{\sqrt{2}vu}\right)(v^2+u^2).
\eea
The field that is orthogonal to $H_5$,
$G^\pm_W = \frac{u\eta^\pm_2 - v\rho^\pm_1}{\sqrt{u^2+v^2}}$, has a zero mass and can be identified as the Goldstone boson of the $W^{\pm}$ gauge boson. Similarly, 
the orthogonal field to $H_4$,
$G^\pm_Y = \frac{\om\chi^\pm_2-v\rho^\pm_3}{\sqrt{v^2+\om^2}}$, is massless and can be identified as the Goldstone boson of the new $Y^{\pm}$ gauge boson.

For the neutral scalar fields, we start with the $A$ group, \bea
V^A_{\mathrm{mass}}&=& A^2_1 \left( \fr{\mu_3^2}{2} +\frac{1}{2}\la_3 u^2+\frac{1}{4}\lambda_5v^2+\frac{1}{4}\lambda_6\om^2+\frac{1}{4}\lambda_{12}\La^2\right)\crn
&&+ A^2_2\left( \fr{\mu_1^2}{2}+\frac{1}{2}\lambda_1 v^2+\frac{1}{4}\lambda_4\om^2+\frac{1}{4}\lambda_5 u^2+\frac{\lambda_{10}}{4}v\La^2\right)\crn
&&+A^2_3\left( \fr{\mu_2^2}{2}+\frac{1}{2}\lambda_2 \om^2 + \frac{\lambda_4}{4} v^2+ \frac{\lambda_6}{4}u^2+\frac{\lambda_{11}}{4}\om\La^2\right)\\
&&+ A^2_4\left(\fr{\mu^2}{2} + \frac{1}{2}\lambda \La^2+\frac{1}{4}\lambda_{10} v^2+\frac{1}{4}\lambda_{11}
 \om^2+ \frac{1}{4}\lambda_{12} u^2\right)\crn
 &&- \fr{fu}{\sqrt2}A_2 A_3 - \fr{fv}{\sqrt2}A_1 A_3 - \fr{f\om}{\sqrt2}A_1 A_2\crn
&=& - \fr{f}{2 \sqrt{2}} \left(\fr{v \om}{u} + \fr{u\om}{v} +\fr{uv}{\om}\right) \left(\fr{v\om A_1 + u \om A_2 + u v A_3}{\sqrt{u^2v^2 + v^2\om^2+ u^2\om^2}}\right)^2,
\eea with the help of the potential minimization conditions. Therefore, we have a physical pseudo-scalar field with corresponding mass, \be
\mathcal{A} \equiv \fr{v \om A_1 + u \om A_2 + u v A_3}{\sqrt{u^2 v^2+v^2 \om^2+u^2 \om^2}},\hs 
 m^2_{\mathcal{A}} = - \fr{f}{\sqrt{2}}\left(\fr{v \om}{u} + \fr{u\om}{v} +\fr{u v}{\om}\right).\ee
If $u,v,\om > 0$ we have $f < 0$ so that the squared mass is always positive. We realize that  the $A_4$ is massless and can be identified as the Goldstone boson of the new neutral gauge boson $C$ of $U(1)_N$. The remaining massless fields are orthogonal to $\mathcal{A}$ as follows
\bea
G_Z &=& \frac{uA_1-vA_2}{\sqrt{u^2+v^2}},\crn 
G_{Z'}&=&\frac{-uv(vA_1+uA_2)+\om(u^2+v^2)A_3}
{\sqrt{(u^2 v^2+v^2\om^2+u^2\om^2)(u^2+v^2)}}.
\eea They are the Goldstone bosons of the neutral gauge bosons $Z$ and $Z'$, respectively (where the $Z$ is standard model like while the $Z'$ is 3-3-1 model like).
 
 For the $A'$ group, we have 
 \bea
 V^{A'}_{\mathrm{mass}}&=& A'^2_1 \left(\fr{\mu_2^2}{2}+\frac{1}{2}\lambda_2 \om^2 + \frac{\lambda_4}{4} v^2+ \frac{\lambda_6}{4}u^2+\frac{\lambda_{11}}{4}\om\La^2\right)\crn
 &&+ A'^2_3\left(\fr{\mu_3^2}{2}+\frac{1}{2}\lambda_2 \om^2 + \frac{\lambda_4}{4} v^2+ \frac{\lambda_6}{4}u^2+\frac{\lambda_{11}}{4}\om\La^2\right)\crn
 &&+ \fr{fv}{\sqrt{2}}A'_1 A'_3 + \fr{\la_9}{4}(\om A'_3 - uA'_1)^2\crn
 &=& \fr{1}{2}\left(\fr{\la_9}{2}- \fr{1}{\sqrt{2}}\fr{fv}{u \om}\right) (u^2+\om^2) \left(\fr{\om A'_3 - u A'_1 }{\sqrt{u^2+\om^2}}\right)^2,\nn
\eea by using the minimization conditions. Hence, a physical $W$-odd pseudo-scalar and its mass follow  
 \be
 A' \equiv \fr{\om A'_3 - u A'_1}{\sqrt{u^2+\om^2}},\hs m^2_{A'} = \left(\fr{\la_9}{2}- \fr{1}{\sqrt{2}}\fr{fv}{u \om}\right) (u^2+\om^2).\ee
Similarly, for the $S'$ group, we obtain  
\be V^{S'}_{\mathrm{mass}}=
\fr{1}{2}\left(\fr{\la_9}{2}- \fr{1}{\sqrt{2}}\fr{fv}{u \om}\right) (u^2+\om^2) \left(\fr{\om S'_3+u S'_1}{\sqrt{u^2+\om^2}}\right)^2,\ee which yields a physical $W$-odd scalar with corresponding mass, 
\bea
S' &\equiv& \fr{\om S'_3 + u S'_1}{\sqrt{u^2+\om^2}},\hs m^2_{S'} = \left(\fr{\la_9}{2}- \fr{1}{\sqrt{2}}\fr{fv}{u \om}\right) (u^2+\om^2).\nn
\eea The remarks are given in order: 
\ben \item We see that the scalar $S'$ and pseudo-scalar $A'$ have the same mass. They can be identified as the real and imaginary components of a physical neutral complex field:
\bea H'^0 &\equiv & \frac{S'+iA'}{\sqrt{2}}=\fr{1}{\sqrt{u^2+\om^2}}(u\chi^{0*}_1+\om\eta^{0}_3),\nn \eea 
with the mass \be m^2_{H'} = \left(\fr{\la_9}{2}- \fr{1}{\sqrt{2}}\fr{fv}{u \om}\right) (u^2+\om^2).\ee 
\item The field that is orthogonal to $H'$, $G^0_{X} = \fr{1}{\sqrt{u^2+\om^2}}(\om\chi^0_1-u\eta^{0*}_3)$, is massless and can be identified as the Goldstone boson of the new neutral non-Hermitian gauge boson $X^0$.
\een 
 
Finally, there remains the $S$ group of the $W$-even, real scalar fields. Using the potential minimization conditions, we have   
 \bea
 V^S_{\mathrm{mass}}&=&\left(\la_3 u^2-\fr{1}{2\sqrt{2}}f\fr{v\om}{u}\right)S^2_1+\left(\la_1 v^2-\fr{1}{2\sqrt{2}}f\fr{u\om}{v}\right)S^2_2+ \left(\la_2 \om^2-\fr{1}{2\sqrt{2}}f\fr{vu}{\om}\right)S^2_3\crn
 &&+\left(\la_5 uv+\fr{1}{\sqrt{2}}f\om\right)S_1S_2+ \left(\la_6 u\om+\fr{1}{\sqrt{2}}fv\right)S_1S_3 + \left(\la_4 \om v+\fr{1}{\sqrt{2}}fu\right)S_2S_3\crn
 && + \la\La^2S^2_4 + \la_{12}u\La S_1S_4 + \la_{10}v\La S_2S_4  + \la_{11}\om\La S_3S_4\crn
&=&
\fr{1}{2}\left(\begin{array}{cccc}
S_1 & S_2 & S_3 & S_4
\end{array}\right) M^2_S \left(\begin{array}{c}
S_1 \\
S_2\\
S_3\\
S_4
\end{array}\right),
\eea where
 \be
M^2_S \equiv \left(\begin{array}{cccc}
 2\la_3 u^2-\fr{1}{\sqrt{2}}f\fr{v\om}{u} & \la_5 uv+\fr{1}{\sqrt{2}}f\om & \la_6 u\om+\fr{1}{\sqrt{2}}fv & \la_{12}u\La\\
\la_5 uv+\fr{1}{\sqrt{2}}f\om & 2\la_1 v^2-\fr{1}{\sqrt{2}}f\fr{u\om}{v} & \la_4 \om v+\fr{1}{\sqrt{2}}fu & \la_{10}v\La\\
\la_6 u\om+\fr{1}{\sqrt{2}}fv & \la_4 \om v+\fr{1}{\sqrt{2}}fu & 2\la_2 \om^2-\fr{1}{\sqrt{2}}f\fr{vu}{\om} &  \la_{11}\om\La\\
\la_{12}u\La & \la_{10}v\La & \la_{11}\om\La & 2\la\La^2
\end{array}\right).\label{moivh}
\ee In \cite{3311}, the physical states have been derived when the $B-L$ breaking scale is large enough as the GUT one, for example, so that the $S_4$ is completely decoupled from the remaining three scalars of the 3-3-1 model. In this work we consider a possibility of the $B-L$ interactions that might happen at a TeV scale like those of the 3-3-1 model, characterized by the $\om$, $f$ scales. Therefore, let us assume that the $\La$ is in the same order with the $f,\ \om$ and all are sufficiently large in comparison to the weak scales $u,\ v$ so that the new physics is safe \cite{3311}, i.e. 
\be -f \sim \om \sim\La \gg u \sim v.\label{dieuk} \ee Notice that all the physical scalar fields which have been found so far are new particles with the corresponding masses given in the $\om$ or $\sqrt{|f\om|}$ scales.  

The mass matrix (\ref{moivh}) will provide a small eigenvalue as the mass of the standard model Higgs boson. Whereas, the remaining eigenvalues will be large to be identified as the corresponding masses of the new neutral scalars. To see this explicitly, it is appropriately to consider the leading order contributions of the mass matrix (\ref{moivh}). Imposing (\ref{dieuk}) and keeping only the terms that are proportional to $(\om,\ \Lambda,\ f)^2$, we have the result, 
  \be
M^2_S|_{\mathrm{LO}} = \left(\begin{array}{cccc}
 -\fr{1}{\sqrt{2}}f\fr{v\om}{u} & \fr{1}{\sqrt{2}}f\om & 0 & 0\\
\fr{1}{\sqrt{2}}f\om & -\fr{1}{\sqrt{2}}f\fr{u\om}{v} & 0 & 0\\
0 & 0 & 2\la_2 \om^2 &  \la_{11}\om\La\\
0 & 0 & \la_{11}\om\La & 2\la\La^2
\end{array}\right).
\ee The $2 \times 2$ matrix at the first diagonal box gives a zero eigenvalue with corresponding eigenstate: \be m^2_H = 0,\hs H \equiv \fr{uS_1 + vS_2}{\sqrt{u^2+v^2}}.\ee This state is identified as the standard model Higgs boson. The remaining eigenvalue is \be m^2_{H_1} = -\fr{f\om}{\sqrt{2}}\left(\frac{u}{v}+\frac{v}{u}\right)\sim \om^2,\ee which corresponds to a new, heavy neutral scalar: \be H_1 \equiv  \fr{-vS_1 + uS_2}{\sqrt{u^2+v^2}}.\ee 
The $2 \times 2$ matrix at the second diagonal box provides two heavy eigenstates with their masses respectively given in the $\om$ scale,  
 \bea H_2 &\equiv & c_\varphi S_3 + s_\varphi S_4,\hs m^2_{H_2}= \la_2\om^2+\la\La^2 - \sqrt{\la^2_2\om^4+(\la^2_{11}-2\la\la_2)\om^2\La^2+\la^2\La^4}\sim \om^2,\crn 
 H_3 &\equiv& -s_\varphi S_3 + c_\varphi S_4,\hs m^2_{H_3}= \la_2\om^2+\la\La^2 + \sqrt{\la^2_2\om^4+(\la^2_{11}-2\la\la_2)\om^2\La^2+\la^2\La^4}\sim \om^2,\nn \eea
where the mixing angle is obtained by  \be t_{2\varphi} = -\fr{\la_{11}\om\La}{\la \La^2-\la_2 \om^2}.\ee We have adopted the notations $s_x = \sin x,\ c_x = \cos x,\ t_x=\tan x $, and so forth, for any $x$ angle like the $\varphi$ and others throughout this text.     

We see that at the leading order, the standard model like Higgs boson has a vanishing mass. Hence, when considering the next-to-leading order contribution, its mass gets generated to be small due to the perturbative expansion. In fact, we can write the general mass matrix $M^2_S$ in a new basis of the states $(H,\ H_1,\ H_2,\ H_3)$. Since the mass of the standard model like Higgs boson is much smaller than those of the new particles, the resulting mass matrix will have a seesaw like form \cite{seesaw} that can transparently be diagonalized. Indeed, putting  
\be
\left(\begin{array}{c}
S_1 \\
S_2\\
S_3\\
S_4
\end{array}\right) = U \left(\begin{array}{c}
H\\
H_1 \\
H_2\\
H_3
\end{array}\right),\hs 
 U \equiv \left(\begin{array}{cccc}
\fr{u}{\sqrt{u^2+v^2}} & -\fr{v}{\sqrt{u^2+v^2}} & 0 & 0\\
\fr{v}{\sqrt{u^2+v^2}} & \fr{u}{\sqrt{u^2+v^2}} & 0 & 0\\
0 & 0 & c_\varphi & -s_\varphi\\
0 & 0 & s_\varphi & c_\varphi
\end{array}\right), 
\ee the mass matrix (\ref{moivh}) in the new basis results \bea 
M'^2_S = U^T M^2_S U = 
\left (\begin{array}{cc}
A_{1\times 1} & B_{1\times 3} \\
B^T_{1\times 3} & C_{3\times3}
\end{array} \right ), \label{neutral2}
\eea where 
 \bea
A&\equiv& 2\fr{v^4 \la_1 + u^4 \la_3 + u^2 v^2 \la_5}{u^2 + v^2},\crn
B^T&\equiv& \left(
 \begin{array}{c}
    \fr{u v [v^2 (2 \la_1 - \la_5) +
   u^2 (-2 \la_3 + \la_5)]}{u^2 + v^2} \\
    \fr{s_\varphi \La (v^2 \la_{10} + u^2 \la_{12}) +
 c_\varphi (\sqrt{2} f u v + v^2 \om \la_4 + u^2 \om \la_6)}{\sqrt{
 u^2 + v^2}} \\
    \fr{-\sqrt2 f s_\varphi u v +
 c_\varphi \La (v^2 \la_{10} + u^2 \la_{12}) -
 s_\varphi \om (v^2 \la_4 + u^2 \la_6)}{\sqrt{u^2 + v^2}} \\
  \end{array}
\right),
\eea
and $C$ is a $3\times3$ matrix with corresponding components given by  
\bea
C_{11}&\equiv&\fr{-\sqrt2 f (u^2 + v^2)^2 \om +
 4 u^3 v^3 (\la_1 + \la_3 - \la_5)}{2 u v (u^2 + v^2)},\crn
C_{12}&=&C_{21}\equiv \fr{2 s_\varphi u v \La (\la_{10} - \la_{12}) +
 c_\varphi [\sqrt2 f (u^2-v^2) +
    2 u v \om (\la_4 - \la_6)]}{2 \sqrt{u^2 + v^2}},\crn
C_{13}&=&C_{31}\equiv \fr{\sqrt2 f s_\varphi (-u^2 + v^2) +
 2 u v [c_\varphi \La (\la_{10} - \la_{12}) +
    s_\varphi \om (-\la_4 + \la_6)]}{2 \sqrt{u^2 + v^2}},\crn
C_{22}&\equiv & 2s_\varphi^2 \la\La^2 +
 2c_\varphi (-\fr{c_\varphi f u v}{2 \sqrt2 \om} + s_\varphi \om \La \la_{11} +
    c_\varphi \om^2 \la_2),\crn
C_{23}&=&C_{32}\equiv  (c_\varphi^2 -s_\varphi^2) \om \La \la_{11} +
 2c_\varphi s_\varphi (\fr {f u v}{2\sqrt2 \om}  +  \la \La^2 -
     \om^2 \la_2),\crn
C_{33}&\equiv&-\fr{f s_\varphi^2 u v}{ \sqrt2 \om} +
 2c_\varphi \La (c_\varphi \la \La - s_\varphi \om \la_{11}) +
  2s_\varphi^2 \om^2 \la_2.
\eea
Because of $-f\sim \om\sim \La \gg u\sim v$, we achieve the seesaw form for $M'^2_S$, where $||C||\sim \om^2 \gg ||B||\sim u\om \gg ||A||\sim u^2$, with $||A||\equiv \sqrt{\mathrm{Tr}(A^T A)}$ and so forth. Therefore, the standard model like Higgs boson obtains a mass given by the seesaw formula \cite{seesaw}, 
\be
\de m^2_H = A - BC^{-1}B^T \sim \mathcal{O}(u^2,v^2),
\ee 
which is realized at the weak scales in spite of the large scales $\om$, $\La$ and $f$ (see below). The standard model like Higgs boson is given by \be 
H+\de H= H-BC^{-1}\left(\begin{array}{c}
H_1\\
H_2\\
H_3\end{array}\right).\ee  The physical heavy scalars are given to be orthogonal to this light state with their masses negligibly changed in comparison to the leading order values, respectively.    

The mass of the standard model like Higgs boson can be approximated as \bea
\de m_{H}^{ 2} &=&2\left( \frac{\la_3u^4+\la_5u^2 v^2+\la_1v^4}{u^2+v^2}+m^2_0+m^2_{1}\frac{f}{\om}+m^2_{2}\frac{f^2}{\om^2}\right),
\eea where the mass parameters $m_0,\ m_1,\ m_2$ are given by
\bea
m^2_0 &\equiv &-\frac{1}{(\la_{11}^2-4\la\la_2)(v^2+u^2)}\left[ -\la_{12}^2 \la_2u^4-\la(\la_6u^2+\la_4 v^2)^2
\right. \crn &&+ \left.
\la_{12}u^2(\la_{11}\la_6 u^2-2\la_{10}\la_2 v^2+\la_{11}\la_4v^2) +\la_{10}v^2(\la_{11}\la_6 u^2- \la_{10} \la_2 v^2+ \la_{11} \la_4 v^2) \right],\label{m00} \\
m^2_1 &\equiv& -\frac{\sqrt{2}uv\left[(\la_{11}\la_{12}-2\la\la_6)u^2+
(\la_{10}\la_{11}-2\la\la_4)v^2 \right]}{(\la_{11}^2-4\la\la_2)(u^2+v^2)},\label{m11} \\
m^2_2 &\equiv &\frac{2 \la u^2 v^2}{(\la_{11}^2-4\la \la_2)(u^2+v^2)}.\label{m22}
\eea Because the quantity $f/\om$ is finite, the Higgs mass $\de m^2_H$ depends on only the weak scales $u^2,\ v^2$ as stated. We will evaluate the Higgs mass and assign $\de m^2_H = (125\ \mathrm{GeV})^2$ as measured by the LHC \cite{atlas,cms}. For the purpose, let us assume $u=v,\ \om=-f$ that leads to \bea
\de m_{H}^{ 2}&=&\left( \la_3+\la_5+\la_1\right)u^2+2m_0^2- 2m_{1}^2+2m_2^2 \equiv \bar{\la}u^2.
\eea Here, $\bar{\la}$ is a function of only the $\la$'s couplings, which can easily be achieved with the help of (\ref{m00}), (\ref{m11}) and (\ref{m22}) for the respective $m^2_{0,1,2}$. In addition, we have $u^2+v^2 = (246\ \mathrm{GeV})^2$, i.e. $u = \frac{246}{\sqrt{2}}\ \mathrm{GeV}$, that is given from the mass of the $W$ boson as shown below. Hence, we identify $\de m^2_H =\bar{\la}\left(\frac{246}{\sqrt{2}}\ \mathrm{GeV}\right)^2 = (125\ \mathrm{GeV})^2$ that yields $\bar{\la} = \left(\frac{125\sqrt{2}}{246}\right)^2 \simeq 0.5$. This is an expected value for the effective self-interacting scalar coupling.  

In summary, we have the eleven Higgs bosons ($H^0$, $\mathcal{A}^0$, $H^0_{1,2,3}$, $H^{\pm}_{4,5}$, $H'^{0,0*}$) as well as the nine Goldstone bosons corresponding to the nine massive gauge bosons ($G^{\pm}_W$, $G^0_{Z}$, $G^{0,0*}_X$, $G^{\pm}_Y$, $G^0_{Z'}$, $G^0_{C}$). Because of the constraints $u,v \ll \om, \La, -f$, the standard model like Higgs boson ($\sim H$) results to be light with the mass at the weak scales, whereas all the new Higgs bosons are heavy with their masses at the $\om$, $\La$ or $-f$ scales. In calculating below, we will ignore the mixing effects of the standard model Higgs boson $H$ with the new particles $H_{1,2,3}$ (where the mixing angles defined by $BC^{-1}$ are typically proportional to $\frac{u}{\om}\ll 1$ which is actually small). Therefore, we have the $H,\ H_1,\ H_2,\ H_3$ as the physical states found out. Denoting $t_\beta=v/u$ and taking the effective limit $u/\om,\ v/\om \ll 1$, the physical scalar states are related to the gauge states as follows 
\bea
\left(
  \begin{array}{c}
    H \\
    H_1 \\
  \end{array}
\right) &\simeq&
\left(
  \begin{array}{cc}
    c_\beta & s_\beta \\
    -s_\beta & c_\beta \\
  \end{array}
\right)
\left(
  \begin{array}{c}
    S_1 \\
    S_2 \\
  \end{array}
\right),
\hs \left(
  \begin{array}{c}
   \mathcal{A} \\
   G_Z \\ 
  \end{array}
\right) \simeq 
\left(\begin{array}{cc}
c_\beta & s_\beta  \\
  -s_\beta &  c_\beta   \\
  \end{array}
\right)
\left(
  \begin{array}{c}
    A_2 \\
    A_1 \\
  \end{array}
\right),\crn
 \left(
  \begin{array}{c}
    H_2 \\
    H_3 \\
  \end{array}
\right) &\simeq& 
\left(
  \begin{array}{cc}
  c_\varphi & s_\varphi \\
  -s_\varphi & c_\varphi\\
  \end{array}
\right)
\left(
  \begin{array}{c}
    S_3 \\
    S_4 \\
  \end{array}
\right),\hs \left(
  \begin{array}{c}
    H_5^- \\
    G_W^- \\
  \end{array}
\right) =
\left(
  \begin{array}{cc}
    c_\beta & s_\beta  \\
    -s_\beta & c_\beta   \\
  \end{array}
\right)
\left(
  \begin{array}{c}
    \rho_1^- \\
    \eta_2^- \\
  \end{array}
\right),\crn
H_4 &\simeq& \rho_3,\ G_Y\simeq \chi_2,\ G_X\simeq \chi_1,\ H'\simeq \eta_3,\ G_{Z'}\simeq A_3,\ G_{C}=A_4. \label{vohuongvl}
\eea   

As mentioned, the mixings of the standard model Higgs boson $H$ with the new scalars $H_{1,2,3}$ are proportional to $u/\om$ where the proportional coefficients depend on the couplings of the scalar potential. Since the strengths of the scalar self-couplings are mostly unknown, those coefficients are undefined too. Therefore, if the coefficients are small as expected, the new physics effects via the mixings can be neglected, in similarity to the gauge boson sector discussed below. Otherwise, it is important to note that the leading-order new-physics effects must include the $\mathcal{O}(\{u,v\}/\{\om,\La,-f\})$ corrections to the couplings of the standard model Higgs boson due to the mixing with the new scalars as well as the modifications of the $H$ interactions to the new physics processes via those new scalars ($H_{1,2,3}$). In this case, the mixing parameters as determined by $BC^{-1}$ have to be taken into account. However, it is also noted that even for the proportional coefficients of order unity like a scalar self-coupling in the large strength regime, the modifications to the standard model Higgs couplings are around $|\Delta \kappa| \equiv u/\om\sim 0.1$ that easily satisfies the $\kappa_{k}$ bounds as presented in \cite{pdg}.          

Let us remind the reader that apart from the $H'$ that will be identified as a viable dark matter candidate, the remaining
scalars in this model would be sufficiently heavy in order to obey the bounds coming from the muon anomalous magnetic moment \cite{muonbound}.

\section{\label{gsec}Gauge sector}

The gauge bosons obtain masses when the scalar fields develop the VEVs. Therefore, their mass Lagrangian is given by 
\be \mathcal{L}^{\mathrm{gauge}}_{\mathrm{mass}}=\sum_\Phi (D^\mu\langle \Phi \rangle)^\dagger(D_\mu \langle \Phi \rangle ).\ee 
Substituting the scalar multiplets $\eta$, $\rho$, $\chi$ and $\phi$ with their covariant derivative, gauge charges and VEVs as given before, we get
\bea \mathcal{L}^{\mathrm{gauge}}_{\mathrm{mass}} &=&  \frac{ g^2u ^2 }{8}\left[ \left( A_{3\mu }  + \frac{A_{8\mu } }{\sqrt{3}} - \frac{2}{3}t_X B_\mu   + \frac{2}{3}t_N C_\mu   \right)^2  + 2W^+_\mu  W^{-\mu}  + 2X^{0*}_\mu X^{0\mu}  \right]\crn
&&+\frac{g^2v ^2 }{8}\left[ \left(  - A_{3\mu }  + \frac{A_{8\mu} }{\sqrt{3}} + \frac{4}{3}t_X B_\mu   + \frac{2}{3}t_N C_\mu \right)^2  + 2W^+_\mu W^{-\mu}  + 2Y^+_\mu Y^{-\mu}   \right]\crn
&&+ \frac{g^2\omega ^2 }{8}\left[ \left(  - \frac{2A_{8\mu } }{\sqrt{3}} - \frac{2}{3}t_X B_\mu   -  \frac{4}{3}t_N C_\mu   \right)^2  + 2Y^+_\mu Y^{-\mu}   + 2X^{0*}_\mu X^{0\mu}  \right]\crn
&& + 2g_N^2  \La ^2 C_\mu ^2,\eea where we have defined $t_X\equiv \fr{g_X}{g}$, $t_N\equiv \fr{g_N}{g}$,
and 
\be W_\mu^{\pm}=\fr{A_{1\mu} \mp i A_{2\mu}}{\sqrt 2},\hs X^{0,0*}_\mu= \fr{A_{4\mu}\mp i A_{5\mu}}{\sqrt 2},\hs
Y_\mu^\mp=\fr{A_{6\mu} \mp i A_{7\mu}}{\sqrt 2}. \label{nonherg}
 \ee 
 
The mass Lagrangian can be rewritten as 
\bea 
 \mathcal{L}_{\mathrm{mass}}^{\mathrm{gauge}}  
 &=& \frac{g^2 }{4}\left(u^2  + v^2\right) W^+ W^-+ \frac{g^2 }{4}  \left(v^2  + \omega ^2 \right) Y^ +  Y^ -  + \frac{g^2 }{4} \left( u^2  + \omega^2  \right) X^{0*}X^0 \crn
 &&+ \frac{1}{2}\left(A_{3} \  A_{8}  \  B \ C  \right) M^2 
 \left(\begin{array}{c}
   A_{3 }   \\
   A_{8 }   \\
   B  \\
   C  
\end{array} \right), \eea where the Lorentz indices have been omitted and should be understood. The squared-mass matrix of the neutral gauge bosons is found to be, 
 \bea M^2=
\fr{g^2}{2}\left(
  \begin{array}{cccc}
    \fr 1 2 (u^2 + v^2) &
     \fr{u^2 - v^2}{2\sqrt 3} &
    -\fr{t_X(u^2 +2 v^2)}{3} &
    \fr{t_N (u^2 - v^2)}{3} \\
    \fr{u^2 - v^2}{2\sqrt 3} &
    \fr 1 6 (u^2 + v^2 + 4 \om^2) &
    -\fr {t_X(u^2 -2(  v^2  +\om^2))}{3\sqrt 3} &
    \fr {t_N(u^2 +v^2 +4\om^2)}{3\sqrt 3} \\
    -\fr{t_X(u^2 +2 v^2)}{3} &
    -\fr {t_X(u^2 -2(  v^2  +\om^2))}{3\sqrt 3} &
    \fr {2}{9}t_X^2 (u^2 +4 v^2  +\om^2) &
    -\fr {2}{9}t_X t_N (u^2 -2( v^2 +\om^2 )) \\
    \fr{t_N (u^2 - v^2)}{3} &
    \fr {t_N (u^2 +v^2 +4\om^2)}{3\sqrt 3} &
    -\fr {2}{9}t_X t_N (u^2 -2( v^2 +\om^2 )) &
    \fr {2}{9}t_N^2 (u^2 + v^2 +4 ( \om^2+9 \La^2))  \\
  \end{array}
\right).\nn
\eea

The non-Hermitian gauge bosons $W^{\pm}$, $X^{0,0*}$ and $Y^\pm$ by themselves are physical fields with corresponding masses,
\be
m^2_W
=\fr 1 4 g^2 (u^2 + v^2 ),\hs m^2_X=\fr 1 4 g^2 (u^2 + \om^2),\hs
m^2_Y =\fr 1 4 g^2 (v^2 + \om^2).
\ee Because of the constraints $u, v\ll \om$, we have $m_W\ll m_{X}\simeq m_{Y}$. The $W$ is identified as the standard model $W$ boson, which implies 
\be u^2+v^2=(246\ \mathrm{GeV})^2.\ee The $X$ and $Y$ fields are the new gauge bosons with the large masses as given in the $\om$ scale.

The neutral gauge bosons $(A_3,\ A_8,\ B,\ C)$ mix via the mass matrix $M^2$. It is easily checked that $M^2$ has a zero eigenvalue with corresponding eigenstate, 
\be
m^2_A=0,\hs A_\mu= \fr {\sqrt3} {\sqrt{3 + 4 t_X^2}}
\left(t_X A_{3\mu} -  \fr{t_X}{\sqrt 3}A_{8\mu} +B_\mu\right),
 \ee which are independent of the VEVs and identified as those of the photon (notice that all the other eigenvalues of $M^2$ are nonzero). The independence of the VEVs for the photon field and its mass is a consequence of the electric charge conservation \cite{tw}. With this at hand, electromagnetic vertices can be calculated that result in the form $-eQ(f)\bar{f}\ga^\mu f A_\mu$, where the electromagnetic coupling constant is identified as $e=gs_W$ in which the sine of Weinberg's angle is given by \cite{tw}
\be
s_W=\fr{\sqrt{3}t_X}{\sqrt{3+4t^2_X}}.
\ee
The photon field can be rewritten as \be \fr{A_\mu}{e}=\fr{A_{3\mu}}{g}-\fr{1}{\sqrt{3}}\fr{A_{8\mu}}{g}+\fr{B_\mu}{g_X},\ee which is identical to the electric charge operator expression in (\ref{ecqbl}) if one replaces its generators by the corresponding gauge bosons over couplings (namely, the $Q$ is replaced by $A_\mu/e$, the $T_i$ by $A_{i\mu}/g$, and the $X$ by $B_\mu/g_X$). Hence, $A_\mu$ can be achieved from $Q$ that need not mention $M^2$. The mass eigenstate $A_\mu$ depends on just
$A_{3\mu}$, $A_{8\mu}$ and $B_\mu$, whereas the new gauge boson $C_\mu$ does not give any contribution, which results from the electric charge conservation too \cite{tw}.  

To identify the physical gauge bosons, we firstly rewrite the photon field in the form of 
\be A=s_W A_3 + c_W\left(-\fr{t_W}{\sqrt{3}}A_8+\sqrt{1-\fr{t^2_W}{3}}B\right),\ee with the aid of $t_X=\sqrt{3}s_W/\sqrt{3-4s^2_W}$. In the above expression, the combination in the parenthesis $(\cdots)$ is just the field that is associated with the weak hyper-charge $Y=-\fr{1}{\sqrt{3}}T_8+X$. The standard model $Z$ boson is therefore identified as 
\be Z=c_W A_3 -s_W\left(-\fr{t_W}{\sqrt{3}}A_8+\sqrt{1-\fr{t^2_W}{3}}B\right),\ee which is orthogonal to the $A$ as usual. The 3-3-1 model $Z'$ boson, which is a new neutral one, is obtained to be orthogonal to the field that is coupled to the hyper-charge $Y$ as mentioned (thus it is orthogonal to both the $A$ and $Z$ bosons), 
\be Z'=\sqrt{1-\fr{t^2_W}{3}}A_8+\fr{t_W}{\sqrt{3}}B. \ee Hence, we can work in a new basis of the form $(A,\ Z,\ Z',\ C)$, where the photon is a physical particle and decoupled while the other fields $Z,\ Z'$ and $C$ mix themselves.     

The mass matrix $M^2$ can be diagonalized via several steps. In the first step, we change the basis to: $(A_3,\ A_8,\ B,\ C)\rightarrow (A,\ Z,\ Z',\ C)$,
\bea \left(\begin{array}{c}
A_3\\
A_8\\
B\\
C
\end{array}\right)=
U_1\left(
\begin{array}{c}
A\\
Z\\
Z'\\
C\end{array}
\right),\hs 
U_1=\left(
\begin{array}{cccc}
s_W & c_W & 0 & 0\\
-\fr{s_W}{\sqrt{3}} & \fr{s_W t_W}{\sqrt{3}} & \sqrt{1-\fr{t^2_W}{3}} & 0 \\
c_W\sqrt{1-\fr{t^2_W}{3}} & -s_W\sqrt{1-\fr{t^2_W}{3}}& \fr{t_W}{\sqrt{3}} & 0\\
0 & 0 & 0 & 1
\end{array} \right). \eea In this new basis, the mass matrix $M^2$ becomes
\bea M'^2=U^T_1 M^2 U_1=
\left(
\begin{array}{cc}
0 & 0 \\
0 & M'^2_{s}\end{array} \right), \eea where the 11 component is the zero mass of the photon which is decoupled, while the $M'^2_{s}$ is a $3\times 3$ mass sub-matrix of $Z,\ Z'$ and $C$,
\bea && M'^2_{s} \equiv
\left(
\begin{array}{ccc}
m^2_{Z} & m^2_{ZZ'} & m^2_{ZC}\\
m^2_{ZZ'} & m^2_{Z'} & m^2_{Z'C}\\
m^2_{ZC} & m^2_{Z'C} & m^2_{C}
\end{array}
\right)=
\fr{g^2}{2}\times \crn
&& \left(
  \begin{array}{ccc}
    \fr {(3 + 4 t_X^2) (u^2 + v^2)}{ 2 (3 + t_X^2)}
    &  \fr{\sqrt{3 + 4 t_X^2} ((3 - 2 t_X^2) u^2 - (3 + 4 t_X^2) v^2)}
    {6 (3 + t_X^2)}
    & \fr {\sqrt{3 + 4 t_X^2} t_N (u^2-v^2)}{3 \sqrt{3 + t_X^2}} \\
    \fr{\sqrt{3 + 4 t_X^2} ((3 - 2 t_X^2) u^2 - (3 + 4 t_X^2) v^2)}
    {6 (3 + t_X^2)}
    & \fr {(3 -2 t_X^2)^2 u^2 + (3 + 4 t_X^2)^2 v^2 + 4(3 + t_X^2)^2 \om^2}{18 (3 + t_X^2)}
    & \fr {t_N ((3-2 t_X^2) u^2 + (3 + 4 t_X^2) v^2 + 4 (3 + t_X^2) \om^2)}{9\sqrt{3 + t_X^2}} \\
    \fr {\sqrt{3 + 4 t_X^2} t_N (u^2-v^2)}{3 \sqrt{3 + t_X^2}}
    & \fr {t_N ((3-2 t_X^2) u^2 + (3 + 4 t_X^2) v^2 + 4 (3 + t_X^2) \om^2)}{9\sqrt{3 + t_X^2}}
    & \fr 2 9 t_N^2 (u^2 + v^2 + 4 (\om^2 +9 \La^2)) \\
  \end{array}
\right). \nn
\eea

Because of the conditions, $u,v\ll \om,\La$, we have $m^2_{Z},\ m^2_{ZZ'},\ m^2_{ZC}\ll m^2_{Z'},\ m^2_{Z'C},\ m^2_C$. Hence, in the second step, the mass matrix $M'^2$ (or $M'^2_{s}$) can be diagonalized by using the seesaw formula \cite{seesaw} to separate the light state ($Z$) from the heavy states ($Z',\ C$). We denote the new basis as $(A,\ Z_{1},\ \mathcal{Z}^\prime,\ \mathcal{C})$ so that the $A,\ Z_1$ are physical fields and decoupled while the rest mix,    
\bea 
\left(
  \begin{array}{c}
  A\\
    Z \\
    Z^\prime \\
    C \\
  \end{array}
\right)
&=& U_2\left(
  \begin{array}{c}
    A\\
    Z_{1} \\
    \mathcal{Z}^\prime \\
    \mathcal{C} \\
  \end{array}
\right),\hs 
M''^2 = U^T_2 M'^2 U_2 =
\left(
  \begin{array}{ccc}
  0 & 0 & 0 \\
    0 & m_{Z_1}^2 & 0 \\
    0 & 0 & M''^2_{s} \\
  \end{array}
\right),\eea 
where $M''^2_s$ is a $2\times 2$ mass sub-matrix of the $\mathcal{Z}',\ \mathcal{C}$ heavy states, while $m_{Z_1}$ is the mass of the $Z_1$ light state. By the virtue of seesaw approximation, we have    
\bea
U_2 &\simeq& \left(
   \begin{array}{ccc}
 1 & 0 & 0  \\
   0 &  1 & \mathcal{E} \\
    0 & -\mathcal{E}^T & 1 \\
   \end{array}
 \right),\hs \mathcal{E} \equiv (m^2_{ZZ'}\ m^2_{ZC})\left(
\begin{array}{cc}
m^2_{Z'} & m^2_{Z'C}\\
m^2_{Z'C} & m^2_C
\end{array}
\right)^{-1}, \\ 
m_{Z_1}^2 &\simeq& m^2_Z-\mathcal{E} \left(
\begin{array}{c}
m^2_{ZZ'}\\
m^2_{ZC}\end{array}\right),\hs 
M''^2_s  \simeq  \left(
\begin{array}{cc}
m^2_{Z'} & m^2_{Z'C}\\
m^2_{Z'C} & m^2_C
\end{array}
\right). \eea The $\mathcal{E}$ is a two-component vector given by 
 \bea
\mathcal{E}_1 &=& -\fr{\sqrt{4 t_X^2+3} \{3 \La^2 [(2 t_X^2-3) u^2+(4 t_X^2+3) v^2]+t_X^2 \om^2 (u^2+v^2)\}}{4 \La^2 (t_X^2+3)^2 \om^2}\ll 1,\crn
\mathcal{E}_2 &=& \fr{t_X^2 \sqrt{4 t_X^2+3} (u^2+v^2)}{8 \La^2 (t_X^2+3)^{3/2} t_N}\ll 1,\nn\eea which are suppressed at the leading order $u,v\ll\om,\La$. The $Z_1$, $\mathcal{Z}'$ and $\mathcal{C}$ fields are the standard model like, 3-3-1 model like and $U(1)_N$ like gauge bosons, respectively. To be concrete, we write $Z_1\simeq Z-\mathcal{E}_1 Z'-\mathcal{E}_2 C$, $\mathcal{Z}'\simeq Z'+\mathcal{E}_1 Z$ and $\mathcal{C}\simeq C+\mathcal{E}_2 Z$ which differ from the $Z$, $Z'$ and $C$ fields by the only small mixing terms, respectively.       

Moreover, with the help of $t_X=\sqrt{3}s_W/\sqrt{3-4s^2_W}$, we have 
\be \mathcal{E}_1=-\fr{\sqrt{3-4s^2_W}}{4c^4_W}\left[\fr{v^2-c_{2W}u^2}{\om^2}+\fr{s^2_W(u^2+v^2)}{9\La^2}\right],\hs \mathcal{E}_2=\fr{s^2_W}{24c^3_W t_N}\fr{u^2+v^2}{\La^2}.\ee We realize that the first term in $\mathcal{E}_1$ is just the mixing angle of $Z$-$Z'$ in the 3-3-1 model with right-handed neutrinos, $t_{\theta}\simeq \sqrt{3-4s^2_W}(c_{2W}u^2-v^2)/(4c^4_W\om^2)$ \cite{tw}, when $\La\gg \om$. With the aid of $v^2_{\mathrm{w}}\equiv u^2+v^2=(246\ \mathrm{GeV})^2$ (that is the weak scale and is fixed) as well as $0<u^2, v^2<v^2_{\mathrm{w}}$, the $\mathcal{E}_1$ parameter is bounded by \be -\fr{\sqrt{3-4s^2_W}}{4c^4_W}\left[\left(\fr{v_{\mathrm{w}}}{\om}\right)^2+\fr{s^2_W}{9}\left(\fr{v_{\mathrm{w}}}{\La}\right)^2\right]<\mathcal{E}_1< -\fr{\sqrt{3-4s^2_W}}{4c^4_W}\left[-c_{2W}\left(\fr{v_{\mathrm{w}}}{\om}\right)^2+\fr{s^2_W}{9}\left(\fr{v_{\mathrm{w}}}{\La}\right)^2\right],\ee where the second terms in the brackets are negligible since $\La \gtrsim \om$.   
Therefore, the $\mathcal{E}_{1}$ bounds as well as the $\mathcal{E}_2$ parameter can be approximated as
\be -3.5\times 10^{-3}<\mathcal{E}_1< 3 \times 10^{-3},\hs \mathcal{E}_2\simeq 0.014\left(\fr{1}{t_N}\right)\left(\fr{v_\mathrm{w}}{\La}\right)^2\sim 10^{-4},\label{e1e2sosanh}\ee provided that $s^2_W\simeq 0.231$, $t_N\sim 1$, $ \La \sim \om$ and $\om>3.198$ TeV as given from the $\rho$-parameter below. With such small values of the $\mathcal{E}_{1,2}$ mixing parameters, their corrections to the couplings of the $Z$ boson such as the well-measured $Zf\bar{f}$ ones (due to the mixing with the new $Z',\ C$ gauge bosons) can be neglected \cite{pdg}. [But, notice that they can be changed due to the one-loop effects of $Z',\ C$ as well as of the non-Hermitian $X,\ Y$ gauge bosons accompanied by the corresponding new fermions, which subsequently give the constraints on their masses and the $g_N$ coupling. A detailed study on this matter is out of the scope of this work and it should be published elsewhere]. Even, the modifications of the $Z$ interactions (due to the mixings) to the new physics processes via the $Z'$, $C$ bosons are negligible, which will be explicitly shown when some of those processes are mentioned at the end of this work. Therefore, except for an evaluation of the mentioned $\rho$-parameter, we will use only the leading order terms below. In other words, the mixing of the $Z$ with the $Z',\ C$ bosons can be neglected so that $m_{Z_1}\simeq m_Z$, $Z_1\simeq Z$, $\mathcal{Z}'\simeq Z'$ and $\mathcal{C}\simeq C$.  

For the final step, it is easily to diagonalize $M''^2$ (or $M''^2_s$) to obtain the remaining two physical states, denoted by $Z_{2}$ and $Z_{N}$, such that  
\bea \left(\begin{array}{c}
A \\
 Z_{1}\\
  \mathcal{Z}^\prime\\
   \mathcal{C}
   \end{array}\right)
   &=& U_3\left(\begin{array}{c}
   A\\
    Z_{1}\\
     Z_2\\
      Z_N
      \end{array}\right),\hs 
U_3=\left(
\begin{array}{cccc}
1 & 0 & 0& 0\\
0 & 1& 0 & 0\\
0 & 0 & c_\xi & -s_\xi \\
0 & 0& s_\xi & c_\xi 
\end{array}
\right),\crn
M'''^2 &=& U^T_3 M''^2 U_3 = \mathrm{diag}(0,m^2_{Z_1},m^2_{Z_2},m^2_{Z_N}).
\eea
The mixing angle and new masses are given by
\bea
t_{2\xi}&\simeq &\fr{4\sqrt{3 + t_X^2} t_N \om^2}
{(3 + t_X^2) \om^2- 4 t_N^2 (\om^2 + 9 \La^2)},\label{goctronxi}\\ 
 m_{Z_N}^2 & \simeq &
\fr{g^2} {18}\left((3 + t_X^2) \om^2+ 4 t_N^2 (\om^2 + 9 \La^2) + \sqrt{((3 + t_X^2) \om^2- 4 t_N^2 (\om^2 + 9 \La^2))^2
+ 16 (3 + t_X^2)t_N^2 \om^4}\right),\crn \label{kluongz2nhe}\\
m_{Z_2}^2& \simeq&
\fr{g^2} {18}\left((3 + t_X^2) \om^2+ 4 t_N^2 (\om^2 + 9 \La^2) -\sqrt{((3 + t_X^2) \om^2- 4 t_N^2 (\om^2 + 9 \La^2))^2
+ 16 (3 + t_X^2)t_N^2 \om^4}\right).\crn \label{kluongznnhe} 
\eea It is noteworthy that the mixing of the 3-3-1 model $Z'$ boson and $U(1)_N$ $C$ boson is finite and may be large since $\om\sim \La$. The $Z_2$ and $Z_N$ are heavy particles with the masses in the $\om$ scale.  

In summary, the physical fields are related to the gauge states as 
\bea \left(\begin{array}{c}
A_{3}\\
A_{8}\\
B\\
C\end{array}
\right)=U \left(
\begin{array}{c}
A\\
Z_1\\
Z_2\\
Z_N\end{array}\right),\label{herg}\eea 
where 
\bea U=U_1U_2 U_3\simeq U_1 U_3 = 
\left(
\begin{array}{cccc}
s_W & c_W & 0 & 0\\
-\fr{s_W}{\sqrt{3}} & \fr{s_W t_W}{\sqrt{3}} & c_\xi \sqrt{1-\fr{t^2_W}{3}} & -s_\xi \sqrt{1-\fr{t^2_W}{3}} \\
c_W\sqrt{1-\fr{t^2_W}{3}} & -s_W\sqrt{1-\fr{t^2_W}{3}}& c_\xi \fr{t_W}{\sqrt{3}} & -s_\xi \fr{t_W}{\sqrt{3}} \\
0 & 0 & s_\xi & c_\xi
\end{array} \right).
\label{thuy}\eea The approximation above is given at the leading order $\{u^2,v^2\}/\{\om^2,\La^2\}\ll 1$ and this means that the standard model $Z$ boson by itself is a physical field $Z\simeq Z_1$ that does not mix with the new neutral gauge bosons, $Z_{2}$ and $Z_N$.

The next-to-leading order term $(\mathcal{E})$ gives a contribution to the $\rho$-parameter obtained by
\be \rho=\fr{m^2_W}{c^2_Wm^2_{Z_1}}=\fr{m^2_Z}{m^2_Z-\mathcal{E}(m^2_{ZZ'}\ m^2_{ZC})^T}\simeq 1+\mathcal{E}(m^2_{ZZ'}\ m^2_{ZC})^T/m^2_Z.\ee Here, notice that $m_W=c_W m_Z$ and $m^2_Z\sim m^2_{ZZ'}\sim m^2_{ZC}$. To have a numerical value, let us put $u= v=(246/\sqrt{2})$ GeV and $\om = \La$. Hence, we get the deviation as 
\be \Delta \rho \equiv \rho-1 \simeq \fr{5s^2_W t^4_W}{18\pi\al}\fr{u^2}{\om^2}\simeq 0.236 \fr{u^2}{\om^2},\ee with the aid of $s^2_W=0.231,\ \al=1/128$ \cite{pdg}. From the experimental data
$\Delta \rho < 0.0007$ \cite{pdg}, we have $u/\om < 0.0544$ or $\om>3.198$ TeV (provided that $u=246/\sqrt{2}$ GeV as mentioned). Therefore, the value of $\om$ results in the TeV scale as expected.

\section{\label{ints}Interactions}

\subsection{Fermion--gauge boson interaction}

The interactions of fermions with gauge bosons are derived from the Lagrangian, 
\be \mathcal{L}_{\mathrm{fermion}}\equiv \bar{\Psi}i\gamma^\mu D_\mu \Psi,\ee where $\Psi$ runs on all the fermion multiplets of the model. The covariant derivative as defined in (\ref{dhhb}) can be rewritten as $D_\mu=\pa_\mu + ig_s G_\mu + ig P_\mu$, where $G_\mu\equiv t_iG_{i\mu}$ and $P_\mu\equiv  T_iA_{i\mu}+t_X X B_\mu +t_NNC_\mu$ (note that $t_X=g_X/g,\ t_N=g_N/g$). Expanding the Lagrangian we find,
\be \mathcal{L}_{\mathrm{fermion}}=\bar{\Psi}i\gamma^\mu \pa_\mu\Psi -g_s\bar{\Psi}\gamma^\mu G_{\mu} \Psi -g\bar{\Psi}\gamma^\mu P_\mu \Psi,\label{gint} \ee where the first term is kinematic whereas the last two give rise to the strong, electroweak and $B-L$ interactions of the fermions.   

Notice that the $SU(3)_C$ generators, $t_i$, equal to $0$ for leptons and $\fr{\la_i}{2}$ for quarks $q$, where $q$ indicates to all the quarks of the model such as $q=u,\ d,\ c,\ s,\ t,\ b,\ D_{1,2},\ U$. Hence, the interactions of gluons with fermions as given by the second term of (\ref{gint}) yield   
\be -g_s\bar{\Psi}\gamma^\mu G_{\mu} \Psi =-g_s \bar{q}_L\gamma^\mu \fr{\la_i}{2}q_L G_{i\mu}-g_s \bar{q}_R\gamma^\mu \fr{\la_i}{2}q_R G_{i\mu}=-g_s \bar{q}\gamma^\mu \fr{\la_i}{2}q G_{i\mu},\ee which takes the form as usual (only the colored particles have the strong interactions).  

Let us separate $P=P^{\mathrm{CC}}+P^{\mathrm{NC}}$, where
\bea 
P^{\mathrm{CC}}&\equiv& T_1 A_1+ T_2 A_2 + T_4 A_4 + T_5 A_5 + T_6 A_6 + T_7 A_7,\crn
P^{\mathrm{NC}}&\equiv& T_3 A_3 + T_8 A_8 + t_X X B + t_N N C. \eea 
Hence, the last term of (\ref{gint}) can be rewritten as 
\be -g\bar{\Psi}\ga^\mu P_\mu \Psi=-g\bar{\Psi}\ga^\mu P^{\mathrm{CC}}_\mu \Psi-g\bar{\Psi}\ga^\mu P^{\mathrm{NC}}_\mu \Psi.\ee Here, the first term provides the interactions of the non-Hermitian gauge bosons $W^\mp$, $X^{0,0*}$, and $Y^{\pm}$ with the fermions, while the last term leads to the interactions of the neutral gauge bosons $A$, $Z_1$, $Z_2$, and $Z_N$ with the fermions. 

Substituting the gauge states from (\ref{nonherg}) into $P^{\mathrm{CC}}$, we get 
\be P^{\mathrm{CC}}=\fr{1}{\sqrt{2}}T^+ W^+ + \fr{1}{\sqrt{2}}U^+ X^{0}+\fr{1}{\sqrt{2}} V^+ Y^-+H.c.,\label{partcc} \ee where the raising and lowering operators are defined as 
\be T^\pm \equiv T_1\pm i T_2,\hs U^\pm\equiv T_4\pm i T_5,\hs V^\pm\equiv T_6\pm i T_7.\ee Notice that $T^\pm$, $U^\pm$ and $V^\pm$ vanish for the right-handed fermion singlets. Therefore, the interactions of the non-Hermitian gauge bosons with fermions are obtained by 
\bea -g\bar{\Psi}\ga^\mu P^{\mathrm{CC}}_\mu \Psi 
&=& -\fr{g}{\sqrt{2}}\bar{\Psi}\ga^\mu (T^+ W^+_\mu + U^+ X^{0}_\mu+ V^+ Y^-_\mu) \Psi +H.c.\crn
&=&-\fr{g}{\sqrt{2}}\bar{\Psi}_L \ga^\mu T^+\Psi_L W^+_\mu -\fr{g}{\sqrt{2}}\bar{\Psi}_L\ga^\mu U^+\Psi_L X^{0}_\mu -\fr{g}{\sqrt{2}}\bar{\Psi}_L\ga^\mu V^+\Psi_L Y^-_\mu  +H.c.\crn
&=& J^{-\mu}_W W^+_\mu + J^{0\mu}_X X^0_\mu + J^{+\mu}_Y Y^-_\mu +H.c, \eea where the currents as associated with the corresponding non-Hermitian gauge bosons are given by 
\bea J^{-\mu}_W&\equiv &-\fr{g}{\sqrt{2}}\bar{\Psi}_L \ga^\mu T^+\Psi_L=-\fr{g}{\sqrt{2}}\left(\bar{\nu}_{aL}\ga^\mu e_{aL}+\bar{u}_{aL}\ga^\mu d_{aL}\right),\crn
J^{0\mu}_X &\equiv& -\fr{g}{\sqrt{2}}\bar{\Psi}_L\ga^\mu U^+\Psi_L = -\fr{g}{\sqrt{2}}\left(\bar{\nu}_{aL}\ga^\mu N^c_{aR}+\bar{u}_{3L}\ga^\mu U_L - \bar{D}_{\al L}\ga^\mu d_{\al L}\right),\\
J^{+\mu}_Y&\equiv&-\fr{g}{\sqrt{2}}\bar{\Psi}_L\ga^\mu V^+\Psi_L=-\fr{g}{\sqrt{2}}\left(\bar{e}_{aL}\ga^\mu N^c_{aR}+\bar{d}_{3L}\ga^\mu U_L +\bar{D}_{\al L}\ga^\mu u_{\al L}\right).\nn \eea The interactions of the $W$ boson are similar to those of the standard model, while the new interactions with the $X$ and $Y$ bosons are like those of the ordinary 3-3-1 model. 

Substituting the gauge states as given by (\ref{herg}) into $P^{\mathrm{NC}}$, we have 
\bea P^{\mathrm{NC}}_\mu &=& s_W Q A_\mu + \fr{1}{c_W} \left(T_3-s^2_W Q\right) Z_\mu\crn
&&+\fr{1}{c_W}\left[c_\xi \left(\sqrt{\fr{3-4s^2_W}{3}}T_8 + \fr{s^2_W}{\sqrt{3-4s^2_W}} X\right)+s_\xi c_W t_N N \right]Z_{2\mu}\crn
&&+\fr{1}{c_W}\left[-s_\xi \left(\sqrt{\fr{3-4s^2_W}{3}} T_8 +  \fr{s^2_W}{\sqrt{3-4s^2_W}} X\right)+c_\xi c_W t_N N\right]Z_{N\mu}.\label{partnc} \eea For this expression, we have used $t_X=\sqrt{3}s_W/\sqrt{3-4s^2_W}$ and $Q=T_3-T_8/\sqrt{3}+X$. The interactions of the neutral gauge bosons with fermions are given by 
\bea -g\bar{\Psi}\ga^\mu P^{\mathrm{NC}}_\mu \Psi &=& -gs_W \bar{\Psi}\ga^\mu Q \Psi A_\mu  -\fr{g}{c_W}\bar{\Psi}\ga^\mu  \left(T_3-s^2_W Q\right)\Psi Z_\mu \crn 
&&-\fr{g}{c_W}\bar{\Psi} \ga^\mu \left[c_\xi \left(\sqrt{\fr{3-4s^2_W}{3}}T_8 + \fr{s^2_W}{\sqrt{3-4s^2_W}} X\right)+s_\xi c_W t_N N \right] \Psi Z_{2\mu}\crn 
&&-\fr{g}{c_W}\bar{\Psi}\ga^\mu\left[-s_\xi \left(\sqrt{\fr{3-4s^2_W}{3}} T_8 +  \fr{s^2_W}{\sqrt{3-4s^2_W}} X\right)+c_\xi c_W t_N N\right]\Psi Z_{N\mu}.\label{ttz2zn} \eea Three remarks are in order
\ben
\item With the help of $e=gs_W$, the interactions of photon with fermions take the normal form \be -gs_W \bar{\Psi}\ga^\mu Q \Psi A_\mu = -eQ(f)\bar{f}\ga^\mu f A_\mu, \ee where $f$ indicates to any fermion of the model. 
\item The interactions of $Z$ with fermions can be rewritten as 
\bea -\fr{g}{c_W}\bar{\Psi}\ga^\mu  \left(T_3-s^2_W Q\right)\Psi Z_\mu &=& -\fr{g}{c_W}\left\{\bar{f}_L\ga^\mu  \left[T_3(f_L)-s^2_W Q(f_L)\right]f_L\right.\crn
&& \left.+\bar{f}_R\ga^\mu  \left[-s^2_W Q(f_R)\right]f_R\right\} Z_\mu,\crn
&=& -\fr{g}{2c_W}\bar{f}\ga^\mu \left[g^Z_V(f)-g^Z_A(f)\ga_5\right]f Z_\mu, \eea
where \be g^Z_V(f)\equiv T_3(f_L)-2s^2_W Q(f),\hs g^Z_A(f)\equiv T_3(f_L).\ee Therefore, the interactions of $Z$ take the normal form. For a convenience in reading, the couplings of  $Z$ with fermions are given in Table \ref{tttzz}.
\begin{table}[htdp]
\bc
\begin{tabular}{|c|c|c|}
\hline
$f$ & $g^Z_V(f)$ & $g^Z_A(f)$ \\ 
\hline
$\nu_a$ &  $\fr 1 2$ & $\fr 1 2$ \\
\hline
$e_a$ & $-\fr 1 2 + 2 s^2_W$ & $-\fr 1 2 $ \\
\hline
$N_a$ & 0 & 0 \\
\hline
$u_a$ & $\fr 1 2 -\fr 4 3 s^2_W$ & $\fr 1 2$ \\
\hline
$d_a$ & $-\fr 1 2 + \fr 2 3 s^2_W $ & $-\fr 1 2$ \\
\hline
$U$ & $-\fr 4 3 s^2_W$ & $0$ \\
\hline
$D_\al$ & $\fr 2 3 s^2_W$ & $0$ \\
\hline
\end{tabular}
\caption{\label{tttzz}The couplings of $Z$ with fermions.}  
\ec
\end{table}  
\item 
It is noteworthy that the interactions of $Z_2$ with fermions are identical to those of $Z_N$ if one makes a replacement in the $Z_2$ interactions by $c_\xi\rightarrow -s_\xi,\ s_\xi \rightarrow c_\xi$, and vice versa. Thus, we need only to obtain the interactions of either $Z_2$ or $Z_N$, the remainders are straightforward.    
\een   

The interactions of $Z_2$ and $Z_N$ with fermions can respectively be rewritten in a common form like that of $Z$. Therefore, the last two terms of (\ref{ttz2zn}) yield
\be -\fr{g}{2c_W}\bar{f}\ga^\mu\left[g^{Z_2}_V(f)-g^{Z_2}_A(f)\ga_5\right]f Z_{2\mu}-\fr{g}{2c_W}\bar{f}\ga^\mu\left[g^{Z_N}_V(f)-g^{Z_N}_A(f)\ga_5\right]f Z_{N\mu},\label{ttva2nd}\ee
where 
\bea 
  g^{Z_2}_A(f) &=& -\frac{c_\xi s^2_W}{\sqrt{3-4s^2_W}}T_3(f_L)+ \left(\frac{\sqrt{3}c_\xi c^2_W}{\sqrt{3-4s^2_W}}+\frac{2 s_\xi c_W t_N}{\sqrt{3}}\right) T_8(f_L),\crn
   g^{Z_2}_V(f) &=& g^{Z_2}_A(f)+ 2 \frac{c_\xi s^2_W}{\sqrt{3-4s^2_W}}Q(f)+2 s_\xi c_Wt_N(B-L)(f),\crn
   g^{Z_N}_{A,V}&=&g^{Z_2}_{A,V}(c_\xi \rightarrow -s_\xi,\ s_\xi \rightarrow c_\xi).
  \eea 
The interactions of $Z_2$ and $Z_N$ with fermions are listed in Table \ref{tttz2} and \ref{tttzn}, respectively. 
\begin{table}[htdp]
\bc
\begin{tabular}{|c|c|c|}
\hline
$f$ & $g^{Z_2}_V(f)$ & $g^{Z_2}_A(f)$ \\
\hline 
$\nu_a $ & $\fr{c_\xi c_{2W}}{2\sqrt{3-4s^2_W}} -\fr 5 3 s_\xi c_W t_N$ & $\fr{c_\xi c_{2W}}{2\sqrt{3-4s^2_W}} +\fr 1 3 s_\xi c_W t_N$ \\
\hline
$e_a$ & $\fr{c_\xi (1-4 s^2_W) }{2\sqrt{3-4s^2_W}} -\fr 5 3 s_\xi c_W t_N$ & $\fr{c_\xi }{2\sqrt{3-4s^2_W}} +\fr 1 3 s_\xi c_W t_N$ \\
\hline 
$N_a$ & $\fr{c_\xi c^2_W}{\sqrt{3-4s^2_W}}+\fr 2 3 s_\xi c_W t_N$ & $-\fr{c_\xi c^2_W}{\sqrt{3-4s^2_W}}-\fr 2 3 s_\xi c_W t_N$ \\
\hline 
$u_\al$ & $-\fr{c_\xi (3-8s^2_W)}{6\sqrt{3-4s^2_W}}+\fr 1 3 s_\xi c_W t_N$ & $- \fr{c_\xi }{2\sqrt{3-4s^2_W}} -\fr 1 3 s_\xi c_W t_N$ \\
\hline 
$u_3$  & $\fr{c_\xi(3+2s^2_W)}{6\sqrt{3-4s^2_W}}+s_\xi c_W t_N$ & $\fr{c_\xi c_{2W}}{2\sqrt{3-4s^2_W}} +\fr 1 3 s_\xi c_W t_N$ \\
\hline 
$d_\al$ & $-\fr{c_\xi (3-2s^2_W)}{6\sqrt{3-4s^2_W}}+\fr 1 3 s_\xi c_W t_N$ & $-\fr{c_\xi c_{2W}}{2\sqrt{3-4s^2_W}} - \fr 1 3 s_\xi c_W t_N$\\
\hline
$d_3$ & $ \fr{c_\xi\sqrt{3-4s^2_W}}{6}+s_\xi c_W t_N$ & $\fr{c_\xi }{2\sqrt{3-4s^2_W}} +\fr 1 3 s_\xi c_W t_N$ \\
\hline 
$U$ & $-\fr{c_\xi (3-7s^2_W)}{3\sqrt{3-4s^2_W}}+2 s_\xi c_W t_N$ & $-\fr{c_\xi c^2_W}{\sqrt{3-4s^2_W}}-\fr 2 3 s_\xi c_W t_N$ \\
\hline
$D_\al$ & $\fr{c_\xi (3-5s^2_W)}{3\sqrt{3-4s^2_W}}-\fr 2 3 s_\xi c_W t_N$ & $\fr{c_\xi c^2_W}{\sqrt{3-4s^2_W}}+\fr 2 3 s_\xi c_W t_N$ \\
\hline
\end{tabular}
\caption{\label{tttz2} The couplings of $Z_2$ with fermions}
\ec
\end{table}  
\begin{table}[htdp]
\bc
\begin{tabular}{|c|c|c|}
\hline
$f$ & $g^{Z_N}_V(f)$ & $g^{Z_N}_A(f)$ \\
\hline 
$\nu_a $ & $-\fr{s_\xi c_{2W}}{2\sqrt{3-4s^2_W}} -\fr 5 3 c_\xi c_W t_N$ & $-\fr{s_\xi c_{2W}}{2\sqrt{3-4s^2_W}} +\fr 1 3 c_\xi c_W t_N$ \\
\hline
$e_a$ & $-\fr{s_\xi (1-4 s^2_W) }{2\sqrt{3-4s^2_W}} -\fr 5 3 c_\xi c_W t_N$ & $-\fr{s_\xi }{2\sqrt{3-4s^2_W}} +\fr 1 3 c_\xi c_W t_N$ \\
\hline 
$N_a$ & $-\fr{s_\xi c^2_W}{\sqrt{3-4s^2_W}}+\fr 2 3 c_\xi c_W t_N$ & $\fr{s_\xi c^2_W}{\sqrt{3-4s^2_W}}-\fr 2 3 c_\xi c_W t_N$ \\
\hline 
$u_\al$ & $\fr{s_\xi (3-8s^2_W)}{6\sqrt{3-4s^2_W}}+\fr 1 3 c_\xi c_W t_N$ & $\fr{s_\xi }{2\sqrt{3-4s^2_W}} -\fr 1 3 c_\xi c_W t_N$ \\
\hline 
$u_3$  & $-\fr{s_\xi(3+2s^2_W)}{6\sqrt{3-4s^2_W}}+c_\xi c_W t_N$ & $-\fr{s_\xi c_{2W}}{2\sqrt{3-4s^2_W}} +\fr 1 3 c_\xi c_W t_N$ \\
\hline 
$d_\al$ & $\fr{s_\xi (3-2s^2_W)}{6\sqrt{3-4s^2_W}}+\fr 1 3 c_\xi c_W t_N$ & $\fr{s_\xi c_{2W}}{2\sqrt{3-4s^2_W}} - \fr 1 3 c_\xi c_W t_N$\\
\hline
$d_3$ & $- \fr{s_\xi\sqrt{3-4s^2_W}}{6}+c_\xi c_W t_N$ & $-\fr{s_\xi }{2\sqrt{3-4s^2_W}} +\fr 1 3 c_\xi c_W t_N$ \\
\hline 
$U$ & $\fr{s_\xi (3-7s^2_W)}{3\sqrt{3-4s^2_W}}+2 c_\xi c_W t_N$ & $\fr{s_\xi c^2_W}{\sqrt{3-4s^2_W}}-\fr 2 3 c_\xi c_W t_N$ \\
\hline
$D_\al$ & $-\fr{s_\xi (3-5s^2_W)}{3\sqrt{3-4s^2_W}}-\fr 2 3 c_\xi c_W t_N$ & $-\fr{s_\xi c^2_W}{\sqrt{3-4s^2_W}}+\fr 2 3 c_\xi c_W t_N$ \\
\hline
\end{tabular}
\caption{\label{tttzn} The couplings of $Z_N$ with fermions}
\ec
\end{table}

\subsection{Scalar--gauge boson interaction}

The interactions of gauge bosons with scalars arise from
\be \mathcal{L}_{\mathrm{scalar}}\equiv (D^\mu \Phi)^\dagger (D_\mu \Phi),\ee where $\Phi$ runs on all the scalar multiplets of the model. From Eqs. (\ref{scl1}) and (\ref{scl2}), $\Phi$ possesses a common form $\Phi=\langle \Phi \rangle + \Phi'$. Moreover, the covariant derivative has the form $D_\mu =\pa_\mu+igP_\mu = \pa_\mu + ig(P^{\mathrm{CC}}_\mu+P^{\mathrm{NC}}_\mu)$ (see the previous subsection for details). Notice that the strong interaction vanishes because the scalars are colorless. Substituting all those into the Lagrangian, we have 
\bea \mathcal{L}_{\mathrm{scalar}}&=&(\pa^\mu \Phi')^\dagger (\pa_\mu \Phi') + \left[ig(\pa^\mu \Phi')^\dagger (P_\mu \langle\Phi\rangle)+H.c.\right]  + g^2\langle \Phi \rangle^\dagger P^\mu P_\mu \langle \Phi \rangle \crn
&&+\left[ig (\pa^\mu \Phi')^\dagger (P_\mu \Phi')+H.c.\right]+\left[g^2\langle \Phi \rangle P^\mu P_\mu \Phi'+H.c.\right] +g^2\Phi'^\dagger P^\mu P_\mu \Phi'. \eea The terms in the first line are respectively realized as the kinematic, scalar-gauge mixing and mass terms which are not relevant to this analysis. The second line includes all the interactions of three and four fields among the scalars and gauge bosons that we are interested in the investigation.  

To calculate the interactions, we need to present $\Phi$ and $P_\mu$ in terms of the physical fields. Indeed, the gauge part takes the form $P_\mu=P^{\mathrm{CC}}_\mu+P^{\mathrm{NC}}_\mu$, where its terms have already been obtained by (\ref{partcc}) and (\ref{partnc}), respectively. On the other hand, the physical scalars are related to the gauge states by (\ref{vohuongvl}). Let us work in a basis that all the Goldstone bosons are gauged away. In this unitary gauge, the scalar multiplets are given by 
\bea \eta &=& \left( {\begin{array}{c}
    \frac{u}{\sqrt{2}} \\
   0 \\
   0 \\
 \end{array}} \right)+\left( {\begin{array}{c}
    \fr{1}{\sqrt{2}}(c_\beta H - s_\beta H_1+is_\beta \mathcal{A})  \\
    s_\beta H^-_5 \\
    H'
 \end{array}} \right),\hs \rho = \left( {\begin{array}{c}
    0\\
   \frac{v}{\sqrt{2}} \\
   0
 \end{array}} \right)+\left( {\begin{array}{c}
  c_\beta H^+_5 \\
  \frac{1}{\sqrt{2}}(s_\beta H + c_\beta H_1 + ic_\beta \mathcal{A})  \\
 H^+_4
 \end{array}} \right),\crn 
 \chi & =&\left( {\begin{array}{c}
   0 \\
   0 \\
   \frac{\omega}{\sqrt{2}} \\
 \end{array}} \right)+\left( {\begin{array}{c}
    0  \\
    0 \\
    \frac{1}{\sqrt{2}}(c_\varphi H_2 - s_\varphi H_3) 
 \end{array}} \right),\hs 
 \phi = \frac{\Lambda}{\sqrt{2}} + \frac{s_\varphi H_2 + c_\varphi H_3}{\sqrt{2}}.\eea Notice that in each expansion above for the multiplet $\Phi = \eta,\ \rho,\ \chi,\ \phi$, the first term is identified to the $\langle \Phi \rangle$ while the second term is the $\Phi'$ with the physical fields explicitly displayed. The denotations for the scalar multiplets including the gauge bosons in this unitary gauge have conveniently been retained unchanged which should be understood.        

The interactions of one gauge boson with two scalars arise from 
\be ig(\pa^\mu \Phi')^\dagger (P_\mu \Phi')+H.c.=ig(\pa^\mu \Phi')^\dagger (P^{\mathrm{CC}}_\mu \Phi')+ig(\pa^\mu \Phi')^\dagger (P^{\mathrm{NC}}_\mu \Phi')+H.c. \ee Substituting all the known multiplets into this expression we have Table \ref{bangdd1} and \ref{bangdd2}. Let us note that $A\overleftrightarrow{\pa}B\equiv A(\pa B) - (\pa A)B$ is frequently used. 
\begin{table}[htdp]
\bc
\begin{tabular}{|c|c|c|c|}
\hline 
Vertex & Coupling & Vertex & Coupling \\ \hline
$W^+_{\mu} H^-_5 \overleftrightarrow{\partial}^\mu H_1$ & $- \dfrac{ig}{2}$&$W^+_{\mu} H^-_5 \overleftrightarrow{\partial}^\mu \mathcal{A}$ & $\frac{g}{2}$\\
$Y^+_{\mu} H'^* \overleftrightarrow{\partial}^\mu H^-_5$ & $-\dfrac{i g s_\beta}{\sqrt{2}}$&$Y^+_{\mu} H^-_4 \overleftrightarrow{\partial}^\mu H$ & $-\dfrac{i g s_\beta}{2}$\\
$Y^+_{\mu} H^-_4 \overleftrightarrow{\partial}^\mu H_1$ & $-\dfrac{i g c_\beta}{2}$&$Y^+_{\mu} H^-_4 \overleftrightarrow{\partial}^\mu \mathcal{A}$ & $\dfrac{g c_\beta}{2}$\\
$X^0_{\mu} H^+_4 \overleftrightarrow{\partial}^\mu H^-_5$ &  $\dfrac{i g c_\beta}{\sqrt{2}}$&$X^0_{\mu} H' \overleftrightarrow{\partial}^\mu H$ &  $\dfrac{i g c_\beta}{2}$\\
$X^0_{\mu} H' \overleftrightarrow{\partial}^\mu H_1$ &  $-\dfrac{i g s_\beta}{2}$&$X^0_{\mu} H'\overleftrightarrow{\partial}^\mu \mathcal{A}$ &  $\dfrac{g s_\beta}{2}$\\
\hline
\end{tabular}
\caption{\label{bangdd1}
The interactions of a non-Hermitian gauge boson with two scalars.}
\ec
\end{table}

\begin{table}[htdp]
\bc
\begin{tabular}{|c|c|c|c|}
\hline 
Vertex & Coupling & Vertex & Coupling \\ \hline
$A_\mu H^+_5 \overleftrightarrow{\partial}^\mu H^-_5$ & $ie$ & $A_\mu H^+_4 \overleftrightarrow{\partial}^\mu H^-_4$ & $ie$ \\ \hline 
$Z_\mu H^+_4 \overleftrightarrow{\partial}^\mu H^-_4$ & $- \frac{ig s^2_W}{c_W}$ & $Z_\mu H^+_5 \overleftrightarrow{\partial}^\mu H^-_5$ & $  \frac{ig c_{2W}}{2c_W}$ \\ \hline
$Z_\mu \mathcal{A} \overleftrightarrow{\partial}^\mu H_1$ & $\frac{g}{2c_W} $ & $Z_{2\mu}H_1 \overleftrightarrow{\partial}^\mu \mathcal{A} $ & $ g[\frac{c_\xi(c^2_\beta - c_{2W}s^2_\beta)}{2c_W \sqrt{3-4s^2_W}}+\frac{t_N s_\xi c_{2\beta}}{3}]$ \\ \hline 
$Z_{2\mu} H^+_4 \overleftrightarrow{\partial}^\mu H^-_4$&$ig(\frac{-c_{2W}c_\xi}{c_W\sqrt{3-4s^2_W}}+\frac{t_N s_\xi}{3})$&$Z_{2\mu} H^-_5 \overleftrightarrow{\partial}^\mu H^+_5$&$ig[\frac{c_\xi (c^2_\beta -c_{2W} s^2_\beta)}{2c_W\sqrt{3-4s^2_W}}+\frac{t_Ns_\xi c_{2\beta}}{3}]$\\ \hline 
$Z_{2\mu}H' \overleftrightarrow{\partial}^\mu H'^*$&$-ig(\frac{c_Wc_\xi}{\sqrt{3-4s^2_W}}-\frac{t_Ns_\xi}{3})$&$Z_{2\mu}H \overleftrightarrow{\partial}^\mu \mathcal{A}$&$\frac{gs_{2\beta}}{2}(\frac{c_W c_\xi}{\sqrt{3-4s^2_W}}+\frac{2t_Ns_\xi}{3})$\\ \hline 
$Z_{N\mu} H^+_4 \overleftrightarrow{\partial}^\mu H^-_4$&$ig(\frac{c_{2W}s_\xi}{c_W\sqrt{3-4s^2_W}}+\frac{t_Nc_\xi}{3})$&$Z_{N\mu} H^-_5 \overleftrightarrow{\partial}^\mu H^+_5$&$ig[\frac{-s_\xi (c^2_\beta - c_{2W} s^2_\beta)}{2c_W\sqrt{3-4s^2_W}}+\frac{t_Nc_\xi c_{2\beta}}{3}]$\\ \hline
$Z_{N\mu}H' \overleftrightarrow{\partial^\mu} H'^*$&$ig(\frac{c_Ws_\xi}{\sqrt{3-4s^2_W}}+\frac{t_Nc_\xi }{3})$&$Z_{N\mu}H \overleftrightarrow{\partial}^\mu \mathcal{A}$&$\frac{gs_{2\beta} }{2}(\frac{-c_W s_\xi }{\sqrt{3-4s^2_W}}+\frac{2t_Nc_\xi }{3})$\\ \hline
$Z_{N\mu}H_1 \overleftrightarrow{\partial}^\mu \mathcal{A} $ & $g[\frac{-s_\xi (c^2_\beta - c_{2W} s^2_\beta)}{2c_W\sqrt{3-4s^2_W}}+\frac{t_Nc_\xi c_{2\beta}}{3}]$&$ $&$ $ \\
\hline
\end{tabular}
\caption{\label{bangdd2}
The interactions of a neutral gauge boson with two scalars.}
\ec
\end{table}

The interactions of one scalar with two gauge bosons are given by 
\bea g^2\langle \Phi \rangle P^\mu P_\mu \Phi'+H.c.&=& g^2\langle \Phi \rangle P^{\mathrm{CC}\mu}P^{\mathrm{CC}}_\mu \Phi'+g^2\langle \Phi \rangle (P^{\mathrm{CC}\mu} P^{\mathrm{NC}}_\mu+P^{\mathrm{NC}\mu} P^{\mathrm{CC}}_\mu) \Phi'\crn
&&+g^2\langle \Phi \rangle P^{\mathrm{NC}\mu}P^{\mathrm{NC}}_\mu \Phi'+H.c. \eea These interactions are listed in Table \ref{bangdd3}, \ref{bangdd4} and \ref{bangdd5} corresponding to the terms in the r.h.s., respectively.  

\begin{table}[htdp]
\bc
\begin{tabular}{|c|c|c|c|}
\hline 
Vertex & Coupling & Vertex & Coupling \\ \hline
$H_2 X^0 X^{0*}$ &$\frac{g^2 \omega}{2}c_\varphi $&$H_3 X^0 X^{0*} $&$-\frac{g^2 \omega}{2}s_\varphi $\\
$H_2 Y^+Y^-$&$\frac{g^2 \omega}{2}c_\varphi$&$H_3 Y^+Y^-$&$-\frac{g^2 \omega}{2}s_\varphi$\\
$HW^+W^-$&$ \frac{g^2 \sqrt{u^2+v^2}}{2} $&$HX^0X^{0*}$&$\frac{g^2 u}{2}c_\beta$\\
$H_4^-W^+X^{0*}$&$\frac{g^2v}{2 \sqrt{2}}$    &$H_1X^0X^{0*}$&$-  \frac{g^2 u}{2}s_\beta$\\
$HY^+ Y^-$&$\frac{g^2 v}{2} s_\beta$&$H_1Y^+Y^-$&$\frac{g^2 v}{2}c_\beta $\\
$H_5^- X^0 Y^+$&$ \frac{g^2\sqrt{u^2+v^2}}{2\sqrt{2}}s_{2\beta}$&$H'^* W^-Y^+$&$ \frac{g^2u}{2\sqrt{2}}$\\
\hline
\end{tabular}
\caption{\label{bangdd3}
The interactions of a scalar with two non-Hermitian gauge bosons.}
\ec
\end{table}

\begin{table}[htdp]
\bc
\begin{tabular}{|c|c|}
\hline 
Vertex & Coupling \\ \hline
$H^+_ 5 W^- Z_{2}$&$g^2 u s_{\beta} ( \frac{c_W c_\xi}{\sqrt{3 - 4 s^2_W}} +\frac{2 t_N  s_\xi}{3} )$\\
$H^+_5 W^- Z_{N}$&$g^2 u s_{\beta} (-\frac{c_W s_{\xi}}{\sqrt{3-4 s^2_W}} +\frac{2 t_N  c_{\xi}}{3} )$\\
$H' X^0 Z  $&$\frac{g^2u}{4c_W}$\\
$H' X^0 Z_{2}  $&$\frac{g^2u}{2} (-\frac{c_{\xi}}{2 c_W \sqrt{3-4 s^2_W}} +\frac{2t_N}{3}s_{\xi})$\\
$H' X^0 Z_{N}  $&$ \frac{g^2u}{2} (\frac{s_{\xi}}{2 c_W \sqrt{3-4 s^2_W}}  +\frac{2t_N}{3}c_{\xi} )$\\
$H^-_4 Y^+A $&$\frac{gve}{2}$\\
$H^-_4 Y^+ Z$&$- \frac{g^2 v}{4c_W}(1+2s^2_W)$\\
$H^-_4 Y^+ Z_{2}$&$ \frac{g^2v}{2} [\frac{(1-2c_{2W}) c_{\xi}}{2 c_W \sqrt{3-4 s^2_W}}  +\frac{2t_N s_\xi}{3}]$\\
$H^-_4 Y^+ Z_{N}$&$ \frac{g^2v}{2} [-\frac{(1-2c_{2W}) s_{\xi}}{2 c_W \sqrt{3-4 s^2_W}}  +\frac{2t_N c_\xi}{3}]$\\
\hline
\end{tabular}
\caption{\label{bangdd4}
The interactions of a scalar with a non-Hermitian gauge boson and a neutral gauge boson.}
\ec
\end{table}

\begin{table}[htdp]
\bc
\begin{tabular}{|c|c|}
\hline 
Vertex & Coupling \\ \hline
$H_2 Z_2Z_2 $ & $4\La g^2 t^2_N s_\varphi s^2_\xi + \omega c_\varphi g^2 (\frac{c_W c_\xi}{\sqrt{3-4s^2_W}}+\frac{2t
_N}{3}s_\xi )^2$\\
$H_2 Z_N Z_N$ & $4\La g^2 t^2_N s_\varphi c^2_\xi + \omega c_\varphi g^2 (-\frac{c_W s_\xi}{\sqrt{3-4s^2_W}}+\frac{2t
_N}{3}c_\xi)^2$\\
$H_2 Z_{2} Z_{N}$ & $4\La g^2 t^2_N s_\varphi s_{2\xi} +2 \omega c_\varphi g^2 (\frac{c_W c_\xi}{\sqrt{3-4s^2_W}}+\frac{2t_N}{3}s_\xi)(-\frac{c_W s_\xi}{\sqrt{3-4s^2_W}}+\frac{2t_N}{3}c_\xi)$\\
$H_3 Z_{2} Z_2$ & $4\La g^2 t^2_N c_\varphi s^2_{\xi} - \omega s_{\varphi} g^2 (\frac{c_W c_{\xi}}{\sqrt{3-4s^2_W}}+\frac{2t_N}{3}s_{\xi})^2$\\
$H_3 Z_{N} Z_N$ & $4 \La g^2 t^2_N c_\varphi c^2_{\xi} - \omega s_\varphi g^2 (-\frac{c_W s_{\xi}}{\sqrt{3-4 s^2_W}}+\frac{2t_N}{3} c_{\xi})^2$\\
$H_3 Z_{2} Z_{N}$ & $4\La g^2 t^2_N c_\varphi s_{2\xi}  - 2 \omega s_{\varphi} g^2 (\frac{c_W c_{\xi}}{\sqrt{3-4s^2_W}}+\frac{2t_N}{3}s_{\xi})(-\frac{c_W s_\xi}{\sqrt{3-4s^2_W}}+\frac{2t_N}{3}c_\xi)$\\
$H Z Z $ & $ \frac{g^2}{4c^2_W}\sqrt{u^2+v^2}$\\
$ H Z_{2} Z_2$ & $g^2[u c_\beta(\frac{c_{2W} c_\xi}{2c_W\sqrt{3-4s^2_W}}+\frac{t_{N}}{3} s_{\xi})^2 +vs_\beta (\frac{c_{\xi}}{2c_W\sqrt{3-4 s^2_W}} + \frac{t_N}{3} s_{\xi} )^2]$\\ 
$H Z_{N} Z_N$ & $g^2[uc_\beta(\frac{-c_{2W} s_\xi}{2c_W \sqrt{3-4s^2_W}}+\frac{t_N}{3}c_\xi)^2+vs_\beta (\frac{-s_\xi}{2c_W\sqrt{3-4s^2_W}}+\frac{t_N}{3}c_\xi)^2] $\\
$H Z Z_2$ & $\frac{g^2}{c_W} [uc_\beta ( \frac{c_{2W}c_\xi}{2c_W \sqrt{3-4s^2_W}}+\frac{t_N}{3}s_\xi) -vs_\beta(\frac{c_\xi}{2c_W \sqrt{3-4s^2_W}}+\frac {t_N}{3}s_\xi)]$\\
$H Z Z_{N} $&$\frac{g^2}{c_W}[uc_\beta(\frac{-c_{2W}s_\xi}{2c_W\sqrt{3-4s^2_W}}+\frac{t_N}{3}c_\xi)-vs_\beta(\frac{-s_\xi}{2c_W\sqrt{3-4s^2_W}}+\frac{t_N}{3}c_\xi]$\\
$H Z_{2} Z_{N}$ & $2g^2[uc_\beta (\frac{c_{2W}c_\xi}{2c_W\sqrt{3-4s^2_W}}+\frac{t_N}{3}s_\xi)(\frac{-c_{2W}s_\xi}{2c_W\sqrt{3-4s^2_W}}+\frac{t_N}{3}c_\xi)$\\
$ $&$+ vs_\beta(\frac{c_\xi}{2c_W\sqrt{3-4s^2_W}}+\frac{t_N}{3}s_\xi)(\frac{-s_\xi}{2c_W\sqrt{3-4s^2_W}}+\frac{t_N}{3}c_\xi)]$\\
$H_1 Z_2 Z_2 $ & $g^2[-us_\beta(\frac{c_{2W}c_\xi}{2c_W\sqrt{3-4s^2_W}}+\frac{t_N}{3}s_\xi)^2+vc_\beta(\frac{c_\xi}{2c_W\sqrt{3-4s^2_W}}+\frac{t_N}{3}s_\xi)^2]$\\
$H_1 Z_N Z_N$ & $g^2[-us_\beta(\frac{-c_{2W}s_\xi}{2c_W\sqrt{3-4s^2_W}}+\frac{t_N}{3}c_\xi)^2+vc_\beta(\frac{-s_\xi}{2c_W\sqrt{3-4s^2_W}}+\frac{t_N}{3}c_\xi)^2]$\\
$H_1ZZ_{2}$ & $- \frac{g^2}{c_W}[us_\beta(\frac{c_{2W}c_\xi}{2c_W\sqrt{3-4s^2_W}}+\frac{t_N}{3}s_\xi)+vc_\beta(\frac{c_\xi}{2c_W\sqrt{3-4s^2_W}}+\frac{t_N}{3}s_\xi)]$\\
$H_1 Z Z_{N}$ & $ - \frac{g^2}{c_W}[us_\beta(\frac{-c_{2W}s_\xi}{2c_W\sqrt{3-4s^2_W}}+\frac{t_N}{3}c_\xi)+vc_\beta(\frac{-s_\xi}{2c_W\sqrt{3-4s^2_W}}+\frac{t_N}{3}c_\xi)]$\\
$H_1 Z_{2} Z_{N}$ & $2g^2[-us_\beta(\frac{c_{2W}c_\xi}{2c_W\sqrt{3-4s^2_W}}+\frac{t_N}{3}s_\xi)(\frac{-c_{2W}s_\xi}{2c_W\sqrt{3-4s^2_W}}+\frac{t_N}{3}c_\xi)$\\
$ $&$+vc_\beta(\frac{c_\xi}{2c_W\sqrt{3-4s^2_W}}+\frac{t_N}{3}s_\xi)(\frac{-s_\xi}{2c_W\sqrt{3-4s^2_W}}+ \frac{t_N}{3}c_\xi)]$\\
\hline
\end{tabular}
\caption{\label{bangdd5}
The interactions of a scalar with two neutral gauge bosons.}
\ec
\end{table}

The interactions of two scalars and two gauge bosons are derived from 
\bea g^2\Phi'^\dagger P^\mu P_\mu \Phi' = g^2\Phi'^\dagger P^{\mathrm{CC}\mu} P^{\mathrm{CC}}_\mu \Phi' + g^2\Phi'^\dagger (P^{\mathrm{CC}\mu} P^{\mathrm{NC}}_\mu+P^{\mathrm{NC}\mu} P^{\mathrm{CC}}_\mu) \Phi'+ g^2\Phi'^\dagger P^{\mathrm{NC}\mu} P^{\mathrm{NC}}_\mu \Phi', \eea which result in Table \ref{bangdd6}, \ref{bangdd7}, \ref{bangdd8a} and \ref{bangdd8b}, respectively.

\begin{table}[htdp]
\bc
\begin{tabular}{|c|c|c|c|}
\hline 
Vertex & Coupling & Vertex & Coupling \\ \hline
$ X^0 X^{0*} H_2 H_2 $ & $ \frac{g^2}{4} c^2_\varphi$ &$ X^0 X^{0*} H_3 H_3 $ & $\frac{g^2}{4} s^2_\varphi$ \\ 
$ X^0 X^{0*}H_2 H_3$ & $-\frac{g^2}{4}s_{2\varphi} $ &$Y^+Y^-H_2 H_2 $ &$ \frac{g^2}{4}c_\varphi^2$\\
$ Y^+Y^-H_3 H_3 $ & $ \frac{g^2}{4}s_\varphi^2 $& $Y^+Y^-H_2 H_3$ &$ -\frac{g^2}{4}s_{2\varphi}$ \\
$ W^+W^-H^+_5H^-_5$ & $ \frac{g^2}{2}$&$ X^0X^{0*}H^+_5H^-_5$& $\frac{g^2}{2}c^2_\beta$\\
$X^0X^{0*}H^+_4H^-_4$ & $ \frac{g^2}{2}$ & $Y^+Y^- H^+_4H^-_4 $& $\frac{g^2}{2}$\\
$W^+W^-H H $ & $\frac{g^2}{4}$&$W^+W^-H_1 H_1 $&$\frac{g^2}{4}$\\
$W^+W^-\mathcal{A} \mathcal{A} $ & $\frac{g^2}{4}$&$Y^+Y^-H H $&$\frac{g^2}{4}s^2_\beta$\\
$Y^+Y^-H_1 H_1 $ & $\frac{g^2}{4}c^2_\beta$&$Y^+Y^-HH_1 $&$\frac{g^2}{4}s_{2\beta}$\\
$Y^+Y^-\mathcal{A} \mathcal{A} $ & $\frac{g^2}{4}c^2_\beta$&$X^0Y^+H^-_5H$&$\frac{g^2}{2\sqrt{2}} s_{2\beta}$\\
$X^0Y^+H^-_5H_1$ & $\frac{g^2}{2\sqrt{2}}c_{2\beta}$&$X^0Y^+H^-_5 \mathcal{A} $&$i\frac{g^2}{2\sqrt{2}}c_{2\beta}$\\
$W^+Y^-H^+_4H^-_5$ & $\frac{g^2}{2}c_\beta$&$W^-X^0HH^+_4$&$\frac{g^2}{2\sqrt{2}}s_{\beta} $\\
$W^-X^0H_1H ^+_4$&$\frac{g^2}{2\sqrt{2}} c_\beta$&$W^-X^0 \mathcal{A} H^+_4$&$\frac{-ig^2}{2\sqrt{2}} c_\beta$\\
$X^{0*}X^0 H H $ & $\frac{g^2}{4}c^2_\beta$&$X^{0*} X^0 H_1 H_1 $&$\frac{g^2}{4}s^2_\beta$\\
$X^{0*}X^0H H_1$ & $-\frac{g^2}{4} s_{2\beta}$&$X^{0*}X^0 \mathcal{A} \mathcal{A} $&$\frac{g^2}{4}s^2_\beta$\\
$Y^+Y^-H^+_5H^-_5$ & $\frac{g^2}{2}s^2_\beta$&$X^{0*}X^0H'^*H'$&$\frac{g^2}{2}$\\
$Y^+Y^-H'^*H'$ & $\frac{g^2}{2}$&$W^+Y^-HH'$ & $\frac{g^2}{2\sqrt{2}}c_\beta$\\
$W^+Y^-H_1H'$ & $-\frac{g^2}{2\sqrt{2}}s_\beta$&$W^+Y^-\mathcal{A}H'$ & $\frac{-ig^2}{2\sqrt{2}}s_\beta$\\
$W^-X^0H^+_5H'$ & $\frac{g^2s_\beta}{2}$&$ $&$ $\\
\hline
\end{tabular}
\caption{\label{bangdd6}
The interactions of two non-Hermitian gauge bosons and two scalars.}
\ec
\end{table}

\begin{table}[htdp]
\bc
\begin{tabular}{|c|c|c|c|} 
\hline
 Vertex & Coupling & Vertex & Coupling \\ \hline
$H_1H^-_5W^+A$ &$ ge/2 $&$\mathcal{A}H^-_5W^+A$ & $ige/2$\\
$H_1H^-_5W^+Z$&$\frac{g^2}{4c_W}(c_{2W}-1)$& $\mathcal{A}H^-_5W^+Z$&$\frac{ig^2}{4c_W}(c_{2W}-1)$\\
$HH^-_5W^+Z_{2}$&$\fr 1 2 g^2 s_{2\beta}(\frac{c_\xi c_W}{\sqrt{3-4s^2_W}}+\frac{2t_N s_\xi}{3})$&$HH^-_5W^+Z_{N}$&$\fr 1 2 g^2 s_{2\beta}(\frac{-s_\xi c_W}{\sqrt{3-4s^2_W}}+\frac{2t_N c_\xi}{3})$\\
$H_1H^-_5W^+Z_{2}$&$g^2[\frac{c_\xi(c^2_\beta-s^2_\beta c_{2W})} {2c_W \sqrt{3-4s^2_W}}+\frac{t_N}{3}s_\xi c_{2\beta}]$&$H_1H^-_5W^+Z_{N}$&$g^2[\frac{-s_\xi(c^2_\beta-s^2_\beta c_{2W})} {2c_W \sqrt{3-4s^2_W}}+\frac{t_N}{3}c_\xi c_{2\beta}]$\\
$\mathcal{A}H^-_5W^+Z_{2}$&$ig^2[\frac{c_\xi(c^2_\beta-s^2_\beta c_{2W})}{2c_W\sqrt{3-4s^2_W}}+\frac{t_N}{3}s_\xi c_{2\beta}]$&$\mathcal{A}H^-_5W^+Z_{N}$&$ig^2[\frac{-s_\xi(c^2_\beta-s^2_\beta c_{2W})}{2c_W\sqrt{3-4s^2_W}}+\frac{t_N}{3}c_\xi c_{2\beta}]$\\
$H^-_5H^+_4X^0A$&$\sqrt{2}gec_\beta$&$H^-_5H^+_4X^0Z$&$\frac{g^2c_\beta}{2\sqrt{2}c_W}(4c^2_W-3)$\\
$H^-_5H^+_4X^0Z_{2}$&$\frac{g^2c_\beta}{\sqrt{2}}[\frac{c_\xi(3-4c^2_W)}{2c_W\sqrt{3-4s^2_W}}+\frac{2t_Ns_\xi}{3}]$&$H^-_5H^+_4X^0Z_{N}$&$\frac{g^2c_\beta}{\sqrt{2}}[\frac{-s_\xi(3-4c^2_W)}{2c_W\sqrt{3-4s^2_W}}+\frac{2t_Nc_\xi}{3}]$\\
$HH^+_4Y^-A$&$\frac{ges_\beta}{2}$&$H_1H^+_4Y^-A$&$\frac{gec_\beta}{2}$\\
$\mathcal{A}H^+_4Y^-A$&$\frac{-igec_\beta}{2}$&$HH^+_4Y^-Z$&$\frac{-g^2s_\beta(2-c_{2W})}{4c_W}$\\
$H_1H^+_4Y^-Z$&$\frac{-g^2c_\beta(2-c_{2W})}{4c_W}$&$\mathcal{A}H^+_4Y^-Z$&$\frac{ig^2c_\beta(2-c_{2W})}{4c_W}$\\
$HH^+_4Y^-Z_{2}$&$\frac{g^2s_\beta}{2}[\frac{c_\xi(1-2c_{2W})}{2c_W\sqrt{3-4s^2_W}}+\frac{2t_Ns_\xi}{3}]$&$H_1H^+_4Y^-Z_{2}$&$\frac{g^2c_\beta}{2}[\frac{c_\xi(1-2c_{2W})}{2c_W\sqrt{3-4s^2_W}}+\frac{2t_Ns_\xi}{3}]$\\
$\mathcal{A}H^+_4Y^-Z_{2}$&$\frac{-ig^2c_\beta}{2}[\frac{c_\xi(1-2c_{2W})}{2c_W\sqrt{3-4s^2_W}}+\frac{2t_Ns_\xi}{3}]$&$HH^+_4Y^-Z_{N}$&$\frac{g^2s_\beta}{2}[\frac{-s_\xi(1-2c_{2W})}{2c_W\sqrt{3-4s^2_W}}+\frac{2t_Nc_\xi}{3}]$\\
$H_1H^+_4Y^-Z_{N}$&$\frac{g^2c_\beta}{2}[\frac{-s_\xi(1-2c_{2W})}{2c_W\sqrt{3-4s^2_W}}+\frac{2t_Nc_\xi}{3}]$&$\mathcal{A}H^+_4Y^-Z_{N}$&$\frac{-ig^2c_\beta}{2}[\frac{-s_\xi(1-2c_{2W})}{2c_W\sqrt{3-4s^2_W}}+\frac{2t_Nc_\xi}{3}]$\\
$HH'X^0Z$&$\frac{g^2c_\beta}{4c_W}$&$H_1H'X^0Z$&$\frac{-g^2s_\beta}{4c_W}$\\
$\mathcal{A}H'X^0Z$&$\frac{-ig^2s_\beta}{4c_W}$&$HH'X^0Z_{2}$&$\frac{g^2c_\beta}{2}(\frac{-c_\xi}{2c_W\sqrt{3-4s^2_W}}+\frac{2t_Ns_\xi}{3})$\\
$H_1H'X^0Z_{2}$&$\frac{-g^2s_\beta}{2}(\frac{-c_\xi}{2c_W\sqrt{3-4s^2_W}}+\frac{2t_Ns_\xi}{3})$&$\mathcal{A}H'X^0Z_{2}$&$-\frac{ig^2s_\beta}{2}(\frac{-c_\xi}{2c_W\sqrt{3-4s^2_W}}+\frac{2t_Ns_\xi}{3})$\\
$HH'X^0Z_{N}$&$\frac{g^2c_\beta}{2}(\frac{s_\xi}{2c_W\sqrt{3-4s^2_W}}+\frac{2t_Nc_\xi}{3})$&$H_1H'X^0Z_{N}$&$\frac{-g^2s_\beta}{2}(\frac{s_\xi}{2c_W\sqrt{3-4s^2_W}}+\frac{2t_Nc_\xi}{3})$\\
$\mathcal{A}H'X^0Z_{N}$&$-\frac{ig^2s_\beta}{2}(\frac{s_\xi}{2c_W\sqrt{3-4s^2_W}}+\frac{2t_Nc_\xi}{3})$&$H^+_5H'Y^-A$&$\frac{-ges_\beta}{\sqrt{2}}$\\
$H^+_5H'Y^-Z$&$\frac{-g^2s_\beta c_{2W}}{2\sqrt{2}c_W}$&$H^+_5H'Y^-Z_{2}$&$\frac{g^2s_\beta}{\sqrt{2}}(\frac{-c_\xi}{2c_W\sqrt{3-4s^2_W}}+\frac{2t_Ns_\xi}{3})$\\
$H^+_5H'Y^-Z_{N}$&$\frac{g^2s_\beta}{\sqrt{2}}(\frac{s_\xi}{2c_W\sqrt{3-4s^2_W}}+\frac{2t_Nc_\xi}{3})$&$ $&$ $\\
\hline
\end{tabular}
\caption{\label{bangdd7} The interactions of two scalars with a non-Hermitian gauge boson and a neutral gauge boson.}
\ec
\end{table}

\begin{table}[htdp]
\bc
\begin{tabular}{|c|c|} 
\hline
  
Vertex & Coupling \\ \hline
$H_2H_2 Z_{2}Z_{2}$&$g^2[2t^2_Ns^2_\varphi s^2_\xi +\frac{1}{2}c^2_\varphi (\frac{c_Wc_\xi}{\sqrt{3-4s^2_W}}+\frac{2t_Ns_\xi}{3})^2]$\\
$H_2H_2 Z_{N}Z_{N}$&$g^2[2t^2_Ns^2_\varphi c^2_\xi +\frac{1}{2}c^2_\varphi (\frac{-c_Ws_\xi}{\sqrt{3-4s^2_W}}+\frac{2t_Nc_\xi}{3})^2]$\\
$H_2H_2 Z_{2}Z_{N}$&$g^2[2t^2_Ns^2_\varphi s_{2\xi} +c^2_\varphi (\frac{c_Wc_\xi}{\sqrt{3-4s^2_W}}+\frac{2t_Ns_\xi}{3})(\frac{-c_Ws_\xi}{\sqrt{3-4s^2_W}}+\frac{2t_Nc_\xi}{3})]$\\ 
$H_3H_3 Z_{2}Z_{2}$&$g^2[2t^2_Nc^2_\varphi s^2_\xi +\frac{1}{2}s^2_\varphi (\frac{c_Wc_\xi}{\sqrt{3-4s^2_W}}+\frac{2t_Ns_\xi}{3})^2]$\\
$H_3H_3 Z_{N}Z_{N}$&$g^2[2t^2_Nc^2_\varphi c^2_\xi +\frac{1}{2}s^2_\varphi (\frac{-c_Ws_\xi}{\sqrt{3-4s^2_W}}+\frac{2t_Nc_\xi}{3})^2]$\\
$H_3H_3 Z_{2}Z_{N}$&$g^2[2t^2_Ns_{2\xi} c^2_\varphi+s^2_\varphi (\frac{c_Wc_\xi}{\sqrt{3-4s^2_W}}+\frac{2t_Ns_\xi}{3})(\frac{-c_Ws_\xi}{\sqrt{3-4s^2_W}}+\frac{2t_Nc_\xi}{3})]$\\
$H_2H_3Z_{2}Z_{2}$&$g^2[2t^2_Ns_{2\varphi} s^2_\xi-\fr{s_{2\varphi}}{2} (\frac{c_Wc_\xi}{\sqrt{3-4s^2_W}}+\frac{2t_Ns_\xi}{3})^2]$\\
$H_2H_3Z_{N}Z_{N}$&$g^2[2t^2_Ns_{2\varphi} c^2_\xi-\fr{s_{2\varphi}}{2} (\frac{-c_Ws_\xi}{\sqrt{3-4s^2_W}}+\frac{2t_Nc_\xi}{3})^2]$\\
$H_2H_3 Z_{2}Z_{N}$&$g^2[2t^2_Ns_{2\varphi} s_{2\xi}-s_{2\varphi} (\frac{c_Wc_\xi}{\sqrt{3-4s^2_W}}+\frac{2t_Ns_\xi}{3})(\frac{-c_Ws_\xi}{\sqrt{3-4s^2_W}}+\frac{2t_Nc_\xi}{3})]$\\
$H^+_5H^-_5AA $&$e^2$\\
$H^+_5H^-_5ZZ$&$\frac{g^2c^2_{2W}}{4c^2_W}$\\
$H^+_5H^-_5Z_{2}Z_{2}$&$g^2 [c^2_\beta (\frac{c_\xi}{2c_W \sqrt{3-4s^2_W}}+\frac{t_Ns_\xi}{3})^2+s^2_\beta(\frac{c_\xi c_{2W}}{2c_W \sqrt{3-4s^2_W}}+\frac{t_Ns_\xi}{3})^2]$\\
$H^+_5H^-_5Z_{N}Z_{N}$&$g^2 [c^2_\beta (\frac{-s_\xi}{2c_W \sqrt{3-4s^2_W}}+\frac{t_Nc_\xi}{3})^2+s^2_\beta(\frac{-s_\xi c_{2W}}{2c_W \sqrt{3-4s^2_W}}+\frac{t_Nc_\xi}{3})^2]$\\
$H^+_5H^-_5A Z$&$\frac{egc_{2W}}{c_W}$\\
$H^+_5H^-_5Z Z_{2}$&$\frac{g^2c_{2W}}{c_W}[c^2_\beta (\frac{c_\xi}{2c_W \sqrt{3-4s^2_W}}+\frac{t_Ns_\xi}{3})-s^2_\beta(\frac{c_\xi c_{2W}}{2c_W \sqrt{3-4s^2_W}}+\frac{t_Ns_\xi}{3})]$\\
$H^+_5H^-_5Z Z_{N}$&$\frac{g^2c_{2W}}{c_W}[c^2_\beta (\frac{-s_\xi}{2c_W \sqrt{3-4s^2_W}}+\frac{t_Nc_\xi}{3})-s^2_\beta(\frac{-s_\xi c_{2W}}{2c_W \sqrt{3-4s^2_W}}+\frac{t_Nc_\xi}{3})]$\\
$H^+_5H^-_5Z_{2} Z_{N}$&$2g^2[c^2_\beta ( \frac{c_\xi}{2c_W \sqrt{3-4s^2_W}}+\frac{t_Ns_\xi}{3})(\frac{-s_\xi}{2c_W \sqrt{3-4s^2_W}}+\frac{t_Nc_\xi}{3})$\\
$ $&$+s^2_\beta(\frac{c_\xi c_{2W}}{2c_W \sqrt{3-4s^2_W}}+\frac{t_Ns_\xi}{3})(\frac{-s_\xi c_{2W}}{2c_W \sqrt{3-4s^2_W}}+\frac{t_Nc_\xi}{3})]$\\
$H^+_4H^-_4 AA $&$e^2$\\
$H^+_4H^-_4 ZZ $&$\frac{g^2 s^4_W}{c^2_W}$\\
$H^+_4H^-_4 Z_{2}Z_{2} $&$g^2(\frac{-c_{2W} c_\xi}{c_W\sqrt{3-4s^2_W}}+\frac{t_Ns_\xi}{3})^2$\\
$H^+_4H^-_4 Z_{N}Z_{N} $&$g^2(\frac{c_{2W} s_\xi}{c_W\sqrt{3-4s^2_W}}+\frac{t_Nc_\xi}{3})^2$\\
$H^+_4H^-_4 A Z $&$\frac{-2egs^2_W}{c_W}$\\
$H^+_4H^-_4 A Z_{2} $&$2eg(\frac{-c_{2W} c_\xi}{c_W\sqrt{3-4s^2_W}}+\frac{t_Ns_\xi}{3})$\\
$H^+_4H^-_4 A Z_{N} $&$2eg(\frac{c_{2W} s_\xi}{c_W\sqrt{3-4s^2_W}}+\frac{t_Nc_\xi}{3})$\\
$H^+_4H^-_4 Z Z_{2} $&$\frac{-2g^2s^2_W}{c_W}(\frac{-c_{2W} c_\xi}{c_W\sqrt{3-4s^2_W}}+\frac{t_Ns_\xi}{3})$\\
$H^+_4H^-_4 Z Z_{N} $&$\frac{-2g^2s^2_W}{c_W}(\frac{c_{2W} s_\xi}{c_W\sqrt{3-4s^2_W}}+\frac{t_Nc_\xi}{3})$\\
$H^+_4H^-_4 Z_{2} Z_{N} $&$2g^2(\frac{-c_{2W} c_\xi}{c_W\sqrt{3-4s^2_W}}+\frac{t_Ns_\xi}{3})(\frac{c_{2W} s_\xi}{c_W\sqrt{3-4s^2_W}}+\frac{t_Nc_\xi}{3})$\\
$HH ZZ$&$\frac{g^2}{8c^2_W}$\\
$HH Z_{N}Z_N$&$\frac{g^2}{2}[s^2_\beta (\frac{-s_\xi}{2c_W \sqrt{3-4s^2_W}}+\frac{t_Nc_\xi}{3})^2+c^2_\beta(\frac{-s_\xi c_{2W}}{2c_W \sqrt{3-4s^2_W}}+\frac{t_Nc_\xi}{3})^2]$\\ \hline
\end{tabular}
\caption{\label{bangdd8a}
The interactions of two scalars with two neutral gauge bosons.}
\ec
\end{table}

\begin{table}[htdp]
\bc
\begin{tabular}{|c|c|}
 \hline
 $H_1 H_1 Z Z $&$\frac{g^2}{8c^2_W}$\\
$H_1 H_1 Z_{2} Z_{2} $&$\frac{g^2}{2}[c^2_\beta (\frac{c_\xi}{2c_W \sqrt{3-4s^2_W}}+\frac{t_Ns_\xi}{3})^2+s^2_\beta(\frac{c_\xi c_{2W}}{2c_W \sqrt{3-4s^2_W}}+\frac{t_Ns_\xi}{3})^2]$\\
$H_1 H_1 Z_{N} Z_{N} $&$\frac{g^2}{2}[c^2_\beta (\frac{-s_\xi}{2c_W \sqrt{3-4s^2_W}}+\frac{t_Nc_\xi}{3})^2+s^2_\beta(\frac{-s_\xi c_{2W}}{2c_W \sqrt{3-4s^2_W}}+\frac{t_Nc_\xi}{3})^2]$\\
$H_1HZ_{2} Z_{2} $&$\fr 1 2 g^2s_{2\beta} (\frac{c^2_\xi s^2_{2W}}{4c^2_W(3-4s^2_W)}+\frac{t_N s_{2\xi} s^2_W}{3c_W\sqrt{3-4s^2_W}})$\\
$H_1HZ_{N}Z_{N} $&$\fr 1 2 g^2s_{2\beta}  (\frac{s^2_\xi s^2_{2W}}{4c^2_W(3-4s^2_W)}-\frac{t_N s_{2\xi} s^2_W}{3c_W\sqrt{3-4s^2_W}})$\\
$H H Z Z_{2}$&$\frac{g^2}{2c_W}[c^2_\beta (\frac{c_\xi c_{2W}}{2c_W \sqrt{3-4s^2_W}}+\frac{t_Ns_\xi}{3})-s^2_\beta(\frac{c_\xi}{2c_W \sqrt{3-4s^2_W}}+\frac{t_Ns_\xi}{3})]$\\  
$H H  Z Z_{N}$&$\frac{g^2}{2c_W}[c^2_\beta (\frac{-s_\xi c_{2W}}{2c_W \sqrt{3-4s^2_W}}+\frac{t_Nc_\xi}{3})-s^2_\beta(\frac{-s_\xi}{2c_W \sqrt{3-4s^2_W}}+\frac{t_Nc_\xi}{3})]$\\
$H H Z_{2} Z_{N}$&$g^2[s^2_\beta ( \frac{c_\xi}{2c_W \sqrt{3-4s^2_W}}+\frac{t_Ns_\xi}{3})(\frac{-s_\xi}{2c_W \sqrt{3-4s^2_W}}+\frac{t_Nc_\xi}{3})$\\
$ $&$+c^2_\beta(\frac{c_\xi c_{2W}}{2c_W \sqrt{3-4s^2_W}}+\frac{t_Ns_\xi}{3})(\frac{-s_\xi c_{2W}}{2c_W \sqrt{3-4s^2_W}}+\frac{t_Nc_\xi}{3})]$\\
$H_1 H_1 Z Z_{2}$&$\frac{g^2}{2c_W}[s^2_\beta (\frac{c_\xi c_{2W}}{2c_W \sqrt{3-4s^2_W}}+\frac{t_Ns_\xi}{3})-c^2_\beta(\frac{c_\xi}{2c_W \sqrt{3-4s^2_W}}+\frac{t_Ns_\xi}{3})]$\\
$H_1 H_1 Z Z_{N}$&$\frac{g^2}{2c_W}[s^2_\beta (\frac{-s_\xi c_{2W}}{2c_W \sqrt{3-4s^2_W}}+\frac{t_Nc_\xi}{3})-c^2_\beta(\frac{-s_\xi}{2c_W \sqrt{3-4s^2_W}}+\frac{t_Nc_\xi}{3})]$\\
$H_1 H_1 Z_{2} Z_{N}$&$g^2[c^2_\beta ( \frac{c_\xi}{2c_W \sqrt{3-4s^2_W}}+\frac{t_Ns_\xi}{3})(\frac{-s_\xi}{2c_W \sqrt{3-4s^2_W}}+\frac{t_Nc_\xi}{3})$
\\
$ $&$+s^2_\beta(\frac{c_\xi c_{2W}}{2c_W \sqrt{3-4s^2_W}}+\frac{t_Ns_\xi}{3})(\frac{-s_\xi c_{2W}}{2c_W \sqrt{3-4s^2_W}}+\frac{t_Nc_\xi}{3})]$\\
$H_1HZ Z_{2}$&$\frac{-g^2 s_{2\beta}}{2c_W}(\frac{c_\xi c_W}{\sqrt{3-4s^2_W}}+\frac{2t_N s_\xi}{3})$\\
$H_1HZ Z_{N}$&$\frac{-g^2 s_{2\beta}}{2c_W}(\frac{-s_\xi c_W}{\sqrt{3-4s^2_W}}+\frac{2t_N c_\xi}{3})$\\
$H_1HZ_{2} Z_{N}$&$g^2s_{2\beta} [( \frac{c_\xi}{2c_W \sqrt{3-4s^2_W}}+\frac{t_Ns_\xi}{3})(\frac{-s_\xi}{2c_W \sqrt{3-4s^2_W}}+\frac{t_Nc_\xi}{3})$\\
$ $&$-(\frac{c_\xi c_{2W}}{2c_W \sqrt{3-4s^2_W}}+\frac{t_Ns_\xi}{3})(\frac{-s_\xi c_{2W}}{2c_W \sqrt{3-4s^2_W}}+\frac{t_Nc_\xi}{3})]$\\
$H'^*H'Z_{2}Z_{2} $&$g^2(\frac{c_Wc_\xi}{\sqrt{3-4s^2_W}}-\frac{t_Ns_\xi}{3})^2$\\
$H'^*H'Z_{N}Z_{N} $&$g^2(\frac{-c_Ws_\xi}{\sqrt{3-4s^2_W}}-\frac{t_Nc_\xi}{3})^2$\\
$H'^*H'Z_{2}Z_{N}$&$2g^2(\frac{c_Wc_\xi}{\sqrt{3-4s^2_W}}-\frac{t_Ns_\xi}{3})(\frac{-c_Ws_\xi}{\sqrt{3-4s^2_W}}-\frac{t_Nc_\xi}{3})$\\
$\mathcal{A}\mathcal{A}ZZ$&$\frac{g^2}{8c^2_W}$\\
$\mathcal{A}\mathcal{A}Z_{2}Z_2$&$\frac{g^2}{2}[s^2_\beta (\frac{c_\xi c_{2W}}{2c_W \sqrt{3-4s^2_W}}+\frac{t_Ns_\xi}{3})^2+c^2_\beta(\frac{c_\xi}{2c_W \sqrt{3-4s^2_W}}+\frac{t_Ns_\xi}{3})^2]$\\
$\mathcal{A}\mathcal{A}Z_{N}Z_{N}$&$\frac{g^2}{2}[s^2_\beta (\frac{-s_\xi c_{2W}}{2c_W \sqrt{3-4s^2_W}}+\frac{t_Nc_\xi}{3})^2+c^2_\beta(\frac{-s_\xi}{2c_W \sqrt{3-4s^2_W}}+\frac{t_Nc_\xi}{3})^2]$\\
$\mathcal{A}\mathcal{A} Z Z_{2}$&$\frac{g^2}{4c_W}[\frac{c_\xi (c^2_\beta - c_{2W}s^2_\beta)}{c_W\sqrt{3-4s^2_W}}+\frac{2t_N s_\xi c_{2\beta}}{3}]$\\
$\mathcal{A}\mathcal{A} Z Z_{N}$&$\frac{g^2}{4c_W}[\frac{-s_\xi (c^2_\beta-c_{2W}s^2_\beta)}{c_W\sqrt{3-4s^2_W}}+\frac{2t_N c_\xi c_{2\beta}}{3}]$\\
$\mathcal{A}\mathcal{A} Z_{2} Z_{N}$&$\frac{g^2}{2}[s^2_\beta(\frac{c_\xi c_{2W}}{2c_W \sqrt{3-4s^2_W}}+\frac{t_Ns_\xi}{3})(\frac{-s_\xi c_{2W}}{2c_W \sqrt{3-4s^2_W}}+\frac{t_Nc_\xi}{3})$\\
$ $&$+c^2_\beta (\frac{c_\xi}{2c_W \sqrt{3-4s^2_W}}+\frac{t_Ns_\xi}{3})( \frac{-s_\xi}{2c_W \sqrt{3-4s^2_W}}+\frac{t_Nc_\xi}{3})]$\\
\hline
\end{tabular}
  \caption{\label{bangdd8b}
The interactions of two scalars with two neutral gauge bosons (continued).}
\ec
\end{table}

\section{\label{ctsec}New physics effects and constraints}

\subsection{Dark matter: Complex scalar $H'$}
 
The spectrum of scalar particles in the model contains an electrically-neutral particle $H'$ that is odd under the $W$-parity. Because the $W$-parity symmetry is exact and unbroken by the VEVs, the $H'$ is stabilized that cannot decay if it is the lightest particle among the $W$-particles. Under this regime we obtain the relic density of the $H'$ at present day and derive some constraints on its mass. Such scalar is within the context of the so-called Higgs portal which has been intensively exploited in the literature \cite{higgsportal1,higgsportal2} due to its interaction with the standard model Higgs boson via the scalar potential regime. We will show that the $H'$ can be a viable dark matter which yields the right abundance ($\Omega h^2 =0.11-0.12$) as well as obeying the direct detection bounds \footnote{Besides the Higgs, there are additional scalars that play a role in setting the abundance of this scalar. However those scalars are assumed to be much heavier, and therefore the annihilation cross section is set by the Higgs portal for WIMP masses not too heavy}.      

In the early universe, the $H'$ was in thermal equilibrium with the standard model particles. As the universe expanded and cooled down, it reaches a point where the temperature is roughly equal to the $H'$ mass, preventing the $H'$ particles to be produced from the annihilation of the standard model particles, and only the annihilations between the $H'$ particles take place. However, as the universe keeps expanding, there is a point where the $H'$ particles can no longer annihilate themselves into the standard model particles, the so-called freeze-out. Then the $H'$ leftovers from the freeze-out episode populate the universe today. In order to accurately find the relic density of a dark matter particle one would need to solve the Boltzmann equation~\cite{reliccal} as we will do for the fermion dark matter case.
However, since the $H'$ is a scalar dark matter there are only s-wave contributions to the annihilation cross-section and thus the abundance can be approximated as
\be
\Om_{H'}h^2 \simeq \frac{0.1\ \mathrm{pb}}{\langle \sigma v_{rel} \rangle}. 
\ee
Here, the $\langle \sigma v_{rel} \rangle$ is the thermal average over the cross-section for two $H'$ annihilation into the standard model particles multiplied by the relative velocity between the two $H'$ particles. 

For the dark matter masses below the $m_H/2$ the Higgs portal is quite constrained as discussed in Refs. \cite{higgsportal1,higgsportal2}. For the dark matter masses larger than the Higgs mass the annihilation channel $H' H' \rightarrow H H$ plays a major role in determining the abundance. Therefore, we will focus on the Higgs portal below in order to estimate the abundance and derive a bound on the scalar dark matter candidate.
That being said, the interaction of $H'$ with $H$ is obtained as follows 
\be \mathcal{L}_{H'-H} = \left(\frac{\la_5}{2}+\la_3\right)H^2H'^*H'. \ee We have the scattering amplitude for $H'H'\rightarrow HH$,
\be iM(H' H' \rightarrow H H)= i(\la_5 +2\la_3)\equiv i \la'.\ee It is also noted that there may be other contributions to $\la'$ as mediated by the Higgs $H$, the new scalars and new gauge bosons. However, such corrections are subleading with the assumption that the $\la'$ coupling is in order of unity as well as the $H'$ is heavy enough.   
Therefore, the differential cross-section in the center-of-mass frame is given by
\be
\frac{d\sigma}{d\Om} = \frac{|M(H' H' \rightarrow H H)|^2 |\vec{k}|}{64 \pi^2 s |\vec{p}|}.\frac{1}{2},
\ee where the $H'$ has an energy and momentum $H'(E, \vec{p})$ and thus $H'^*(E, -\vec{p})$. Also, the two out-going Higgs bosons possess $H(E, \vec{k})$ and $H(E, -\vec{k})$. The coefficient $\fr 1 2$ is due to the creation of the two identical particles. We have $\sqrt{s} = 2E$. 

From the experimental side, the dark matter is non-relativistic $(v \sim 10^{-3}c)$. We approximate 
\be
E = \frac{m_{H'}}{\sqrt{1-v^2}}\simeq m_{H'}(1+\frac{1}{2}v^2),
\ee where the $v$ is the velocity of the dark matter given in natural units, $v \ll 1$. We have also 
\be s = 4 E^2 \simeq 4 m^2_{H'}(1+v^2),\hs 
|\vec{p}|= \frac{m_{H'} v}{\sqrt{1- v^2}}\simeq m_{H'} v(1+\frac{1}{2}v^2) \simeq m_{H'} v. 
\ee The Einstein relation implies   
\bea
|\vec{k}|&=& \sqrt{E^2 - m^2_H} \simeq \sqrt{m^2_{H'} (1 + v^2) - m^2_H}\crn
&\simeq&  m_{H'}\sqrt{1+v^2-\frac{m^2_H}{m^2_{H'}}}\simeq m_{H'}\left(1+\frac{v^2}{2}-\frac{m^2_H}{2 m^2_{H'}}\right). 
\eea Therefore, the differential cross-section takes the form  
\bea
\frac{d\sigma}{d\Om}\simeq  \frac{\la'^2 m_{H'} \left(1 +\fr{v^2}{2} - \fr{m^2_H}{2 m^2_{H'}}\right)}{64\pi^2 4m^2_{H'}(1+v^2)m_{H'} 2 v }. 
\eea It is clear that the r.h.s is independent of the solid angle, where $d\Om = d\varphi \sin\theta d\theta$. Hence, integrating out over the total space is simply multiplied by $4\pi$,  
 $\sigma = \int\frac{d\sigma}{d\Om}d\Om = 4\pi\frac{d\sigma}{d\Om}$. Because the relative velocity between the two dark matters is $v_{rel} = 2v$, we find out  
\bea
\sigma v_{rel} &\simeq & 4\pi.2v\frac{\la'^2 m_{H'} \left(1 +\fr{v^2}{2} - \fr{m^2_H}{2 m^2_{H'}}\right)}{64\pi^2 4m^2_{H'}(1+v^2)m_{H'} 2 v }\simeq  \frac{\la'^2}{64\pi}\frac{1}{m^2_{H'}}\left(1 -\fr{v^2}{2} - \fr{m^2_H}{2 m^2_{H'}}\right).
\eea Taking the thermal average over both sides, we get
\bea
\langle \sigma v_{rel}\rangle \simeq \frac{\la'^2}{64\pi}\frac{1}{m^2_{H'}}\left(1 - \fr{\langle v^2\rangle}{2} - \fr{m^2_H}{2 m^2_{H'}}\right). 
\eea Notice that $\langle v^2\rangle = \frac{3}{2x_F} $ and $x_F = \frac{m_{H'}}{T_F}\simeq 20$ is given at the freeze-out temperature \cite{reliccal}. 
As aforementioned, we are in the regime $m^2_H \ll m^2_{H'}$ thus  
\bea
\langle \sigma v_{rel}\rangle \simeq \left(\frac{\al}{150\ \mathrm{GeV}}\right)^2 \la'^2 \left(\frac{1.328\ \mathrm{TeV}}{m_{H'}}\right)^2. 
\eea

The relic density of the dark matter $(H')$ satisfies the Boltzmann equation with the solution as given by
$\Om_{H'} h^2 \simeq \frac{0.1\ \mathrm{pb}}{\langle \sigma v_{rel}\rangle} \simeq 0.11$. It follows $\langle \sigma v_{rel}\rangle \simeq 1\ \mathrm{pb}$.
Since $\left(\frac{\al}{150\ \mathrm{GeV}}\right)^2 \simeq 1\ \mathrm{pb}$, we get 
\be \la'^2\left(\frac{1.328\ \mathrm{TeV}}{m_{H'}}\right)^2\simeq 1,\ee which leads to the condition for the dark matter $H'$ mass,  
\be m_{H'} \simeq \la' \times 1.328\ \mathrm{TeV}.\ee To conclude, 
the $H'$ is a dark matter if it has a mass $m_{H'} \simeq 1.328\ \mathrm{TeV}$, provided that $\la' \simeq 1$. In the context of the Higgs portal, for the couplings of order unity the direct detection bounds demand the dark matter masses of order of TeV (see Refs. \cite{higgsportal1,higgsportal2}). Therefore, this scalar is a viable dark matter candidate for providing the right abundance and obeying the direct detection bounds simultaneously. Hereunder, we will focus our attention on the neutral fermion of the model which is a natural dark matter candidate because it can be easily chosen to be the lightest particle among the $W$-odd particles under the parity symmetry discussed previously.

\subsection{Dark matter: Dirac vs Majorana fermion} 

\begin{figure}[!h]
\centering
\includegraphics[scale=0.7]{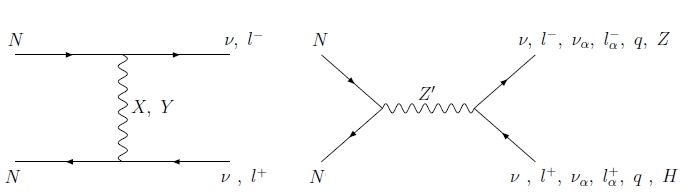}
\caption{Diagrams that contribute to the abundance of the neutral fermion. The neutral fermion scattering off nuclei diagram can be immediately found because it is just the t-channel of the right panel, mediated by the $Z'$-type gauge bosons ($Z_2$ and $Z_N$). The $Z_2$ mediated processes are the most relevant ones though, as we shall see further.}
\label{diagram}
\end{figure}

Among the neutral fermions, $N_a$, the lightest one will be denoted as $N$, which should not be confused with the $U(1)_N$ charge as well as the subscripts of this charge to the $Z_N$ gauge boson, the $g_N$ gauge coupling and the $t_N$ parameter. The neutrino and charged lepton that directly couple to this neutral fermion ($N$) via the $X$ and $Y$ gauge bosons are defined by $\nu$ and $l$, respectively. There remain two other flavors of the neutrinos and charged leptons to be put as $\nu_\al$ and $l_\al$, respectively. In this section we will not dwell on unnecessary details regarding the abundance and direct detection computation. Although we would like to show in Fig.\ref{diagram} the diagrams that contribute the abundance and direct detection signals of the fermion candidate $N$. Surely, the diagram that contributes to the direct detection signal is actually the t-channel diagram of Fig. \ref{diagram} right panel.   

As explicitly shown at the end of Subsection \ref{ctsec1}, the modifications to the couplings of the $Z$ and $Z_{2,N}$ gauge bosons with fermions due to the mixing effects ($Z$ with $Z_{2,N}$) are so small that can be neglected by this analysis. Similarly, the modifications to the $Z_{2,N} Z H$ couplings due to those mixings as well as the neutral scalar mixings ($H$ with $H_{1,2,3}$) are negligible.  

In addition, it is well-known that the interactions of $Z_2$ and $Z_N$ are exchangeable which are only differed by a replacement $(c_\xi\rightarrow -s_\xi;s_\xi \rightarrow c_\xi)$, respectively. Therefore, given that these massive gauge bosons ($Z_{2,N}$) are active particles (i.e. their scales and couplings are equivalent), they play quite the same role in new physical processes (some of these can also be seen obviously in the subsequent subsections). Hence, to keep a simplicity we might consider one particle ($Z_2$) to be active that dominantly sets the dark matter observables while the other one ($Z_N$) almost decouples (which gives negligible contributions). For this aim, we firstly assume $\La>\om$ but not so much larger than the $\om$ so that our postulate of the $\La$ scale, that is comparable to $\om$, is unbroken (still correlated). Hence, choose $\La=10$ TeV and vary $\om$ below this value so that $0.1<\om/\La<1$ (detailedly shown in the cases below). Besides the $\om$ and $\La$ as determined, the $Z_{2,N}$ masses as well as their mixing angle ($\xi$) still depend on their gauge couplings, respectively. The $g,\ g_X$ were fixed via the electromagnetic coupling $e$ and the Weinberg angle, whereas the $g_N$ is unknown. But, we could demand $\al_N\equiv \fr{g^2_N}{4\pi}<1$ or $|g_N|<2\sqrt{\pi}$ so that this interaction to be perturbative. Without loss of generality, we set $0<t_N<2\sqrt{\pi}/g=\fr{s_W}{\sqrt{\al}}\simeq 5.43$. When $t_N$ is large, $t_N \lesssim 5.43$, we have $m_{Z_N}\gg m_{Z_2}$ and the mixing is so small, $t_{2\xi}\simeq -\fr{c_W}{3\sqrt{3-4s^2_W}t_N}\fr{\om^2}{\La^2}\simeq -\fr{0.146}{t_N}\fr{\om^2}{\La^2}\ll 1$, as given from (\ref{goctronxi}). This is the case considered for the relic density of the fermion candidate as a function of its mass ($m_f$), and $t_N=5.43$ is taken into account. Notice that the dark matter annihilation is via s-channels mediated by $Z_{2,N}$. The contribution of $Z_2$ is like $\fr{g^2}{s-m^2_{Z_2}}$, while that of $Z_N$ is $\fr{g_N^2}{s-m^2_{Z_N}}\simeq -\fr{g_N^2}{m^2_{Z_N}}\sim -\fr{1}{\La^2}$ where $s\equiv 4m^2_{f}\sim m^2_{Z_2}\ll m^2_{Z_N}$. Therefore, the $Z_N$ gives a smaller contribution of $\om^2/\La^2$ order which almost vanishes, whereas the relic density is sensitive to the $Z_2$.            

Provided that the relic density of the dark matter gets the right value, we consider both the contributions of $Z_{2,N}$. This is done by varying $0<t_N<5.43$, and respectively $-\pi/2<\xi<0$ as derived from (\ref{goctronxi}). When $t_N\lesssim 5.43$, the $Z_2$ dominates the annihilation as given above. But, when $t_N$ is decreased to $t_N\simeq \fr{c_W}{2\sqrt{3-4s^2_W}}\fr{\om}{\La}\simeq 0.219\fr{\om}{\La}$ or $\xi\simeq -\pi/4$, which is the pole of $t_{2\xi}$ as obtained from (\ref{goctronxi}), the $m_{Z_N}$ becomes comparable to $m_{Z_2}$ as well as the $Z_2$ and $Z_N$ possess the equivalent gauge couplings due to the large mixing. In this case, the $Z_{2}$ and $Z_N$ bosons simultaneously give dominant contributions to the dark matter annihilation despite the fact that $\om \ll \La$. Finally, when $t_N$ approximates zero, $t_N\approx 0$, the $Z_N$ boson governs the annihilation cross-section, while the contribution of $Z_2$ is negligible. The regime that the $Z_N$ dominantly contributes to the dark matter annihilation is very narrow since it is bounded by the maximal mixing value at $t_N\simeq 0.219\om/\La$ which is close to zero due to $\om<\La$. On the other hand, the regime that the $Z_2$ dominates the dark matter annihilation is mostly given in the total $t_N$-range. This is the reason why the $Z_2$ was predicted to govern the dark matter observables while the $Z_N$ is almost neglected, provided that $\om<\La$. It is also clear from all the above analysis that the $Z_2$ and $Z_N$ can be large mixing in spite of small $\om/\La$, given that $t_N\simeq 0.219 \om/\La$. Vice versa, the large regime $t_N\lesssim 5.43$ implies that those gauge bosons can slightly mix $t_{2\xi}\simeq -\fr{0.146}{t_N}\fr{\om^2}{\La^2}\ll1$ even if $\om/\La$ is close to one. Below, we will display the detailed computations for all the cases mentioned.                         

\begin{figure*}[!h]
\centering
\mbox{\includegraphics[scale=0.35]{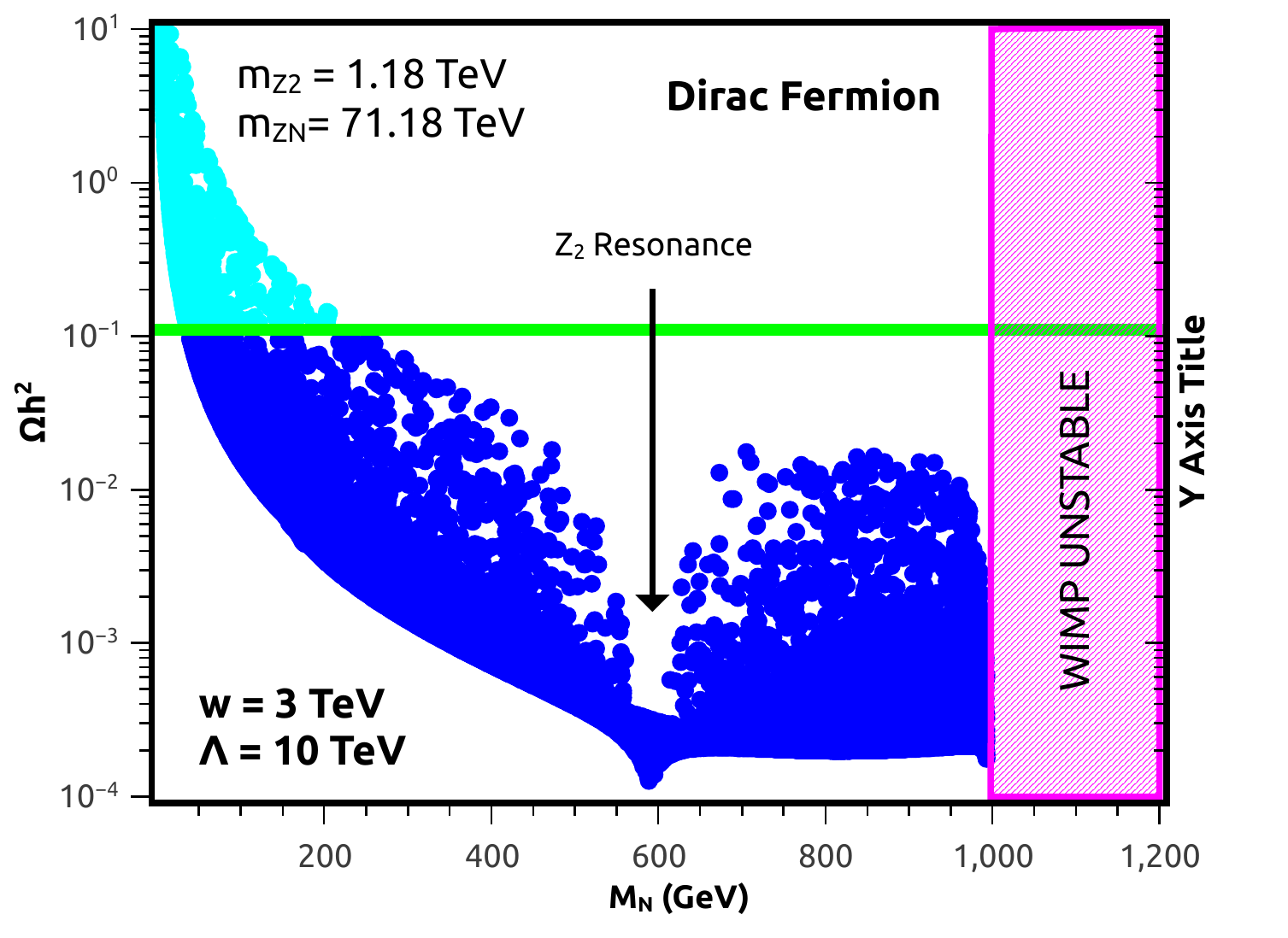}\quad\includegraphics[scale=0.35]{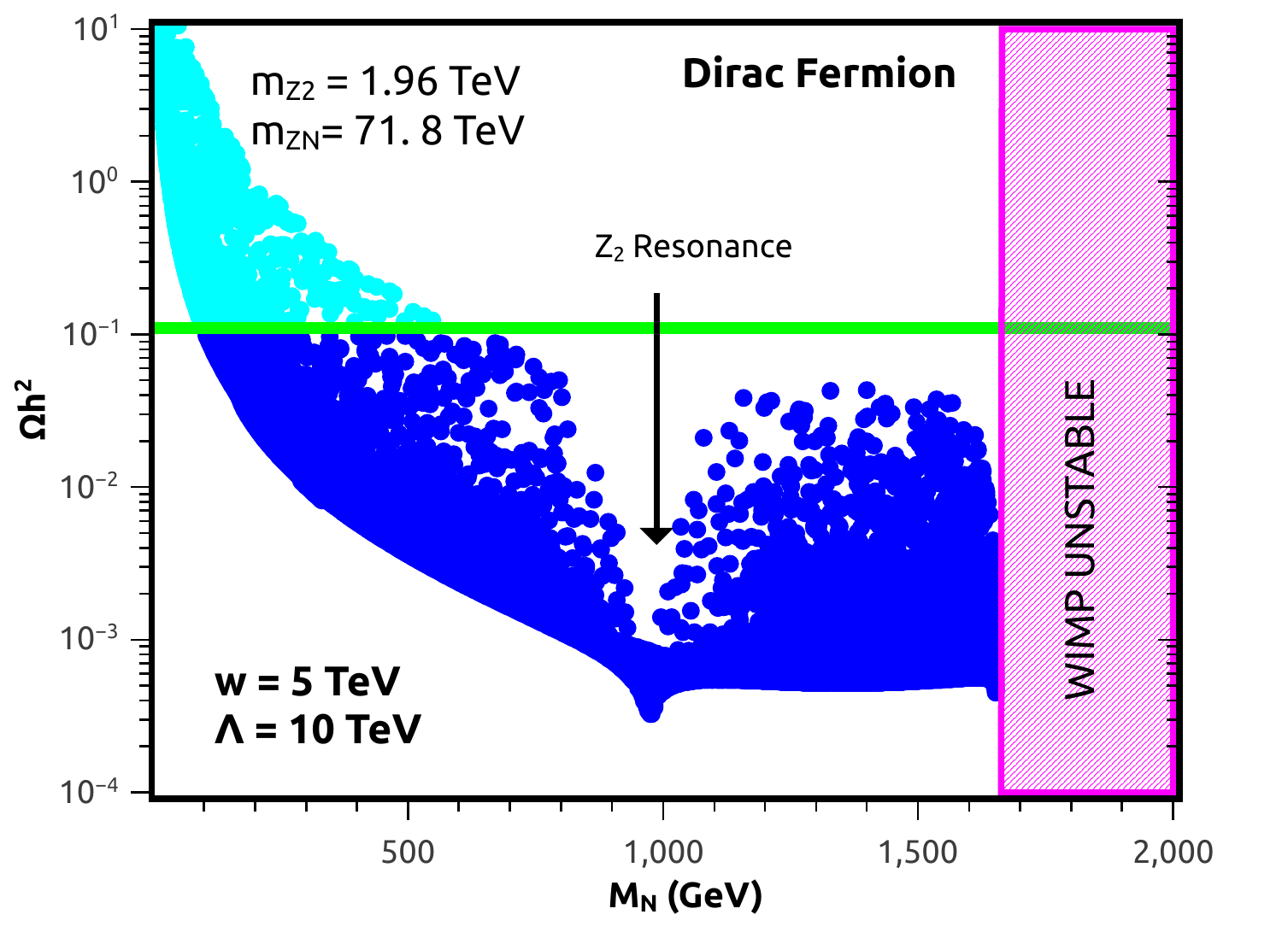}\quad\includegraphics[scale=0.35]{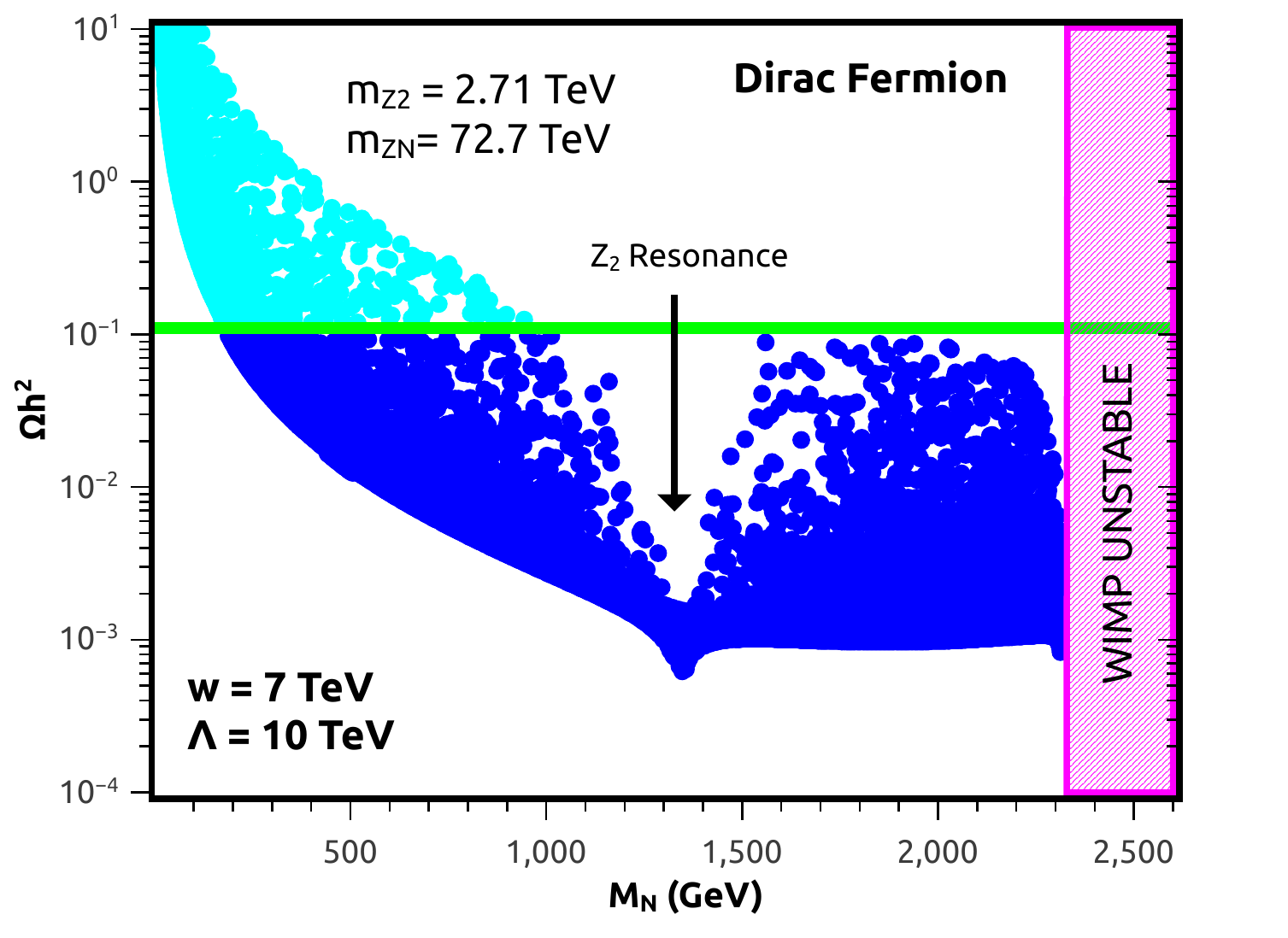}}
\caption{Abundance of the Dirac fermion $N$ as a function of its mass for different scales of the symmetry breaking. The shaded region is excluded for inducing the WIMP decay such as $N \rightarrow X \nu$. One can clearly see that the $Z_2$ resonance plays a major role in the annihilation computation. See text for more detail.}
\label{Graph1}
\end{figure*}

\begin{figure*}[!h]
\centering
\mbox{\includegraphics[scale=0.5]{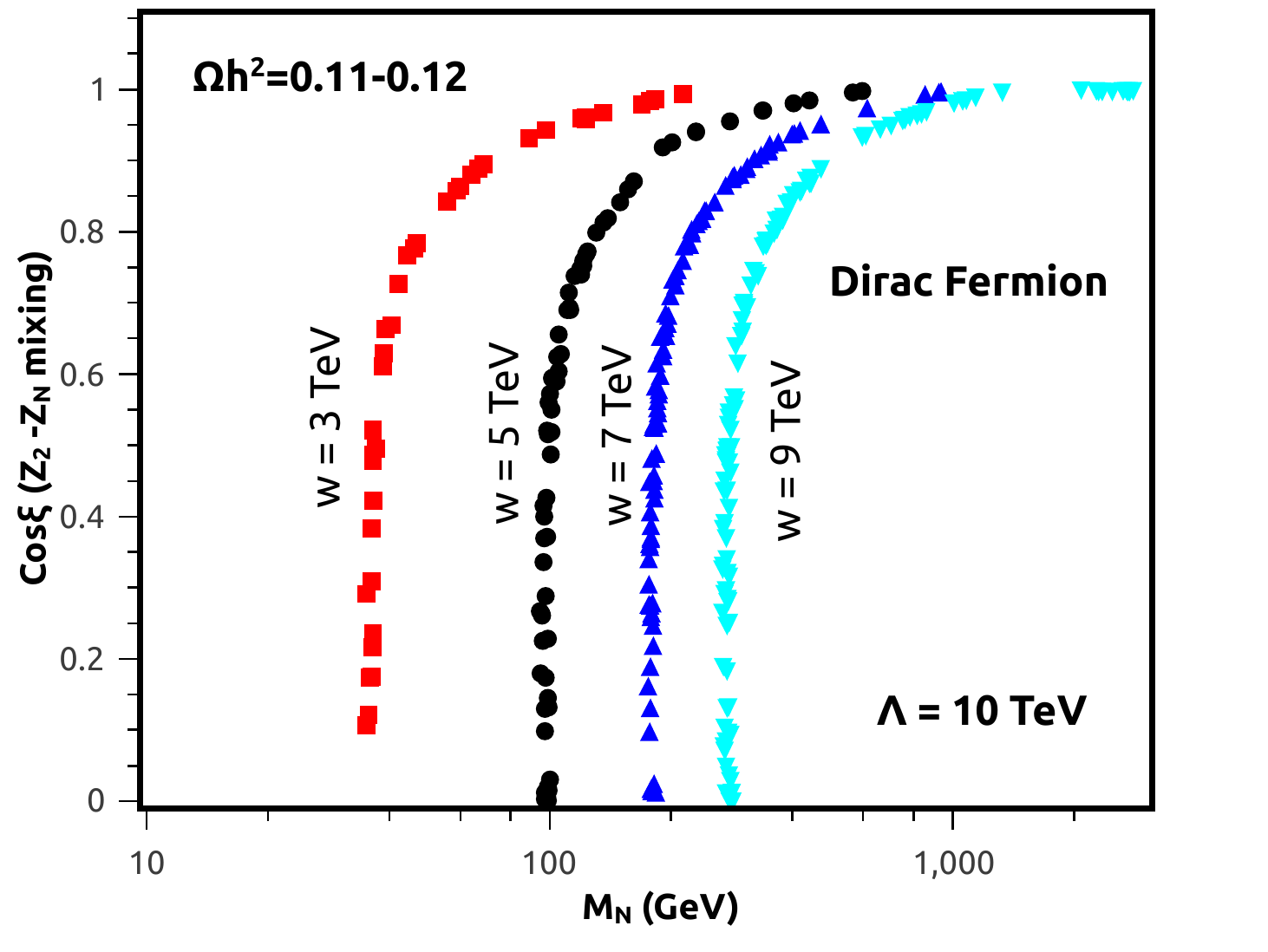}\quad\includegraphics[scale=0.5]{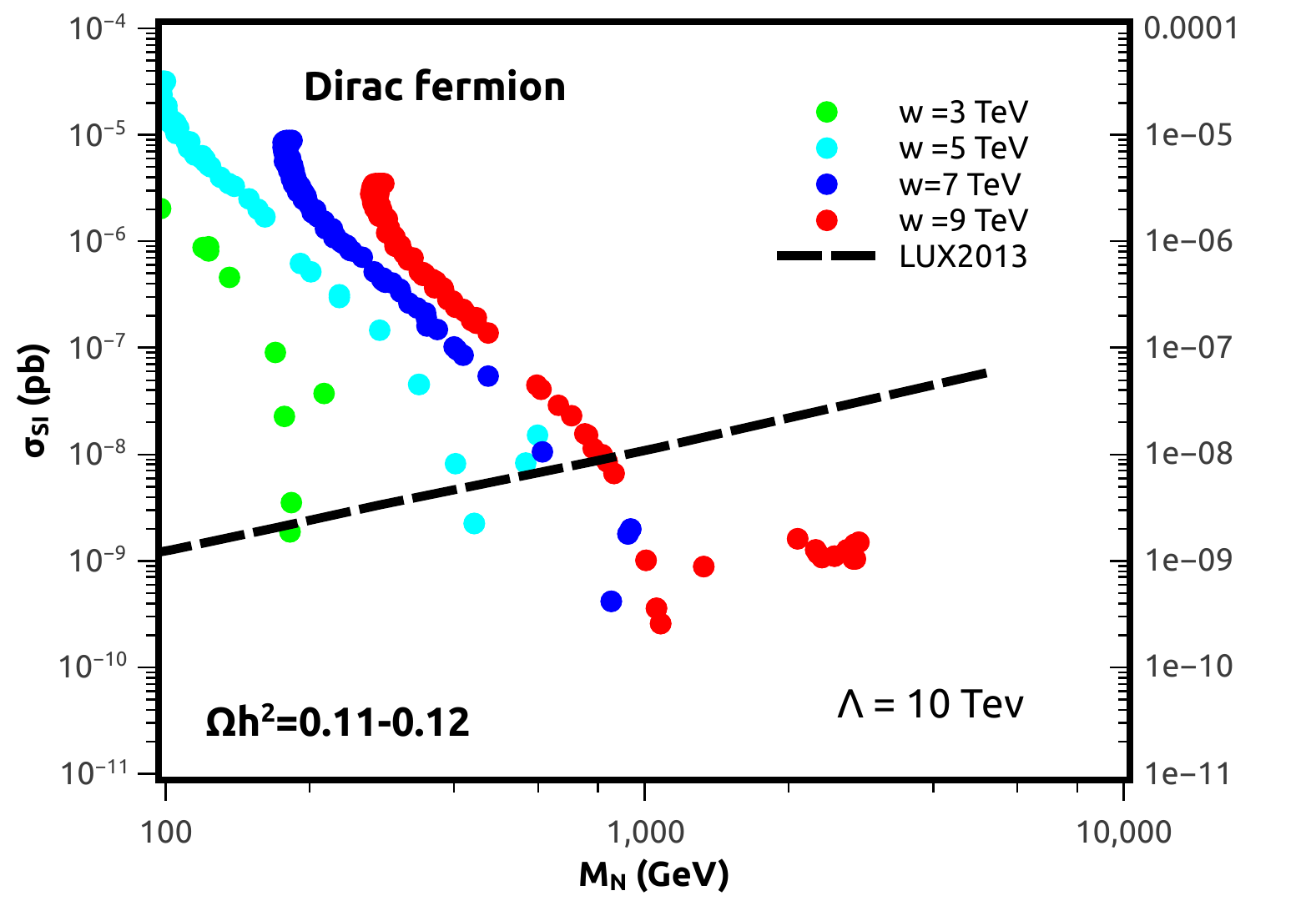}}
\caption{{\it Left}: Mixing angle $\times$ fermion mass plane which yields the right abundance for a Dirac fermion. The discontinuity in the plots has to do with the $Z_2$ resonance that pushes down the overall abundance. {\it Right}: Spin-independent scattering cross-section in terms of the Dirac fermion mass for different values of symmetry breaking.
One can easily conclude that the current LUX bounds require $\om \gtrsim 5$~TeV. We have let the mixing angle $\xi$ free to float in our analyses. As the mixing angle goes to zero ($\cos\xi \rightarrow 1$) the coupling $Z_2$-quarks decreases as seen from Table IV.}
\label{Graph3}
\end{figure*}

\begin{figure*}[!h]
\centering
\mbox{\includegraphics[scale=0.35]{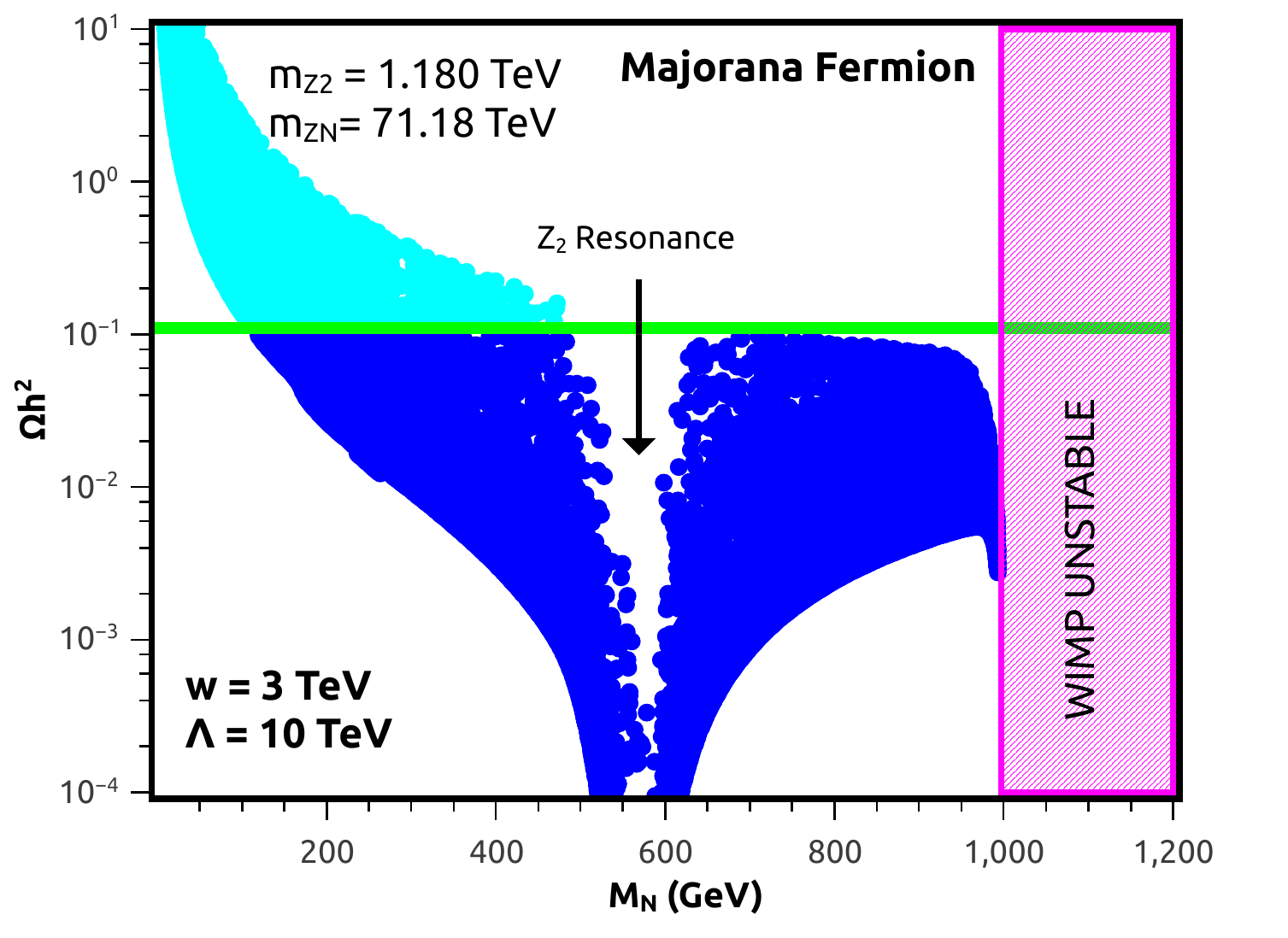}\quad\includegraphics[scale=0.35]{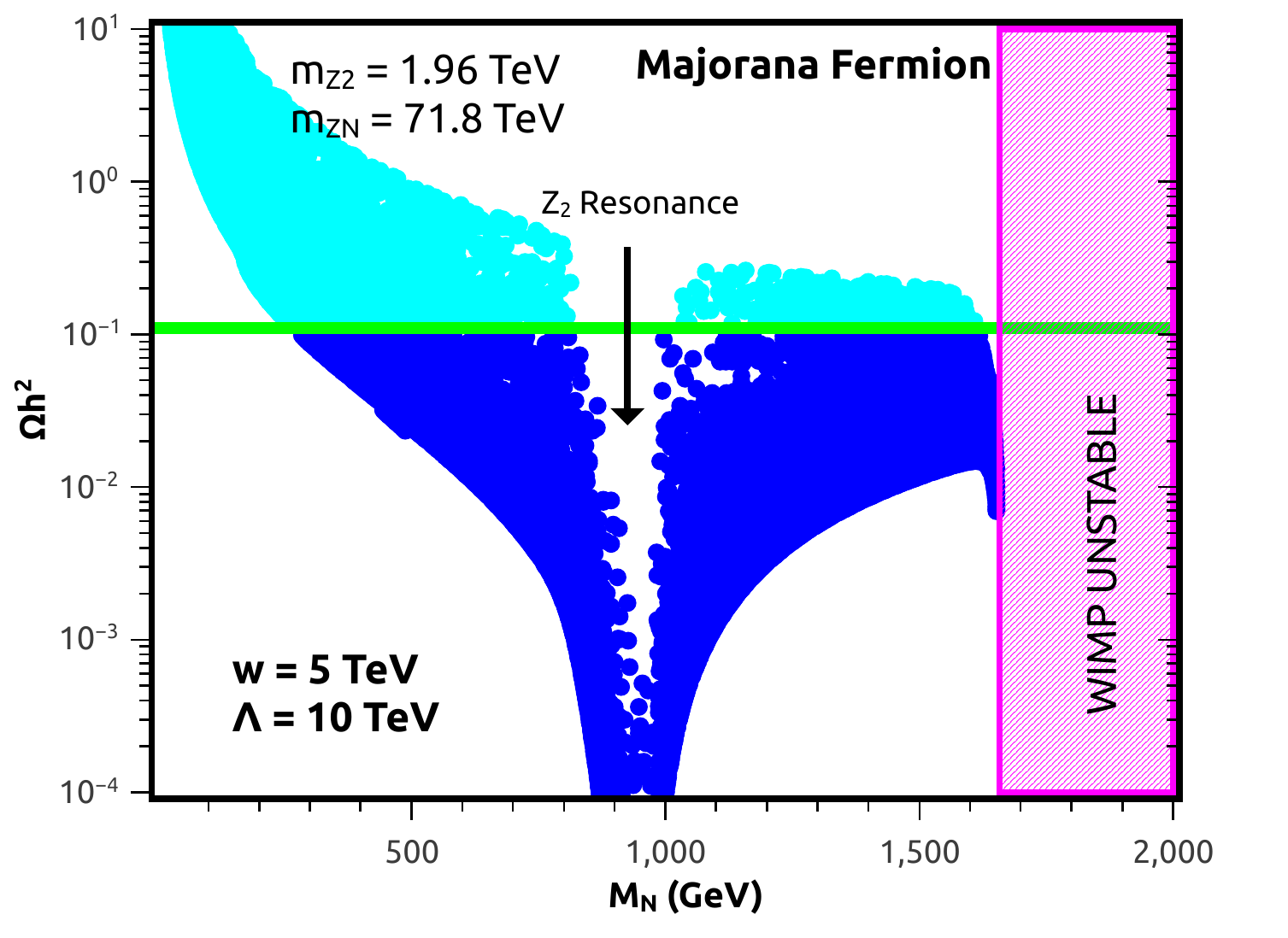}\quad\includegraphics[scale=0.35]{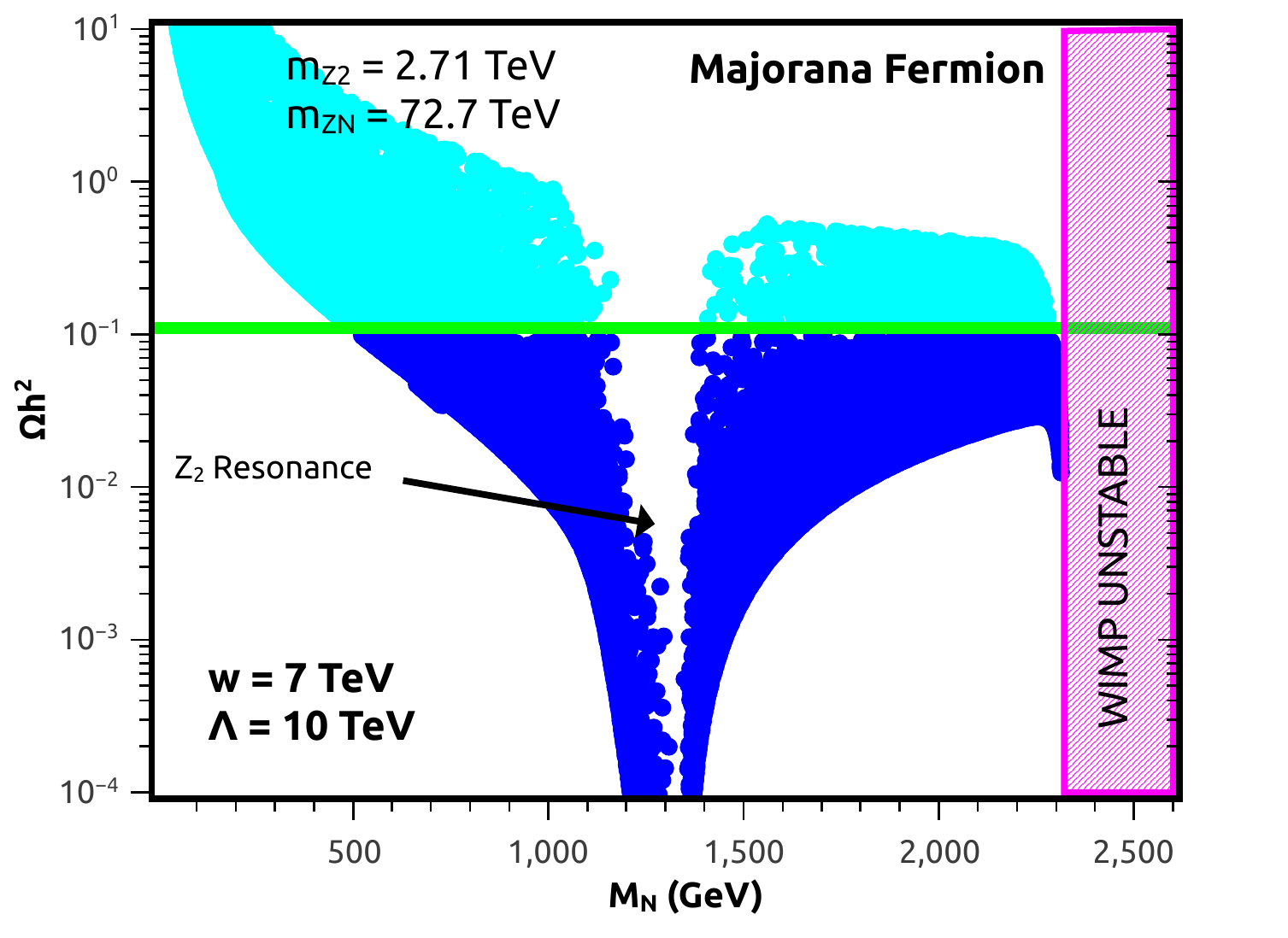}}
\caption{Abundance of the Majorana fermion $N$ as a function of its mass for different scales of the symmetry breaking. The shaded region is excluded for inducing the WIMP decay such as $N \rightarrow X \nu$. One can clearly see that the $Z_2$ resonance plays a major role in the annihilation computation. See text for more detail.}
\label{Graph2}
\end{figure*}

In case the candidate $N$ is a Dirac fermion, it has both vector and axial-vector couplings with the neutral gauge bosons. The abundance is shown in Fig. \ref{Graph1}. [In this figure and the following ones, the $\om$ is sometimes denoted as $w$ instead that should not be confused]. It is clear from Fig.~\ref{Graph1} that the gauge boson $Z_2$ overwhelms the remaining annihilation channels in agreement with Ref. \cite{farinaldoDM1}, and the resonance at the $m_{Z_2}/2$ is crucial in determining the abundance. Moreover, we see that the mass region $100-200$~GeV for $\om=3$~TeV, $100-500$~GeV for $\om=5$~TeV, or $100-1000$~GeV for $\om=7$~TeV provides the right abundance. Additionally, we exhibit in the left panel of Fig. \ref{Graph3} the region of the parameter space $\cos(\xi)\ \times$ the neutral fermion mass that yields the right abundance, where $\xi$ is the $Z_2$ and $Z_N$ mixing angle. When this angle goes to zero the coupling $Z_2$-quarks decreases and for this reason the scattering cross section rapidly decreases as shown in the right panel of Fig. \ref{Graph3}. There, and throughout this work we let cosine of this mixing angle free to float from zero to unity. [Correspondingly, the $\xi$ ($t_N$) run from $-\pi/2$ (0) to 0 (5.43)]. As for the Majorana case, the overall abundance is enhanced and hence we find a larger region of the parameter space that yields the right abundance as can seen in Fig. \ref{Graph2}.

\begin{figure*}[!h]
\centering
\mbox{\includegraphics[scale=0.5]{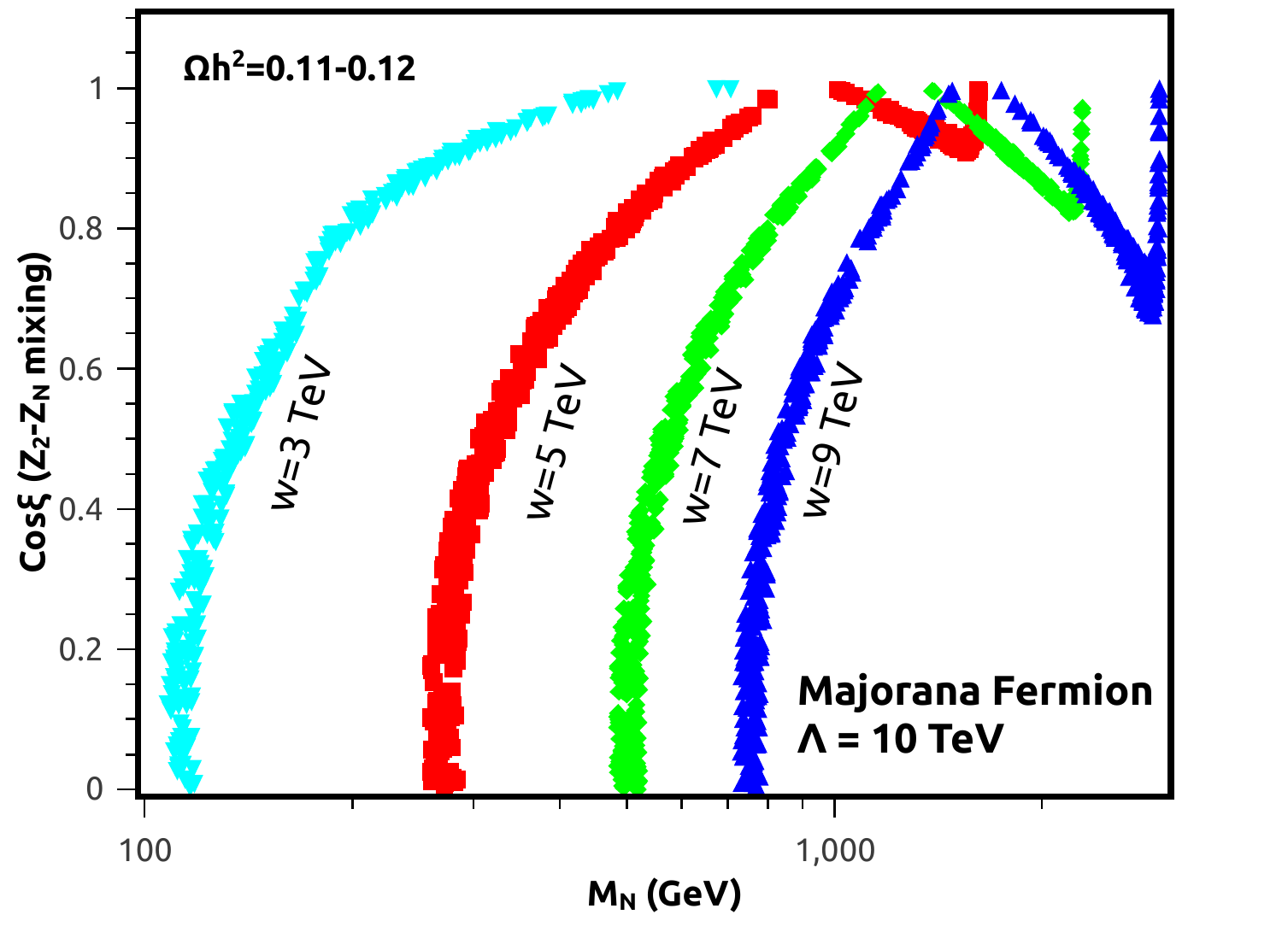}\quad\includegraphics[scale=0.5]{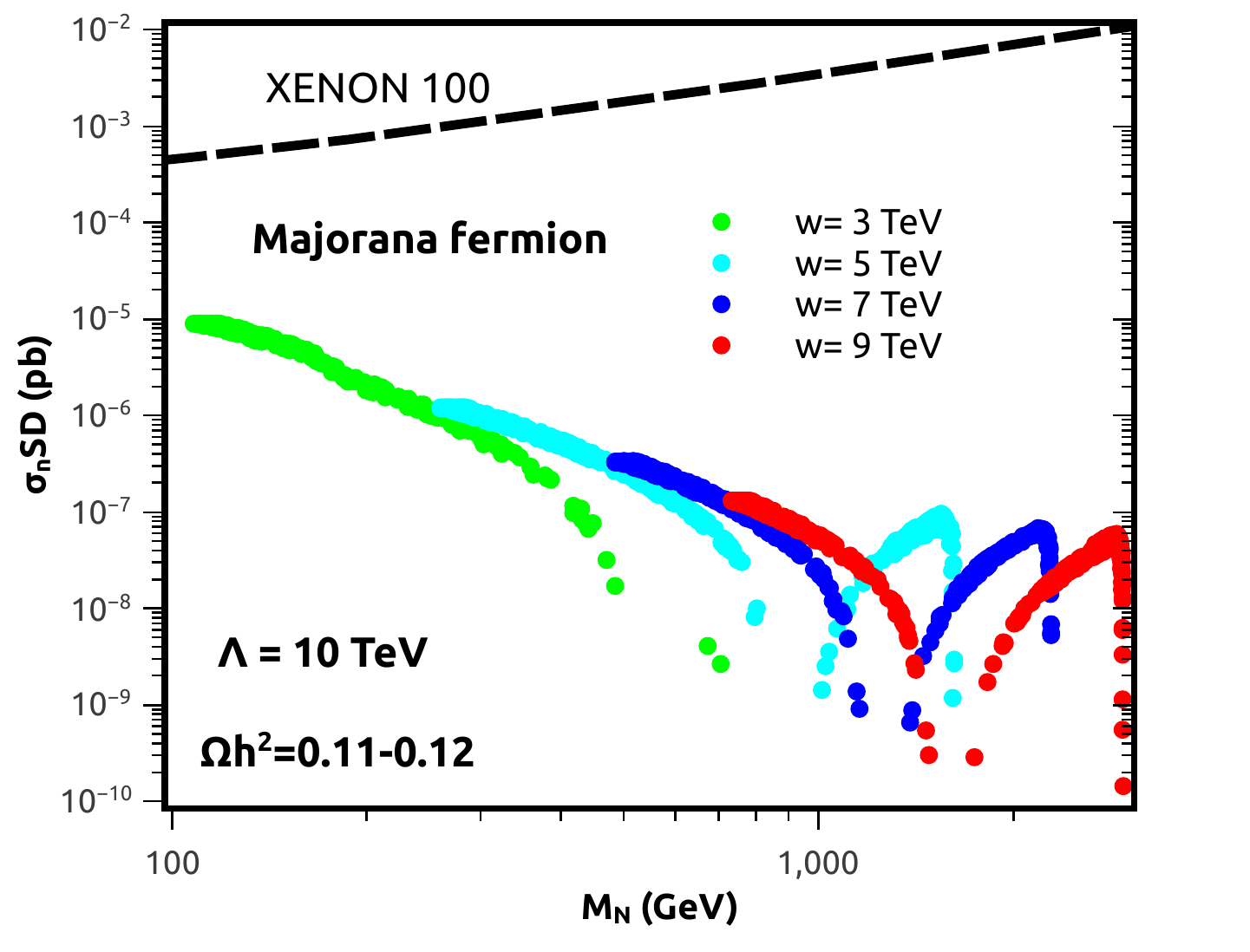}}
\caption{{\it Left}: Mixing angle $\times$ fermion mass plane which yields the right abundance for a Majorana fermion. {\it Right}: Spin-dependent scattering cross section in terms of the Majorana fermion mass for different values of the symmetry breaking. One can easily conclude that the current XENON100 bounds are rather loose.}
\label{Graph4}
\end{figure*}

As for the direct detection signal, the Dirac fermion dark matter candidates give rise to spin-independent (vector) and spin-dependent (axial-vector) scattering cross-sections. But, due to the $A^2$ enhancement that is typical of heavy targets used in direct detection experiments, the spin-independent bounds are the most stringent ones. One can see in Fig. \ref{Graph3}. On the other hand, the Majorana fermions have zero vector current. This is because the current of a fermion is equal to the current of an anti-fermion, but if one applies the Majorana condition ($\psi =\psi^c$) one find that the vector current must vanish (which has also been used for the abundance computation aforementioned). Therefore, only the spin-dependent bounds apply. In Fig. \ref{Graph4} we show those bounds. The LUX collaboration has not reported their spin-dependent bounds yet, so the strongest constraints come from XENON100 \cite{Aprile:2013doa}. One should conclude from Fig. \ref{Graph4} that the XENON100 bounds are quite loose for the Majorana fermion.

\subsection{Monojet and dijet bounds}

Monojet and dijet resonances have been searched at Tevatron, ATLAS and CMS with null results so far. Such signals have been intensively exploited in the literature. In particular, the dijet bounds are neither sensitive to the dark matter mass nor to the $Z_2$-dark matter couplings, but on the other hand it is quite sensitive to the $Z_2$-quarks couplings. In Ref. \cite{dijetbounds} lower bounds namely $M_{Z^{\prime}} \sim 1.7$~TeV have been found under the assumption that the $Z^{\prime}$ boson couples similarly to the standard model $Z$ boson and for the dark matter masses smaller than $500$~GeV. One might notice in fact that the $Z_2$ gauge boson couples similarly to the $Z$ boson. Therefore, the bounds found in Ref. \cite{dijetbounds} apply here up to some extent since the couplings are not precisely identical. That being said, the result shown in the leftmost panel of Fig. \ref{Graph1} might be in tension with the existing dijet bounds. The remaining plots do obey the dijet bounds since they are obtained at the $Z_2$ masses greater than $1.7$~TeV. It is important to keep in mind that the collider bounds derived from simplified models are more comprehensive than the ones using an effective operator approach, because the production cross-sections using the effective operator either over-estimate or under-estimate the collider bounds as discussed in Refs. \cite{An:2012va,Alves:2014yha}. Concerning the monojet bounds, it has been shown that the current direct detection limits coming from LUX are typically more stringent. Therefore, we will not refer to the monojet bounds hereafter.

\subsection{FCNCs}

The fermions get masses from the Yukawa interactions when the scalar fields develop VEVs as presented in \cite{3311}. Due to the $W$-parity conservation, the up quarks ($u_a$) do not mix with $U$ and the down quarks ($d_a$) do not mix with $D_\al$ (remind that the exotic quarks are $W$-odd while the ordinary quarks are $W$-even). The exotic quarks gain large masses in $\om$ scale and decoupled, whereas the ordinary quarks concerned mix by themselves via a mass Lagrangian of the form,
\be \mathcal{L}^{u,d}_{\mathrm{mass}}=-\bar{u}_{aL} m^u_{ab} u_{bR} - \bar{d}_{aL} m^d_{ab} d_{bR}+H.c.,\ee where
\bea m^u_{\al a} &=& \fr{1}{\sqrt{2}}h^u_{\al a} v,\hs m^u_{3a}= -\fr{1}{\sqrt{2}} h^u_{a} u, \crn 
m^d_{\al a} &=& -\fr{1}{\sqrt{2}}h^d_{\al a} u,\hs m^d_{3a}= -\fr{1}{\sqrt{2}} h^d_{a} v. \eea The mass matrices $m^u=\{m^u_{ab}\}$ and $m^d=\{m^d_{ab}\}$ can be diagonalized to yield physical states and masses such as 
\be  u_L = V_{uL}(u\ c\ t)^T_{L},\hs u_R=V_{uR}(u\ c\ t)^T_{R},\hs d_L = V_{dL}(d\ s\ b)^T_{L},\hs d_R=V_{dR}(d\ s\ b)^T_{R},\ee
\be V^\dagger_{uL} m^u V_{uR} = \mathrm{diag}(m_u,\ m_c,\ m_t),\hs V^\dagger_{dL} m^d V_{dR}=\mathrm{diag}(m_d,\ m_s,\ m_b), \ee
where $u=\{u_a\}$ and $d=\{d_a\}$. The CKM matrix \cite{ckm} is defined as $V_{\mathrm{CKM}}=V^\dagger_{uL}V_{dL}$. 

All the mixing matrices $V_{uL},\ V_{dL},\ V_{uR},\ V_{dR}$ including $V_{\mathrm{CKM}}$ are unitary. The GIM mechanism \cite{gim} of the standard model works in this model, which is a consequence of the $W$-parity conservation. Let us note that in the 3-3-1 model with right-handed neutrinos, the ordinary quarks and exotic quarks that have different $T_3$ weak isospins mix by contrast (which results from the unwanted nonzero VEVs of $\eta^0_3$ and $\chi^0_1$ as well as the lepton-number violating interactions $\bar{Q}_{3L}\chi u_{a R}$, $\bar{Q}_{3L}\eta U_{R}$, $\bar{Q}_{3L}\rho D_{\al R}$, $\bar{Q}_{\al L}\chi^* d_{a R}$, $\bar{Q}_{\al L}\eta^* D_{\beta R}$, $\bar{Q}_{\al L}\rho^* U_{R}$ and their Hermitian conjugation, that directly couple ordinary quarks to exotic quarks via mass terms \cite{dkek}). Hence, in that model, the dangerous tree-level FCNCs of $Z$ boson happen due to the non-unitarity of the mixing matrices as listed above ($V_{uL},\ V_{dL},\ V_{uR},\ V_{dR}$). Even, the dangerous FCNCs also come from one-loop contributions of $W$ boson due to the non-unitarity of the CKM matrix ($V_{\mathrm{CKM}}$). Therefore, the standard model GIM mechanism does not work. This will particularly be analyzed at the end of this subsection.          

In this model, the tree level FCNCs happen only with the new gauge bosons $Z_2$ and $Z_N$ (notice that there is a negligible contribution coming from the $Z$ boson due to the mixing with $Z_{2,N}$ as explicitly shown below). This is due to the non-universal property of quark representations under $SU(3)_L$ that the third quark generation differs from the first two generations. Indeed, from (\ref{ttz2zn}) for the interactions of $Z_{2,N}$, the right-handed flavors ($\Psi_R$) are conserved since $T_8=0$, $X=Q$ and $N=B-L$ which are universal for ordinary up- and down-quarks. But, the left-handed flavors ($\Psi_L$) are changing due to the fact that $T_8$ differs for quark triplets and antitriplets [note that $X$ and $N$ are related to $T_8$ by (\ref{ecqbl}); the source for the FCNCs is due to the $T_8$ only since $T_3$ is also universal for ordinary up-quarks and down-quarks as the same reason of the flavor-conserved $Z$ current]. The interactions that lead to flavor changing can be derived from (\ref{ttz2zn}) as          
\bea \mathcal{L}_{T_8} &=& \bar{\Psi}_L\ga^\mu T_8 \Psi_L (g_2 Z_{2\mu} + g_N Z_{N\mu}),\label{tdtronz}\\ 
g_2 &\equiv& -g\left(c_\xi \fr{1}{\sqrt{1-t^2_W/3}}+s_\xi \fr{2t_N}{\sqrt{3}}\right),\crn 
g_N &\equiv& g_2(c_\xi \rightarrow -s_\xi;s_\xi \rightarrow c_\xi),\nn \eea where $\Psi_L$ indicates to all ordinary left-handed quarks. We can rewrite 
\bea \mathcal{L}_{T_8} &=& (\bar{u}_L\ga^\mu T_{u} u_L +\bar{d}_L\ga^\mu T_{d} d_L)(g_2 Z_{2\mu} + g_N Z_{N\mu})\crn
&=&[\bar{u}'_L\ga^\mu (V^\dagger_{uL}T_{u}V_{uL}) u'_L +\bar{d}'_L\ga^\mu (V^\dagger_{dL}T_{d}V_{dL}) d'_L](g_2 Z_{2\mu} + g_N Z_{N\mu}), \eea where $u'=(u,c,t)$, $d'=(d,s,b)$ and $T_u=T_d=\fr{1}{2\sqrt{3}}\mathrm{diag}(-1,-1,1)$. Hence, the tree-level FCNCs are described by the Lagrangian,    
\be \mathcal{L}_{\mathrm{FCNC}}=\bar{q}'_{iL}\ga^\mu q'_{jL}\fr{1}{\sqrt{3}}(V^*_{qL})_{3i} (V_{qL})_{3j} (g_2 Z_{2\mu} + g_N Z_{N\mu})\hs (i\neq j),\ee where we have denoted $q$ as $u$ either $d$.  

The FCNCs lead to hadronic mixings such as $K^0-\bar{K}^0$, $D^0-\bar{D}^0$, $B^0-\bar{B}^0$ and $B^0_s-\bar{B}^0_s$, caused by pairs $(q'_i,q'_j)=(d,s),\ (u,c),\ (d,b),\ (s,b)$, respectively. These mixings are described by the effective interactions as obtained from the above Lagrangian via $Z_{2,N}$ exchanges as 
\be \mathcal{L}^{\mathrm{eff}}_{\mathrm{FCNC}}=(\bar{q}'_{iL}\ga^\mu q'_{jL})^2 \fr 1 3 [(V^*_{qL})_{3i} (V_{qL})_{3j}]^2\left(\fr{g^2_2}{m^2_{Z_2}} + \fr{g^2_N} {m^2_{Z_N}}\right).\label{hutronhd} \ee The strongest constraint comes from the $K^0-\bar{K}^0$ mixing \cite{pdg} that 
\be \fr 1 3 [(V^*_{dL})_{31} (V_{dL})_{32}]^2\left(\fr{g^2_2}{m^2_{Z_2}} + \fr{g^2_N} {m^2_{Z_N}}\right) < \fr{1}{(10^4\ \mathrm{TeV})^2}.\ee   
Assuming that $u_a$ is flavor-diagonal, the CKM matrix is just $V_{dL}$ (i.e. $V_{\mathrm{CKM}}=V_{dL}$). Therefore, $|(V^*_{dL})_{31} (V_{dL})_{32}|\simeq 3.6\times 10^{-4}$ \cite{pdg} and we have 
\be \sqrt{\fr{g^2_2}{m^2_{Z_2}} + \fr{g^2_N} {m^2_{Z_N}}}< \fr{1}{2\ \mathrm{TeV}}.\label{chinhdie1}\ee This gives constraints on the mass and coupling of the new neutral gauge bosons, that is 
\be m_{Z_{2,N}}>g_{2,N}\times 2\ \mathrm{TeV}.
\label{Z2bound}
\ee 

There is another bound coming from the $B^0_s-\bar{B}^0_s$ mixing that is given by \cite{pdg} 
 \be \fr 1 3 [(V^*_{dL})_{32} (V_{dL})_{33}]^2\left(\fr{g^2_2}{m^2_{Z_2}} + \fr{g^2_N} {m^2_{Z_N}}\right) < \fr{1}{(100\ \mathrm{TeV})^2}.\ee In this case, the CKM factor is $|(V^*_{dL})_{32} (V_{dL})_{33}|\simeq 3.9\times 10^{-2}$ \cite{pdg}. Therefore, we have 
\be \sqrt{\fr{g^2_2}{m^2_{Z_2}} + \fr{g^2_N} {m^2_{Z_N}}}< \fr{1}{2.25\ \mathrm{TeV}},\label{chinhdie2}\ee which implies
\be m_{Z_{2,N}}>g_{2,N}\times 2.25\ \mathrm{TeV}. \ee   

To be concrete, suppose that $Z_2$ and $Z_N$ have approximately equal masses and $t_N=g_N/g=1$ so that the $B-L$ interaction strength is equivalent to that of the weak interaction. From (\ref{chinhdie1}), we get \be m_{Z_{2}}\approx m_{Z_N} > 2.037\ \mathrm{TeV},\ee while the relation (\ref{chinhdie2}) yields \be m_{Z_2}\approx m_{Z_N}> 2.291\ \mathrm{TeV}.\ee Here, we have used $g^2=4\pi\al/s^2_W$ with $s^2_W=0.231$ and $\al=1/128$. This is in good agreement with the recent bound \cite{masszp}. Notice that we have used $m_{Z_N} \gg m_{Z_2}$ in the dark matter subsections though which translates to $m_{Z_{2}} \gtrsim 1$~TeV.

Finally, let us give some remarks on the FCNCs due to the mixing effect of the neutral gauge bosons. In this case, the Lagrangian (\ref{tdtronz}) is changed with the replacement by  
\be g_2 Z_{2\mu}+g_N Z_{N\mu}\longrightarrow g_1 Z_{1\mu}+g_2 Z_{2\mu}+g_N Z_{N\mu},\ee where \be g_1\equiv g_2(c_\xi \rightarrow -\mathcal{E}_1;s_\xi\rightarrow -\mathcal{E}_2)=-\fr{\sqrt{3}g}{4c^3_W}\fr{v^2-c_{2W}u^2}{\om^2}.\ee Correspondingly, the effective interactions for the FCNCs given by (\ref{hutronhd}) is also changed with the replacement as follows 
\be \fr{g^2_2}{m^2_{Z_2}}+\fr{g^2_N}{m^2_{Z_N}}\longrightarrow \fr{g^2_1}{m^2_{Z_1}}+\fr{g^2_2}{m^2_{Z_2}}+\fr{g^2_N}{m^2_{Z_N}}.\ee Let us compare the new contribution with the existing one, \be R\equiv \fr{g^2_1/m^2_{Z_1}}{(g^2_2/m^2_{Z_2})+(g^2_N/m^2_{Z_N})}. \ee It is sufficient to consider two cases, $\La \gg \om$ and $\La \sim \om$. For the first case, the $R$ is similar to (becomes) the 3-3-1 model with right-handed neutrinos that 
\be R\simeq \fr{g^2_1/m^2_{Z_1}}{g^2_2/m^2_{Z_2}}\simeq \fr{1}{4c^4_W}\fr{(v^2-c_{2W}u^2)^2}{\om^2(u^2+v^2)} < \fr{1}{4c^4_W}\left(\fr{v_{\mathrm{w}}}{\om}\right)^2 < 0.0025, \ee which is very small. Above, we have used $m^2_{Z_1}\simeq g^2(u^2+v^2)/(4c^2_W)$, $m^2_{Z_2}\simeq g^2 c^2_W \om^2/(3-4s^2_W)$, $v^2_{\mathrm{w}}=u^2+v^2=(246\ \mathrm{GeV})^2$, and $\om>3.198$ TeV as derived from the $\rho$ parameter. For the second case, the contributions of $Z_2$ and $Z_N$ are equivalent. So, the first remark is $R\sim (g^2_1/g^2_{2,N})(m^2_{Z_{2,N}}/m^2_{Z_1}) \sim \mathcal{E}^2_{1,2}(m^2_{Z_{2,N}}/m^2_{Z_1})\sim (u^4/\om^4)(\om^2/u^2)=u^2/\om^2$, which starts from the $(u/\om)^2$ order and must be small too. Indeed, let us show this explicitly   
\bea R\leq \fr{g^2_1/m^2_{Z_1}}{2|g_2 g_N|/(m_{Z_2}m_{Z_N})}&=&\fr{1}{8 c^3_W t_N |s_{2\xi}|\sqrt{3-4s^2_W}} \fr{(v^2-c_{2W}u^2)^2}{\om\La (u^2+v^2)}\crn
&<&\fr{1}{8 c^3_W t_N |s_{2\xi}|\sqrt{3-4s^2_W}}\fr{v^2_{\mathrm{w}}}{\om\La}\simeq 0.00076,\eea provided that $t_N=1$, $\xi=-\pi/4$ ($s_{2\xi}$ is finite due to the large mixing of $Z_2$ and $Z_N$, thus such value could be chosen), and $\La=\om=3.198$ TeV. Above, we have also used $m_{Z_2}m_{Z_N}=2g^2 c_W t_N \om \La/\sqrt{3-4s^2_W}$, which can be derived from (\ref{kluongz2nhe}) and (\ref{kluongznnhe}), the expression (\ref{goctronxi}) for the $\xi$ mixing angle, and the $m^2_{Z_1}$ as approximated before. In summary, the mixing effects with the $Z$ boson do not affect to the FCNCs.           

For the sake of completeness, let us point out the dangerous FCNCs of $Z$ boson due to the mixing of the ordinary quarks and exotic quarks that happens in the 3-3-1 model with right-handed neutrinos, which should be suppressed. The mixing matrices are redefined as $(u_1\ u_2\ u_3\ U)^T_{L,R}=V_{uL,R} (u\ c\ t\ T)^T_{L,R}$ and $(d_1\ d_2\ d_3\ D_1\ D_2)^T_{L,R}=V_{dL,R}(d\ s\ b\ D\ S)^T_{L,R}$ so that the $4\times 4$ mass matrix of up-quarks $(u_a, U)$ and the $5\times 5$ mass matrix of down-quarks $(d_a,\ D_\al)$ are diagonalized, respectively~\cite{dkek}. The Lagrangian that describes the FCNCs of $Z$ boson is given by $(\pm)\fr{g}{2c_W} \bar{q}'_{iL}\ga^\mu q'_{jL} (V^*_{qL})_{Ii} (V_{qL})_{Ij} Z_\mu$, where $I=4$ for $V_u$ and the plus sign is applied, but $I=4,5$ for $V_d$ and the minus sign is taken (note, however, that the right chiral currents of $Z_\mu$ do not flavor-change since $T_3=0$ for any right-handed fermion). All these lead to the effective interactions for the hadronic mixings due to the exchange of $Z$ boson as 
\be (\bar{q}'_{iL}\ga^\mu q'_{jL})^2 [(V^*_{qL})_{Ii} (V_{qL})_{Ij}]^2\fr{1}{u^2+v^2},\label{fcncdz}\ee where we have used $m^2_Z=g^2(u^2+v^2)/(4c^2_W)$ and notice that $v^2_{\mathrm{w}}\equiv u^2+v^2=(246\ \mathrm{GeV})^2$. In the 3-3-1 model with right-handed neutrinos, the Lagrangian for the FCNCs of $Z'$ boson is easily obtained as $\fr{-g}{\sqrt{1-t^2_W/3}}\bar{q}'_{iL}\ga^\mu q'_{jL}\fr{1}{\sqrt{3}}[V^\dagger_{qL} V_{qL}]_{ij} Z'_\mu$, where $[V^\dagger_{uL} V_{uL}]_{ij}\equiv (V^*_{uL})_{3i}(V_{uL})_{3j}-\fr 1 2 (V^*_{uL})_{4i}(V_{uL})_{4j}$ and $[V^\dagger_{dL} V_{dL}]_{ij}\equiv (V^*_{dL})_{3i}(V_{dL})_{3j}+\fr 3 2 (V^*_{dL})_{Ii}(V_{dL})_{Ij}$. Hence, the effective interactions for the hadronic mixings due to the $Z'$ contribution is given by 
\be (\bar{q}'_{iL}\ga^\mu q'_{jL})^2 [V^\dagger_{qL} V_{qL}]^2_{ij} \fr{1}{\om^2}, \label{fcncdzp} \ee where we have adopted $m^2_{Z'}\simeq \fr{g^2c^2_W}{3-4s^2_W}\om^2$ \cite{tw}. Since the weak scale $v_{\mathrm{w}}$ in (\ref{fcncdz}) is too low in comparison to the 3-3-1 scale $\om$ in (\ref{fcncdzp}), it is clear that if the mixing of the ordinary quarks and exotic quarks is similar in size to that of the ordinary quarks, $(V^*_{qL})_{Ii} (V_{qL})_{Ij}\sim (V^*_{qL})_{3i}(V_{qL})_{3j}$, the FCNCs due to the $Z$ boson (\ref{fcncdz}) is too large ($\sim \om^2/v^2_{\mathrm{w}}\sim 10^2$ times the one coming from $Z'$ or the bound for the $K^0-\bar{K}^0$ mixing) as such the theory is invalid. Hence, the FCNCs due to the ordinary and exotic quark mixing are more dangerous than those coming from the non-universal interactions of $Z'$ boson. To avoid the large FCNCs, one must assume $|(V^*_{qL})_{Ii} (V_{qL})_{Ij}|\ll |(V^*_{qL})_{3i}(V_{qL})_{3j}|$ (and the FCNCs of $Z'$ are dominated by the ordinary quark mixing, $[V^\dagger_{qL} V_{qL}]_{ij}\simeq (V^*_{qL})_{3i}(V_{qL})_{3j}$). Indeed, the $K^0-\bar{K}^0$ mixing constrains (\ref{fcncdz}) to be,     
\be |(V^*_{dL})_{I1} (V_{dL})_{I2}| \lesssim 10^{-5}. \ee This mixing of the exotic and ordinary quarks is much smaller than the smallest mixing element (about $5\times 10^{-3}$) of the ordinary quark flavors by the CKM matrix \cite{pdg}. Therefore, the 3-3-1-1 gauge symmetry as well as the resulting $W$-parity provide a more natural framework that not only solves those problems (including the large FCNCs, the unitarity of the CKM matrix, the lepton and baryon number symmetries and the CPT theorem that have strictly been proved by the experiments \cite{pdg}), but also gives the neutrino small masses and the dark matter candidates. 

\subsection{\label{ctsec1} LEPII searches for $Z_2$ and $Z_N$}
 
LEPII searches for new neutral gauge bosons via the channel $e^+e^{-}\rightarrow f \bar{f}$, where $f$ is any ordinary fermion \cite{LEPII}. In this model, the new physics effect in such process is due to the dominant contribution of $Z_2$ and $Z_N$ gauge bosons, which is s-channel exchanges for $f\neq e$. The effective interaction for these contributions can be derived with the help of (\ref{ttva2nd}) as 
\be \mathcal{L}^{\mathrm{eff}}_{\mathrm{LEP2}}=\fr{g^2}{c^2_W m^2_{I}}[\bar{e}\ga^\mu(a^I_L(e)P_L+a^I_R(e)P_R)e][\bar{f}\ga_\mu(a^I_L(f)P_L+a^I_R(f)P_R)f] \hs (I=Z_2,\ Z_N), \ee
where the chiral couplings are given by 
\be a^I_L(f)=\fr{g^I_V(f)+g^I_A(f)}{2},\hs a^I_R(f)=\fr{g^I_V(f)-g^I_A(f)}{2}.\ee 

Let us study a particular process for $f=\mu$, $e^+e^-\rightarrow \mu^+\mu^-$. The chiral couplings can be obtained from Table \ref{tttz2} and \ref{tttzn} as 
\bea a^{Z_2}_L(e_a) &=& \fr{c_\xi c_{2W}}{2\sqrt{3-4s^2_W}}-\fr 2 3 s_\xi c_W t_N, \hs a^{Z_2}_R(e_a)= -\fr{c_\xi s^2_W}{\sqrt{3-4s^2_W}}-s_\xi c_W t_N,\crn a^{Z_N}_{L,R}&=&a^{Z_2}_{L,R}(c_\xi\rightarrow -s_\xi;s_\xi\rightarrow c_\xi). \eea The effective interaction can be rewritten by 
\be \mathcal{L}^{\mathrm{eff}}_{\mathrm{LEP2}}= \fr{g^2}{c^2_W}\left(\fr{[a^{Z_2}_L(e)]^2}{m^2_{Z_2}}+ \fr{[a^{Z_N}_L(e)]^2}{m^2_{Z_N}}  \right)(\bar{e}\ga^\mu P_L e) (\bar{\mu}\ga_\mu P_L \mu) + (LR)+(RL)+(RR), \ee where the last three terms differ the first one only in chiral structures. 

Notice that LEPII searches for such chiral interactions and gives several constrains on the respective couplings, which are commonly given in the order of a few TeV \cite{LEPII}. Therefore, let us choose a typical value \be \fr{g^2}{c^2_W}\left(\fr{[a^{Z_2}_L(e)]^2}{m^2_{Z_2}}+ \fr{[a^{Z_N}_L(e)]^2}{m^2_{Z_N}}  \right) < \fr{1}{(6\ \mathrm{TeV})^2}.\ee It is noted that this value, $6\ \mathrm{TeV}$, is also a bound derived for the case of $U(1)_{B-L}$ gauge boson~\cite{carena}.          

Similarly to the previous subsection, we suppose that $Z_2$ and $Z_N$ have approximately equal masses ($m_{Z_2}\approx m_{Z_N}$) and $t_N=1$. The above constraint leads to 
\be m_{Z_2}\approx m_{Z_N} > 2.737\ \mathrm{TeV}. \ee  This bound is in good agreement with the limit in the previous subsection via the FCNC and the ones given in the literature \cite{masszp}. As we previously emphasized, in the dark matter subsections we have adopted $m_{Z_N}\gg m_{Z_2}$ and therefore in this regime a bound in $m_{Z_2}\sim$~TeV rises. 

Finally, let us discuss the contribution of the mixing effects of the neutral gauge bosons to the above process. When the mixing is turned on, the interacting Lagrangian of the neutral gauge bosons takes the form, $-\fr{g}{c_W}\bar{f}\ga^\mu[\tilde{a}^{Z_i}_L(f)P_L+\tilde{a}^{Z_i}_R(f)P_R]f Z_{i\mu}$, where $i=1,2,N$ and the (chiral) couplings of the neutral gauge bosons are correspondingly changed as follows 
\bea a^{Z}_{L,R}(f)&\longrightarrow& \tilde{a}^{Z_1}_{L,R}(f)\equiv a^{Z}_{L,R}(f)+a^{Z_2}_{L,R}(f)(c_\xi\rightarrow -\mathcal{E}_1;s_\xi \rightarrow -\mathcal{E}_2),\crn
a^{Z_2}_{L,R}(f)&\longrightarrow& \tilde{a}^{Z_2}_{L,R}(f)\equiv a^{Z_2}_{L,R}(f)+a^{Z}_{L,R}(f)\times (\mathcal{E}_1 c_\xi+ \mathcal{E}_2 s_\xi),\\
a^{Z_N}_{L,R}(f)&\longrightarrow& \tilde{a}^{Z_N}_{L,R}(f)\equiv a^{Z_N}_{L,R}(f)+a^{Z}_{L,R}(f)\times (-\mathcal{E}_1 s_\xi+ \mathcal{E}_2 c_\xi).\nn
\eea We realize that all the second terms are the $\mathcal{E}_{1,2}$ corrections corresponding to the existing couplings due to the mixing, which can be neglected because of the so small $\mathcal{E}_{1,2}$ values as given in (\ref{e1e2sosanh}). Indeed, for the concerned process $e^+e^-\rightarrow \mu^+\mu^-$, let us consider the ratios of the corrections to the respective existing couplings for $f=e_a$ (the charged leptons). With the $Z_1$ couplings, we have 
\bea && \left|\fr{a^{Z_2}_{L}(e_a)(c_\xi\rightarrow -\mathcal{E}_1;s_\xi \rightarrow -\mathcal{E}_2)}{a^{Z}_{L}(e_a)}\right|=\left|\fr{\mathcal{E}_1}{\sqrt{3-4s^2_W}}-\fr{4c_W t_N}{3c_{2W}}\mathcal{E}_2\right| < 2.43\times 10^{-3},\\     
 &&\left| \fr{a^{Z_2}_{R}(e_a)(c_\xi\rightarrow -\mathcal{E}_1;s_\xi \rightarrow -\mathcal{E}_2)}{a^{Z}_{R}(e_a)}\right|=\left| \fr{\mathcal{E}_1}{\sqrt{3-4s^2_W}}+\fr{c_W t_N}{s^2_W}\mathcal{E}_2\right| < 2.43\times 10^{-3}, \eea which are easily obtained with the help of (\ref{e1e2sosanh}), $s^2_W=0.231$ and $\La \sim \om>3.198$ TeV. Similarly, for the $Z_2$ couplings, we have 
\bea && \left| \fr{a^{Z}_{L}(e_a)\times (\mathcal{E}_1 c_\xi+ \mathcal{E}_2 s_\xi)}{a^{Z_2}_L(e_a)}  \right|=\left|\fr{\mathcal{E}_1 c_\xi + \mathcal{E}_2 s_\xi}{\fr{c_\xi}{\sqrt{3-4s^2_W}}-\fr{4c_W}{3c_{2W}}t_N s_\xi}\right|<5.04\times 10^{-3},\\
&& \left| \fr{a^{Z}_{R}(e_a)\times (\mathcal{E}_1 c_\xi+ \mathcal{E}_2 s_\xi)}{a^{Z_2}_R(e_a)}  \right|=\left|\fr{\mathcal{E}_1 c_\xi + \mathcal{E}_2 s_\xi}{\fr{c_\xi}{\sqrt{3-4s^2_W}}+\fr{c_W}{s^2_W}t_N s_\xi}\right|<5.04\times 10^{-3},\eea where notice that the mixing angle of the $Z'$, $C$ gauge bosons is bounded by $-\pi/4<\xi<0$ if $t_N>0$ either $0<\xi<\pi/4$ if $t_N<0$. The corrections to the $Z_N$ couplings are so small too. Therefore, the mixing effects of the neutral gauge bosons do not affect to the standard model $e^+e^-\rightarrow \mu^+\mu^-$ process as well as our results given above with the $Z_{2,N}$ exchanges in the absence of the mixing.            

\subsection{Radiative $\beta$ decays involving $Z_{2,N}$ and the violation of CKM unitarity}
The CKM unitarity implies $\sum_{d'=d,s,b}V^*_{u'd'}V_{u''d'} =\de_{u'u''}$ and $\sum_{u'=u,c,t}V^*_{u'd'}V_{u'd''}=\de_{d'd''}$, where the elements of the CKM matrix $V_{u'd'}\equiv (V^{\dagger}_{uL}V_{dL})_{u'd'}$ ($u'=u,c,t$ and $d'=d,s,b$) are defined as before. The standard model calculations have provided a very good agreement with the above relations \cite{pdg}. However, if there is a possible deviation, it is the sign for the violation of the CKM unitarity. Taking for the first row, the experimental bound yields \cite{pdg} 
\be \Delta_{\mathrm{CKM}}=1-\sum_{d'=d,s,b}|V_{ud'}|^2 <10^{-3}. \label{ckmubo}\ee This violation can give the constraints on the new neutral $Z_{2,N}$ gauge bosons as a result of their loop effects that contribute to $\Delta_{\mathrm{CKM}}$. 

Indeed, the $\Delta_{\mathrm{CKM}}$ deviation is derived from the one-loop radiative corrections via the new $Z_{2,N}$ and $W$ bosons to quark $\beta$ decay amplitudes from which the $V_{ud}$, $V_{us}$ and $V_{ub}$ elements are extracted, including muon decay which normalizes the quark $\beta$ decay amplitudes. These have previously been studied in other theories \cite{haisch-sirlin} with the respective diagrams to quark and muon $\beta$ decays similarly displayed therein. Generalizing the results in \cite{haisch-sirlin}, the deviation is obtained as 
\be \Delta_{\mathrm{CKM}}\simeq -\fr{3}{4\pi^2}\sum_{I=Z_2,Z_N}\fr{m^2_W}{m^2_I}\ln\left(\fr{m^2_W}{m^2_I}\right)(\mathcal{G}^I_{e_L})_{11}
\left[(\mathcal{G}^I_{e_L})_{11}-\fr{(\mathcal{G}^I_{d_L})_{11}+(\mathcal{G}^I_{u_L})_{11}}{2}\right], \ee where the lepton and quark couplings are given in the physical basis of the left chiral fields when coupled to $Z_{2,N}$, i.e. $\bar{f}'_L\ga^\mu \mathcal{G}^I_{f_L} f'_L I_\mu$ with $\mathcal{G}^I_{f_L}\equiv -\fr{g}{c_W}V^\dagger_{fL} a^I_L(f)V_{fL}$, that results   
\bea &&(\mathcal{G}^I_{e_L})_{11}=(\mathcal{G}^I_{\nu_L})_{11}=-\fr{g}{c_W}a^{I}_{L}(e_a),\hs (\mathcal{G}^{Z_2}_{u_L})_{11}=\fr{gc_\xi\sqrt{3-4s^2_W}}{6c_W},\hs (\mathcal{G}^{Z_N}_{u_L})_{11}=\fr{-gs_\xi\sqrt{3-4s^2_W}}{6c_W},\crn
&& (\mathcal{G}^{Z_2}_{d_L})_{11}=\fr{gc_\xi\sqrt{3-4s^2_W}}{6c_W}-\fr{g}{c_W}\left(\fr{c_\xi c^2_W}{\sqrt{3-4s^2_W}}+\fr 2 3 s_\xi c_W t_N\right)|(V_{dL})_{31}|^2,\crn
&&(\mathcal{G}^{Z_N}_{d_L})_{11}=(\mathcal{G}^{Z_2}_{d_L})_{11}(c_\xi\rightarrow -s_\xi;s_\xi\rightarrow c_\xi). \eea Notice that the mixing effect of the neutral gauge bosons ($Z$ with $Z_{2,N}$) do not affect to the considering processes as explicitly pointed out in the previous subsection. 

Therefore, we have 
\bea \Delta_{\mathrm{CKM}} &\simeq& -\fr{3g^2}{4\pi^2}\fr{m^2_W}{m^2_{Z_2}}\ln\left(\fr{m^2_W}{m^2_{Z_2}}\right)\left[\fr 2 3 s_\xi t_N-\fr{c_\xi c_{2W}}{2c_W\sqrt{3-4s^2_W}}\right]\left[\fr 2 3 s_\xi t_N-\fr{c_\xi(3-5s^2_W)}{3c_W\sqrt{3-4s^2_W}}\right]\crn
&&+(Z_2\rightarrow Z_N;c_\xi\rightarrow -s_\xi;s_\xi \rightarrow c_\xi).\eea     
We consider two typical cases, $\La\gg \om$ and $\La \sim \om$. In the first case, the $Z_N$ does not contribute, i.e. the second term above vanishes, and $\xi=0$. Therefore, this is the case of the 3-3-1 model with right-handed neutrinos. We have 
\bea \Delta_{\mathrm{CKM}}\simeq -0.0033 \fr{m^2_W}{m^2_{Z_2}}\ln \left(\fr{m^2_W}{m^2_{Z_2}}\right).\eea Using the bound (\ref{ckmubo}) and   $m_W=80.4$ GeV, the $Z_2$ mass is constrained by $m_{Z_2}>200$ GeV. In fact, the $Z_2$ mass should be in TeV range due to the other constraints as given above. For example, taking $m_{Z_2}>1$ TeV, we get $\Delta_{\mathrm{CKM}}<10^{-4}$. Consequently, this case gives a very small contribution to the violation of the CKM unitarity and thus the model is easily to evade the experimental bound. In the second case, assuming that the new neutral gauge bosons have approximately equal masses ($m_{Z_2}\simeq m_{Z_N}$) and $t_N=1$, we derive 
\be \Delta_{\mathrm{CKM}}\simeq -0.0143 \fr{m^2_W}{m^2_{Z_{2,N}}}\ln \left(\fr{m^2_W}{m^2_{Z_{2,N}}}\right). \ee        
Using the bound (\ref{ckmubo}) we have $m^2_{Z_2}\simeq m^2_{Z_N}> 600$ GeV. The model in this case is easily to evade the experimental bound too. To conclude, the new neutral gauge bosons $Z_{2,N}$ give the negligible contribution to the violation of the CKM unitarity.  

\section{\label{concl}Discussion and conclusion}

In the standard model, the fermions come in generations, the subsequent generation is a replication of the former. The gauge anomaly is cancelled out over every generation. Thus, on this theoretical ground the number of the generations can be left arbitrarily. This may be due to the fact that the $SU(2)_L$ anomaly trivially vanishes for any chiral fermion representation. If the $SU(2)_L$ is minimally extended to $SU(3)_L$ with a corresponding enlargement of the lepton and quark representations (the doublets enlarged to triplets/antitriplets while the singlets retain, but for some cases the lepton singlets are put in the corresponding triplets/antitriplets as well), the new $SU(3)_L$ anomaly generally does not vanish for each nontrivial representation. Subsequently, this constrains the generation number to be an integer multiple of three---the fundamental color number---in order to cancel that anomaly over the total fermion content, which provides a partial solution to the number of the generation paradigms. Besides this feature, some very fundamental aspects of the standard model can also be understood by the presence of the $SU(3)_L$ that causes the electric charge quantization \cite{ecq}, the Peccei-Quinn like symmetry for the strong $CP$ \cite{palp} and the oddly-heavy top-quark \cite{tquark}. On the other hand, the $B-L$ number and $Q$ electric charge operators do not commute and also nonclose algebraically with the $SU(3)_L$ generators. Supposing that the $B-L$ is conserved similarly to the $Q$, such $SU(3)_L$ theory is only manifest if it includes two extra Abelian factors so that all the algebras are closed, and the resulting gauge symmetry $SU(3)_L\otimes U(1)_X\otimes U(1)_N$ yields an unification of the weak, electromagnetic and $B-L$ interactions (apart from the strong interaction by the $SU(3)_C$ gauge group). Besides the $B$, $L$ symmetries, some very fundamental matters of the 3-3-1 model can also be understood by this setup. 

Firstly, the breakdown of the 3-3-1-1 gauge symmetry produces a conserved $Z_2$ subgroup (as a remnant) named the $W$-parity similar to the $R$-parity in supersymmetry that plays an important role as well as yielding insights in the present model. The lightest wrong-lepton particle is stabilized due to the $W$-parity conservation, which is responsible for dark matter. The two dark matter particles have been recognized, a neutral complex scalar $H'$ and a neutral fermion $N$ of either Dirac or Majorana nature. The GIM mechanism for the standard model currents works in this model due to the $W$-parity conservation, while the new FCNCs are strictly suppressed. In fact, the experimental bounds can be easily evaded with the expected masses for the new neutral gauge bosons $Z_{2,N}$ in a few TeV. Because of the $W$-parity conservation, the new neutral non-Hermitian gauge boson $X$ does not mix with the neutral $Z_{1,2,N}$ gauge bosons. Hence, there is no mass splitting within the real and imaginary components of the $X$ that ensures the conservation of $CPT$ symmetry. Those problems of the 3-3-1 model with right-handed neutrinos have been solved.   

We have shown that the $B-L$ interactions can coexist with the new 3-3-1 interactions at the TeV scale. To realize this, the scales of the 3-3-1-1 and 3-3-1 breakings are taken to lie in the same energy scale $\La \sim \om$. In this regime, the scalar potential has been diagonalized. The number of Goldstone bosons matches the number of the massive gauge bosons. There are eleven physical scalar fields, one of them is identified as the standard model Higgs boson. The new physical scalar fields $H^0_{1,2,3}$, $\mathcal{A}^0$, $H^\pm_{4,5}$, and $H'^{0,0*}$ are heavy with their masses in the $\om$, $\La$ or $\sqrt{|\om f|}$ scales. There is a finite mixing between the Higgs scalars---the $S_4$ for the $U(1)_N$ breaking and the $S_3$ for the 3-3-1 breaking---that results two physical fields the $H_{2,3}$. The standard model Higgs boson is light with a mass given in the weak scale due to the seesaw-type mechanism associated with the little hierarchy $u,v\ll \om,\La,-f$. The Higgs mass gets a right value of $125$ GeV provided that the effective coupling $\bar{\la}\simeq 0.5$ with the assumption $u=v$, $\om=-f$. All the physical scalar fields are $W$-even except for the $H'$ and $H_4$ that are $W$-odd, known as the $W$-particles. 

In the proposed regime $\La\sim \om$, the gauge sector has been diagonalized with a recognition of the standard model gauge bosons $W^\pm$, $A$ and $Z$. Moreover, we have six new gauge bosons $X^{0,0*}$, $Y^\pm$, $Z_{2,N}$. Although the $Z$ boson mixes with the new neutral gauge bosons, it is realized to be light due to a seesaw-type mechanism in the gauge sector. In order to reproduce the standard model $W$ boson mass, we have constrained $u^2+v^2=(246\ \mathrm{GeV})^2$. From the experimental bound on the rho parameter, we get $\om > 3.198$ TeV provided that $\La \simeq \om$ and $u\simeq v$. There is a finite mixing between the $U(1)_N$ gauge boson and the $Z'$ of the 3-3-1 model that produces two physical states by the diagonalization  as the 3-3-1 like gauge boson $Z_{2}$ and the $U(1)_N$ like gauge boson $Z_N$. All the gauge bosons are $W$-even except for the $X,\ Y$ that are the $W$-particles. The new neutral complex gauge boson $X$ cannot be a dark matter because it entirely annihilates into the standard model particles before the thermal equilibrium process ended \cite{3311}. 

All the interactions of the gauge bosons with the fermions and scalars have been obtained. The result yields that every interaction conserves the $W$-parity. The corresponding standard model interactions are recovered. The new interactions as well as their implication to the new physics phenomenological processes are rich to be devoted to further studies. In this work, some of them have particularly been used for analyzing the new FCNCs, the LEPII collider, the violation of the CKM unitarity, and the fermionic dark matter observables. Because of the seesaw-type mixing suppression between the light and heavy states, namely between the $Z$ and new $Z_{2,N}$ gauge bosons as well as between the $H$ and new $H_{1,2,3}$ Higgs bosons, the mixing effects are radically small. The new physics effects via those mixings in the gauge sector have explicitly been pointed out to be safely neglected. For the scalar sector, the new physics effects via those mixings are also negligible as disregarded for the most cases of the small scalar self-couplings (see the text in more detail). Only if the scalar self-couplings are more strong, they may give considerable contributions but are still in the current bounds. The accuracy of the standard model Higgs mechanism if it is the case could give some constraints on those mixing effects.                  

Supposing that the scalar dark matter $H'$ dominantly annihilates into the standard model Higgs boson $H$ via the Higgs portal, the relic density of $H'$ has been calculated. It gets the right value compatibly to the experiment data if $m_{H'}=1.328$ TeV assumed that the $H'^* H'\rightarrow HH$ coupling equals to unity, $\la'=1$.  As for the neutral fermion candidate as a Dirac particle we conclude that a $\om$ scale of the symmetry breaking greater than $\sim 5$~TeV is required in order to obey the LUX2013 bounds. Whereas when the neutral fermion is a Majorana particle, the direct detection bounds are quite loose and a larger region of the parameter space has been found that yields the right abundance. The fermion dark matter observables are governed by the $Z_2$ gauge boson provided that $\La > \om$. Only if $g_N\ll g$ with $\La\sim \om$ either the $\La$ is rare smaller than the $\om$ with $g_N\sim g$, the $Z_N$ contribution becomes comparable to that of the $Z_2$ boson.         

We have shown that the CKM matrix is unitary as well as the ordinary GIM mechanism of the standard model works in this model, due to the $W$-parity conservation. We have also discussed that this mechanism does not work in the 3-3-1 model with right-handed neutrinos, and in such case the tree-level FCNCs due to the ordinary and exotic quark mixing are more dangerous than those coming from the non-universal couplings of the $Z_{2,N}$ gauge bosons. All the FCNCs associated with the $Z$ boson due to the above fermion mixing are prevented because of the $W$-parity conservation. Also, the new FCNCs coupled to the $Z_{2,N}$ are highly suppressed too. In fact, the FCNCs due to the $Z_{2,N}$ can present but they can be easily evaded by the new physics in the TeV range. Using the current bound on the $K^0-\bar{K}^0$ system, we have shown $m_{Z_{2,N}}>2.037$ TeV under the assumption that the $Z_2$ and $Z_N$ have approximately equal masses as well as $t_N=1$ (the $B-L$ interaction strength equals to that of the weak interaction). For the $B^0_s-\bar{B}^0_s$ system, the bound is $m_{Z_{2,N}}>2.291$ under the same assumptions as the previous case. For hierarchical masses of $Z_{2}$ and $Z_N$, the smaller mass will take a smaller bound, e.g $m_{Z_2}> g_2\times 2$ TeV corresponding to the $K^0-\bar{K}^0$ system, where $g_2$ is the reduced gauge coupling that has a natural value smaller than unity.   

The new neutral currents in the model are now under the experimental detections. We have calculated the contributions of $Z_2$ and $Z_N$, which dominate the corrections of the new physics, to the process $e^+e^-\rightarrow \mu^+\mu^-$ at the LEPII collider. From the experimental bounds, we have shown that $m_{Z_{2,N}}>2.737$ TeV provided that these gauge bosons have approximately equal masses and $t_N=1$. Similarly, for the hierarchal $Z_2$ and $Z_N$ masses, the smaller mass will possess a smaller bound than the above result. Moreover, we have also indicated that the violation of the CKM unitarity due to the one-loop effects of the new neutral gauge bosons $Z_{2,N}$ are negligible if the $Z_{2,N}$ masses are given in the TeV range as expected.  

Finally, the 3-3-1-1 model that unifies the electroweak and $B-L$ interactions along with the strong interaction is a self-consistent extension of the standard model that solves the potential problems of the 3-3-1 model in the consistency with the $B,\ L$, and $CPT$ symmetries as well as curing the large FCNCs. The new physics of the 3-3-1-1 model is interesting with the outcomes in the TeV region. For all the reasons aforementioned, we believe that the 3-3-1-1 model is a compelling theory which is called for much experimental attention.

\section*{Acknowledgments}

This research is funded by Vietnam National Foundation for Science and Technology Development (NAFOSTED) under grant number 103.01-2013.43. FSQ is partly supported by the Department of Energy of the United States award SC0010107 and the Brazilian National Counsel for Technological and Scientific Development (CNPq). P.V.D would like to thank Luong Thi Huyen, Nguyen Thi Tuyet Dung, Tran Minh Thu, and Nguyen Ngoc Diem Chi at Can Tho University for their assistance in calculations of Sections III, IV and V of this work.

\end{document}